\documentclass[12pt,twoside,a4paper,dvips]{thesis-tam}
\usepackage{fancyhdr,psfig}
\usepackage{graphicx}
\usepackage{float,alltt,eepic}
\usepackage{hyperref}

\usepackage{natbib}
\bibpunct{(}{)}{;}{a}{}{,}
\newcommand{\clearemptydoublepage}{\newpage{\pagestyle{empty}\cleardoublepage}}

\graphicspath{{figs/}}

\defcitealias{WebbJ_99a}{W99}
\defcitealias{goldhaber01}{G01}

\def \apj{ApJ}

\def \mnras{MNRAS}

\def \prd{Phys. Rev. D}

\def \qjras{Quart. J. R. Astron. Soc.}

\def \cup{Cambridge U. Press}
\def \cupadr{Cambridge U.K.}
\def \pup{Princeton U. Press}
\def \pupadr{Princeton, U.S.A.}

\def \etal{et al.}

\newcommand{\btab}{\begin{tabular}}
\newcommand{\etab}{\end{tabular}}
\newcommand{\btg}{\begin{tabbing}}
\newcommand{\etg}{\end{tabbing}}
\newcommand{\bctr}{\begin{center}}
\newcommand{\ectr}{\end{center}}

\newcommand{\bpic}{\begin{picture}}
\newcommand{\epic}{\end{picture}}

\newcommand{\bdm}{\begin{displaymath}}
\newcommand{\edm}{\end{displaymath}}
\newcommand{\beq}{\begin{equation}}
\newcommand{\eeq}{\end{equation}}
\newcommand{\bea}{\begin{eqnarray}}
\newcommand{\eea}{\end{eqnarray}}
\newcommand{\ben}{\begin{enumerate}}
\newcommand{\een}{\end{enumerate}}
\newcommand{\bit}{\begin{itemize}}
\newcommand{\eit}{\end{itemize}}
\newcommand{\nn}{\nonumber}
\newcommand{\bqt}{\begin{quote} \small}
\newcommand{\eqt}{\end{quote} \normalsize}
\newcommand{\bfig}{\begin{figure}}
\newcommand{\efig}{\end{figure}}

\newcommand{\omol}{(\Omega_{\rm M},\Omega_{\Lambda})}
\newcommand{\om}{\Omega_{\rm M}}
\newcommand{\oll}{\Omega_{\Lambda}}

\newcommand{\zrec}{z_{\rm rec}}

\newcommand{\vrec}{v_{\rm rec}}
\newcommand{\vpec}{v_{\rm pec}}

\newcommand{\vp}{v_{\rm pec}}

\newcommand{\lsim}{\mbox{$\:\stackrel{<}{_{\sim}}\:$} }
\newcommand{\gsim}{\mbox{$\:\stackrel{>}{_{\sim}}\:$} }

\newcommand{\incite}[1]{[App.~B: {#1}]} 
\newcommand{\inciteFirst}[1]{[Appendix~B: {#1}]} 
\newcommand{\xcite}[2]{[#1] {#2}\newline} 
\begin{document}

\thispagestyle{empty}

\begin{center}

\hbox{}
\vspace{1cm}

{\Large {\bf Fundamental Aspects of the Expansion of the Universe and Cosmic Horizons}}

\vspace{1.5cm}

by

\Large
Tamara M. Davis

\vspace{2cm}
\large
A thesis submitted in satisfaction of\\
the requirements for the degree of

\Large
{\bf Doctor of Philosophy}

\large
in the Faculty of Science.

\vspace{1cm}

23rd of December, 2003

\vspace{2cm}

\leavevmode
\includegraphics[width=6cm]{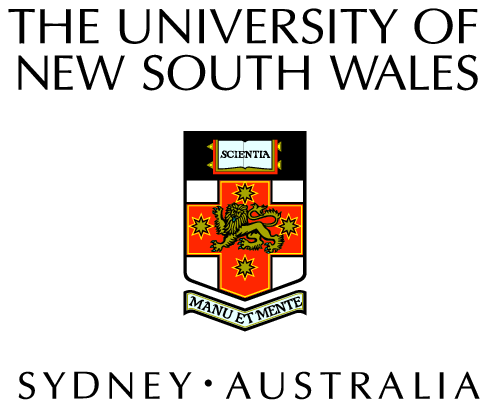}

\end{center}

\clearemptydoublepage

\parskip=0.15cm

\thispagestyle{empty}
\hbox{}
\vspace{2cm}

\begin{center}
{\Large {\bf Statement of Originality}}
\end{center}

\noindent
I hereby declare that this submission is my own work and to the best of my
knowledge it contains no materials previously published or written by
another person, nor material which to a substantial extent has been
accepted for the award of any other degree or diploma at UNSW or any other
educational institution, except where due acknowledgement is made in the
thesis. Any contribution made to the research by others, with whom I have
worked at UNSW or elsewhere, is explicitly acknowledged in the thesis.

I also declare that the intellectual content of this thesis is the product
of my own work, except to the extent that assistance from others in the
project's design and conception or in style, presentation and linguistic
expression is acknowledged.

\vspace{2cm}
\hfill (Signed)\dotfill

\clearemptydoublepage

\thispagestyle{plain}



\bctr
...\vspace{3.5cm}

To Katherine,

who would have made it here herself.
\vspace{1cm}

And to my Grandmother, Dorothy.


\vspace{3.5cm}...
\ectr 

\newpage

\clearemptydoublepage

\pagenumbering{roman}
\thispagestyle{plain}

\addcontentsline{toc}{section}{Abstract}
\markboth{Abstract}{Abstract}

\vspace{-10mm}
\begin{center}
{\Large {\bf Abstract}}
\end{center}

\vspace{-3mm}\noindent
We use standard general relativity to clarify common misconceptions about fundamental aspects of the expansion of the Universe.  
In the context of the new standard $\Lambda$CDM cosmology we resolve conflicts in the literature regarding cosmic horizons and the Hubble sphere (distance at which recession velocity $= c$) 
and we link these concepts to observational tests.   
We derive the dynamics of a non-comoving galaxy and generalize previous analyses to arbitrary FRW universes. 
We also derive the  counter-intuitive result that objects at constant proper distance have a non-zero redshift.  Receding galaxies can be
blueshifted and approaching galaxies can be redshifted, even in an empty universe for which one might expect special relativity to apply.  Using the empty universe model we demonstrate the relationship between special relativity and Friedmann-Robertson-Walker cosmology.

We test the generalized second law of thermodynamics (GSL) and its extension to incorporate cosmological event horizons.  In spite of the fact that cosmological horizons do not generally have well-defined thermal properties, we find that the GSL is satisfied for a wide range of models.   
We explore in particular the relative entropic `worth' of black hole versus cosmological horizon area.
An intriguing set of models show an apparent entropy decrease but we anticipate this apparent violation of the GSL will disappear when solutions are available for black holes embedded in arbitrary backgrounds.

Recent evidence suggests a slow increase in the fine structure constant $\alpha =e^2/\hbar c\;$ over cosmological time scales.  This raises the question of which fundamental quantities are truly constant and which might vary.  We show that black hole thermodynamics may provide a means to discriminate between alternative theories invoking varying constants, because some variations in the fundamental `constants' could lead to a violation of the generalized second law of thermodynamics.

\clearemptydoublepage

\parskip=0.00cm

\tableofcontents
\clearemptydoublepage


\listoffigures
\addcontentsline{toc}{section}{List of figures}
\clearemptydoublepage

\parskip=0.15cm

\thispagestyle{plain}

\addcontentsline{toc}{section}{Acknowledgments}
\markboth{Acknowledgments}{Acknowledgments}

\begin{center}
{\Large {\bf Acknowledgments}}
\end{center}

\noindent

First and foremost it is my pleasure to thank my supervisor, Charles Lineweaver, for all his support and enthusiasm over the course of my PhD.  Charley is an inspirational teacher and tireless campaigner for the cause of scientific rigour.  He has been a fantastic supervisor, who maintains an overwhelming energy throughout his work and carries that through to all around him.  I am very fortunate to have had the chance to work with him so closely.

I have also been very fortunate during my PhD to have the chance to tap the fountain of knowledge that is Paul Davies.  Paul became my unofficial supervisor half way through my PhD research and his input immeasurably strengthened my work.  For all his support and interest I am profoundly grateful.

The key attribute that links all my supervisors is their insight -- their ability to take diverse areas of knowledge and discover how they are linked in order to solve interesting unanswered questions.  My final supervisor, John Webb, exemplifies this skill brilliantly.  It has been enlightening to work with him as exciting project after exciting project appeared.  Thank you for your faith in me.

There are many others who have generously contributed their time and knowledge to my research through informal discussions and correspondence.  
 I am very grateful to John Barrow, Geraint Lewis, Jochen Liske, Jo\~ao Magueijo, Hugh Murdoch, Michael Murphy and Brian Schmidt for many informative discussions.  Although I never met them face to face my correspondences with Tao Kiang, John Peacock, Edward Harrison, Phillip Helbig, and Frank Tipler were invaluable and I thank them all.

For putting food on the table I thank UNSW and the Department of Education, Science and Technology for an Australian Post-graduate Award.  I also gratefully thank UNSW for their support through a Faculty Research Grant.  For various travel bursaries I appreciate the contributions of the University of New South Wales Department of Physics, the University of Michigan and the Templeton Foundation.  For sporting scholarships and travel funding I thank the UNSW Sports Association.

The Department of Astrophysics at the University of New South Wales has been a supportive and enjoyable place to work.  Thanks goes to all my friends and colleagues there, it has been a blast.  Thanks to Jess, Jill and Melinda for their friendship and many a laksa lunch.  I appreciate their efforts in the direction of maintaining my sanity (though they may doubt their success).  I have had a happy time with all my officemates -- thanks to Cormac, Steve, Jess and Michael for putting up with me.  A very special thanks has to go to Melinda for her computer wizardry, and to Michael for continuing the hand-me-down thesis template.
 To all my UNSW colleagues, past and present, all the best for the future.  

Thanks too has to go to the Ultimate crew, for providing me with such a healthy diversion.  

Most importantly I owe a great debt of gratitude to my family for their love and support throughout my schooling, and for encouraging (and teaching) me to excel in every aspect of life, not just academia.  According to Denis Waitley, ``The greatest gifts you can give your children are the roots of responsibility and the wings of independence.''  Thank you for both.

Finally thank you to Piers for your constant support and confidence in me, and for making every day a joy.

\clearemptydoublepage

\thispagestyle{plain}

\addcontentsline{toc}{section}{Preface}
\markboth{Preface}{Preface}

\begin{center}
{\Large {\bf Preface}}
\end{center}

\noindent

The material in this thesis comes from research I have had published over the course of my PhD.  Each Chapter is loosely based on the publications as follows:

\begin{itemize}
\item{{\bf Davis and Lineweaver, 2001}, ``Superluminal recession velocities'', (AIP conference proceedings, 555, New York, p.~348), provides some background for Chapter 1.  
}
\item{{\bf Davis and Lineweaver, 2004}, ``Expanding confusion: common misconceptions of cosmological horizons and the superluminal expansion of the Universe'', (Publications of the Astronomical Society of Australia, in press), forms the basis of Chapter 2.}
\item{{\bf Davis, Lineweaver and Webb, 2003}, ``Solutions to the tethered galaxy problem in an expanding Universe and the observation of receding blueshifted objects'',  (American Journal of Physics, {\bf 71}, 358), forms the basis of Chapter 3. } 
\item{{\bf Davis, Davies and Lineweaver, 2003}, ``Black hole versus cosmological horizon entropy'', (Classical and Quantum Gravity, {\bf 20}, 2753), along with,}
\item{{\bf Davies and Davis, 2002}, ``How far can the generalized second law be generalized'', (Foundations of Physics, {\bf 32}, 1877), form the basis of Chapter 5.}
\item{{\bf Davies, Davis and Lineweaver, 2002}, ``Black holes constrain varying constants'', (Nature, {\bf 418}, 602), forms the basis of Chapter 6.}
\end{itemize}

Although I have used the first person plural throughout, the work presented in this thesis is my own.  The work was largely carried out in close contact between myself and one or both of my two primary collaborators, Charles H. Lineweaver and Paul C. W. Davies.  They are both fountains of ideas, and much of the work I completed arose from investigating their insights.

Charles Lineweaver initiated my investigation into the topics in Part I when he asked ``Can recession velocities exceed the speed of light?''.  This evolved into a research program to elucidate some of the common misconceptions that surround the expansion of the Universe, and turned up some suprising new implications of the general relativistic picture of cosmology.   In this research I was also supported by John Webb.

Paul Davies offered his vast knowledge of event horizon physics, initiating in turn the bulk of Part II.  Paul is largely responsible for the theoretical derivations of the ``Small departures from de Sitter space'' criterion in Sections~\ref{sect:rad-small-dep} and~\ref{sect:bh-small-dep}.  It was Paul's initial insight that led to our investigation, and the subsequent lively debate, of how black hole event horizons may provide constraints on varying contants.  For this we are also endebted to the pioneering observational work of John Webb, Michael Murphy and collaborators.  Sharing an office with Michael and watching the observational data supporting a variation in the fine structure constant unfold, undoubtably fueled my interest in this area.

During the course of my PhD I was also involved in another line of research, in the field of Astrobiology, that does not appear in this thesis.  Papers resulting from this work are:
\begin{itemize}
\item{{\bf Lineweaver and Davis, 2002}\nocite{lineweaver02rapidlife}, ``Does the rapid appearance of life on Earth suggest that life is common in the Universe?'', (Astrobiology, {\bf 2}, 293).  }
\item{{\bf Lineweaver and Davis, 2003}\nocite{lineweaver03recentbio}, ``On the non-observability of recent biogenesis'', (Astrobiology, {\bf 3}, 241). }

\end{itemize}

Throughout this thesis I endeavour to lay credit where credit is due, and provide the background references demonstrating the work of the giants upon whose shoulders we stand.

\clearemptydoublepage

\pagenumbering{arabic}

\part{Detailed Examination of Cosmological Expansion}

\begin{flushright}
\begin{minipage}{5cm}
Most people prefer certainty to truth.
\begin{flushright}
{\it Fortune cookie}
\end{flushright}
\end{minipage}
\end{flushright}

\chapter{Introduction}\label{chap:intro}

\vspace{5mm}
The Big Bang model and the expansion of the Universe are now well established.   Yet there remain many fundamental points that are still under examination.  It is less than a decade ago that the first evidence arrived suggesting the Universe is accelerating, and the nature of the dark energy remains uncertain.   The many unanswered questions, and the precision observational tools emerging to study them, make modern cosmology an exciting and vibrant field of study.

This  thesis has two parts.  Throughout we follow the theme of achieving a better understanding of the expansion of the Universe and cosmic horizons.  
Part I addresses a variety of fundamental questions regarding the  expansion of the Universe, recession velocities and the extent of our observable Universe.  We resolve some key conflicts in the literature, derive some counter-intuitive new results and link theoretical concepts to observational tests.
Part II looks into the details of horizon entropy.  Firstly, we examine horizon entropy in the cosmological context, and test the generalized second law of thermodynamics as it applies to the cosmological event horizon.  Secondly, we assess whether black hole thermodynamics can be used to place any constraints on theories in which the constants of nature vary.
 
Part I begins with an analysis of conflicting views in the literature regarding the Big Bang model of the Universe.  This analysis reveals a wide range of misconceptions, the most important of which we discuss in Chapter~\ref{chap:misconcept}.  
The misconceptions we clarify appear not only in text books, but also in the scientific literature, and they are often being expressed by the researchers making the most significant advances in modern cosmology\footnote{This view is expressed in \cite{peebles93}, preface: ``The full extent and richness of [the hot big bang model of the expanding Universe] is not as well understood as I think it ought to be, even among those making some of the most stimulating contributions to the flow of ideas.  In part this is because the framework has grown so slowly, over the course of some seven decades, and sometimes in quite erratic ways...''}.  
  These misconceptions can be dangerous, because once a feature has become common knowledge, little thought is put into questioning it. 

Having dealt with several fundamental misconceptions, we use Chapter~\ref{chap:chain} to elucidate the effect of the expansion of the Universe on non-comoving objects.  As a result of this analysis we demonstrate that receding objects can appear blueshifted and approaching objects can appear redshifted.  In general zero velocity does not give zero redshift in the expanding Universe.

Many of the misconceptions and conflicts in the literature arise from misapplications of special relativity (SR) to situations in which general relativity is appropriate.  We therefore spend some time in Chapter~\ref{chap:coord} to detail how SR fits into the general relativistic description of the expansion of the Universe.  
Most importantly we show how special relativistic velocities and the Doppler shift relate to recession velocities and the cosmological redshift.  Many of the aspects discussed are conceptual, so we have included observational consequences of these concepts wherever possible.  In Sect.~\ref{sect:data} we provide a new analysis of supernovae data providing observational evidence against the special relativistic interpretation of cosmological redshifts.  This analysis has only recently become possible thanks to the pioneering observations of the two supernovae teams: the Supernova Cosmology Project and the High-redshift Supernova team.

In Part II the detailed knowledge of the expansion of the Universe developed in Part I is used to further investigate properties of event horizons, and in particular their associated entropy.  In Chapter~\ref{chap:gsl}, we question whether the cosmological event horizon has an entropy proportional to its area, as suggested by  an extension of the generalized second law of thermodynamics. 
We compare the entropic worth of competing event horizons by calculating the trade off in event horizon area as black holes disappear over the cosmological event horizon.  In all but a few cases the total horizon area increases, upholding the generalized second law of thermodynamics.  However, there are cases in which a total entropy decrease occurs.  We believe that this apparent violation of the generalized second law of thermodynamics is a limitation of our current understanding of black holes and will disappear when black hole solutions are available in an arbitrary, evolving background.  We provide analytical solutions for small departures from de Sitter space and use numerical results to investigate a wide range of cosmological models.  Using the same calculation for a radiation filled universe we find that entropy always increases in all models tested.  
In Chapter~\ref{chap:varies} we use black hole entropy to suggest possible constraints on varying constant theories. 

Throughout we assume basic knowledge of the general relativistic description of the expansion of the Universe.  To provide a firm foundation from which to proceed we use the remainder of this chapter to review some of the key results, and to introduce our notation.  To further clarify notation and usage we provide a more detailed mathematical summary in Appendix 1.  

\newpage
\section{Standard general relativistic cosmology}\label{sect:standard}

We assume a homogeneous, isotropic universe and use the standard
Robertson-Walker metric,
\beq ds^2 = -c^2dt^2 + R(t)^2[d\chi^2+S_k^2(\chi)d\psi^2].\label{eq:frwmetric}\eeq
Observers with a constant comoving coordinate, $\chi$, recede with the expansion of the Universe and are known as comoving observers.
The time, $t$, is the proper time of a comoving observer, also known as cosmic time (see Section~\ref{sect:conversion}).   The proper distance, $D = R\chi$, is the distance (along a constant time surface, $dt=0$) between us and a galaxy with comoving coordinate, $\chi$.  This is the distance a series of comoving observers would measure if they each lay their rulers end to end at the same cosmic instant \citep{weinberg72,rindler77}.  The evolution of the scalefactor, $R$, is determined by the rate of expansion, density and composition of the Universe according to Friedmann's equation, Eq.~\ref{eq:dotR}, as summarized in Appendix~\ref{sect:math}.  Friedmann's equation together with the Robertson-Walker metric define Friedmann-Robertson-Walker (FRW) cosmology.
Present day quantities are given the subscript zero.   We use two expressions for the scalefactor.  When denoted by $R$, the scalefactor has dimensions of distance.  The dimensionless scalefactor, normalized to 1 at the present day, is denoted by $a=R/R_0$.  Our analysis centres around the behaviour of the Universe after inflation.  We defer a discussion of inflation to Sect.~\ref{sect:inflationissuperluminal}.

We define total velocity to be the derivative of proper distance with respect to proper time, $v_{\rm tot} = \dot{D}$,
\bea
\dot{D} &=& \dot{R}\chi + R\dot{\chi},\label{eq:dotd}\\
v_{\rm tot}&=& v_{\rm rec} + v_{\rm pec}.
\label{eq:vrecvpec}
\eea
Peculiar velocity, $v_{\rm pec}$, is measured with respect to comoving observers coincident with the object in question.   Peculiar velocity $v_{\rm pec}=R\dot{\chi}$ corresponds to our normal, local notion of velocity and must be less than the speed of light.  
The recession velocity $v_{\rm rec}$ is the velocity of the Hubble flow at proper distance $D$ and can be arbitrarily large \citep{murdoch77,stuckey92,harrison93,kiang97,gudmundsson02}.  With the standard definition of Hubble's constant, $H=\dot{R}/R$, Eq.~\ref{eq:dotd} above gives Hubble's law, $v_{\rm rec}=HD$.

Since this thesis deals frequently with recession velocities and the expansion of the Universe it is worth taking a moment to assess their observational status.  Even though distances are notoriously hard to measure in astronomy, modern cosmology has developed an impressive model of the Universe as an expanding, evolving structure.  That model has been developed through an extensive set of observations, combining to give a consistent picture of the expanding Universe, and ever more precise estimates of its rate of expansion and acceleration.  We can now put error bars of about $\pm 6\%$ on our calculations of distant recession velocities\footnote{Based on the $H_0=71^{+4}_{-3}$ km\,s$^{-1}$Mpc$^{-1}$ accuracy of the Hubble constant quoted by WMAP (Bennett \etal~2003), but neglecting peculiar velocities (whose relative effect diminishes with distance).}.  However, all this has been done without ever measuring a recession velocity directly.  It is not possible to send out a single observer with a stopwatch to watch distant galaxies rush past\footnote{This is not just a limitation of our spaceships, it is an intrinsic limitation because we can not define an extended inertial frame in which both us, and the distant observer, could sit.  The required procedure is an infinite set of infinitesimal observers set up along the line of sight to take a synchronized measurement (\citeauthor{weinberg72}, 1972, p.~415; \citeauthor{rindler77}, 1977, p.~218).  This is therefore a measurement we are not likely to make in the foreseeable future.}.  
Even our indirect distance measures are not yet accurate enough to observe galaxies receding over human timescales.  (Although, in a few hundred years it is likely we will be able to measure a change in redshift, and thus directly measure cosmic acceleration, see Sect.~\ref{sect:futuretests-z}).  So despite the fact that expansion is crucial to our modern conception of the Universe, we have never directly measured a recession velocity.  This does not remove the conceptual utility of the expansion picture, nor the accuracy of the description.  

\begin{figure*}
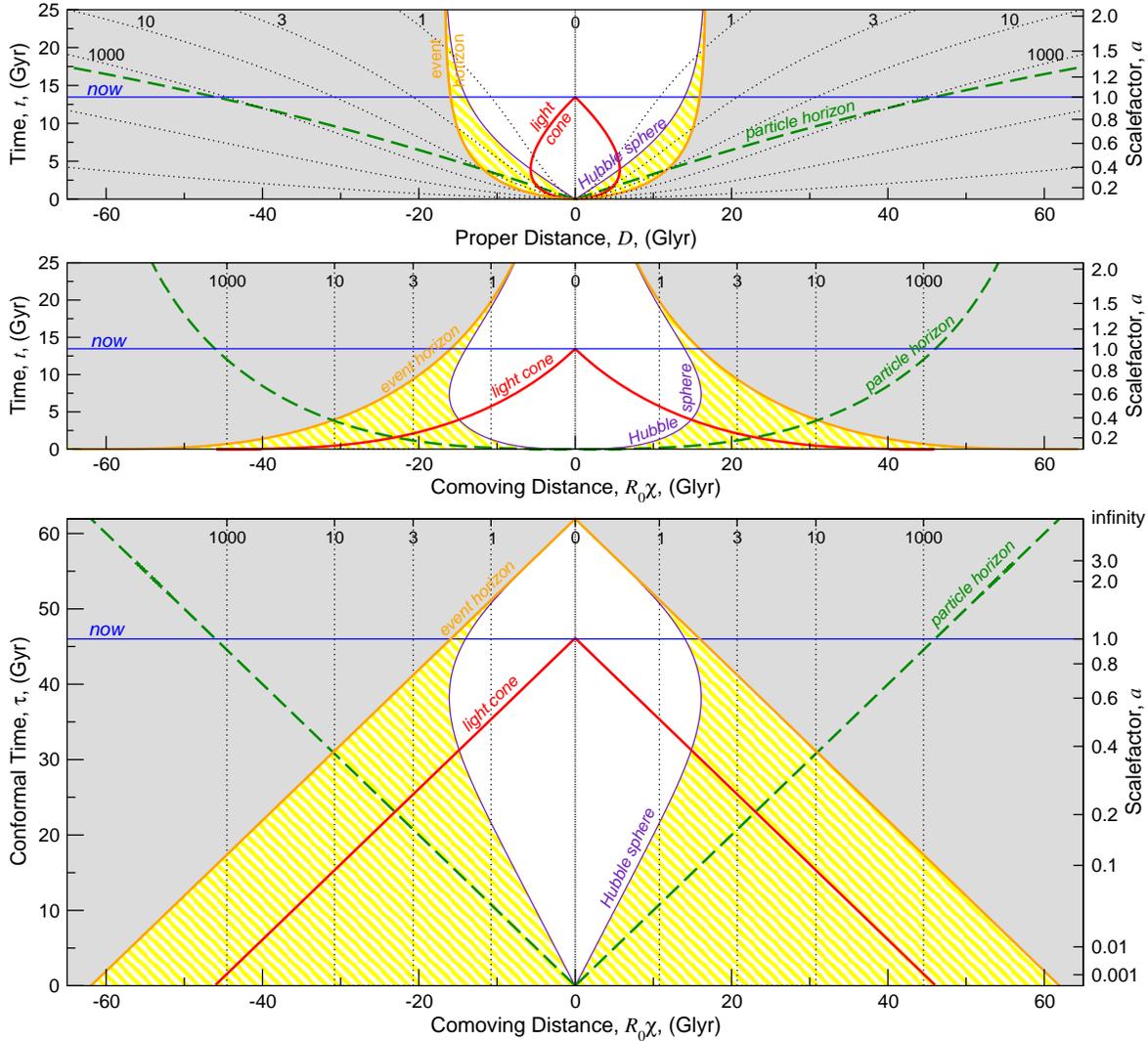
\bctr
\psfig{file=./figs/0-dist0307-colour2-resized3-z-H70top.eps,width=155mm}
\psfig{file=./figs/0-comov0307-colour2-resized3-z-H70top.eps,width=155mm}
\psfig{file=./figs/0-conf0307-colour2-resized-z-H70top.eps,width=155mm}
\caption[Spacetime diagrams for $\omol=(0.3,0.7)$.]{\small{Spacetime diagrams for the $\omol=(0.3,0.7)$ universe with $H_0= 70\,km\; s^{-1} Mpc^{-1}$.  Dotted lines show the worldlines of comoving objects.  The current redshifts of the comoving galaxies shown (Eq.~\ref{eq:z}) appear labeled on each comoving worldline.  The normalized scalefactor, $a=R/R_0$, is drawn as an alternate vertical axis.  
Our comoving coordinate is the central vertical worldline. All events that we currently observe are on our past light cone (cone or ``teardrop'' with apex at $t={\rm now}$, Eq.~\ref{eq:lightconet}).
All comoving objects beyond the Hubble sphere (thin solid line) are receding faster than the speed of light.
%
The speed of photons on our past light cone relative to us (the slope of the light cone) is not constant, but is rather $v_{\rm rec}-c$.  
Photons we receive that were emitted by objects beyond the Hubble sphere were initially receding from us (outward sloping lightcone at $t\lsim 5$ Gyr, upper panel).  {\em Caption continues on next page.}}}
\label{fig:dist}\ectr
\end{figure*}
\newpage\noindent{\em Figure~\ref{fig:dist} caption, continued:}  

\noindent {\small Only when they passed from the region of superluminal recession $v_{\rm rec}>c$ (yellow crosshatching and beyond) to the region of subluminal recession (no shading) could the photons approach us.
More detail about early times and the horizons is visible in comoving coordinates (middle panel) and conformal coordinates (lower panel).  Our past light cone in comoving coordinates appears to approach the horizontal ($t=0$) axis asymptotically, however it is clear in the lower panel that the past light cone reaches only a finite distance at $t=0$ (about $46\, Glyr$, the current distance to the particle horizon).  Light that has been travelling since the beginning of the Universe was emitted from comoving positions which are now $46\, Glyr$ from us.  
The distance to the particle horizon as a function of time is represented by the dashed green line, (Eq.~\ref{eq:chipht}).  
Our event horizon is our past light cone at the end of time, $t=\infty$ in this case.  It asymptotically approaches $\chi=0$ as $t\rightarrow \infty$.  
Many of the events beyond our event horizon (shaded solid gray) occur on galaxies we have seen before the event occurred (the galaxies are within our particle horizon).  We see them by light they emitted billions of years ago but we will never see those galaxies as they are today.  
The vertical axis of the lower panel shows conformal time (Eq.~\ref{eq:conformalt}).  An infinite proper time is transformed into a finite conformal time so this diagram is complete on the vertical axis.
The aspect ratio of $\sim 3/1$ in the top two panels represents the ratio between the size of the Universe and the age of the Universe, $46\, Glyr/13.5\, Gyr$~\citep[c.f.][]{kiang97}. }

\newpage

\section{Features of general relativistic expansion}\label{sect:spacetime}
Figure~\ref{fig:dist} shows three spacetime diagrams drawn using the standard general relativistic formulae for an homogeneous, isotropic universe based on the Robertson-Walker metric, and Friedmann's equation, as summarized in Appendix~\ref{sect:math}.   They show the relationship between comoving objects, light, the Hubble sphere and cosmological horizons. 
These spacetime diagrams are based on the observationally favored $\Lambda$CDM concordance model of the universe: $(\Omega_{\rm M},\Omega_{\Lambda})= (0.3,0.7)$ and use $H_0=70\,kms^{-1}Mpc^{-1}$~\citep[to one significant figure]{bennett03}.
%
The upper diagram plots time versus proper distance, $D$.
The middle diagram plots time versus comoving distance, $R_0\chi$. 
The lower diagram plots conformal time $d\tau = dt/R(t)$ (Eq.~\ref{eq:conformalt}) versus comoving distance.  

Two types of horizon are shown in Fig.~\ref{fig:dist}.  The particle horizon is the distance light can have travelled from $t=0$ to a given time $t$ (Eq.~\ref{eq:chipht}), whereas the event horizon is the distance light can travel from a given time $t$ to $t=\infty$ (Eq.~\ref{eq:eventhorizont}).  Using Hubble's law ($v_{\rm rec}=HD$), the Hubble sphere is defined to be the distance beyond which the recession velocity exceeds the speed of light, $D_{\rm H}=c/H$.  As we will see, the Hubble sphere is not an horizon.  Redshift does not go to infinity for objects on our Hubble sphere (in general) and for many cosmological models, including $\Lambda$CDM, we can see beyond it.

Recession velocities are given by the slopes of the worldlines in the upper diagram. At any given time, the slopes of these world lines are proportional to their distance from us according to Hubble's law, $\vrec = H D=\dot{R}\chi$. 
 One of the clearest aspects of the proper distance diagram is that the slope of comoving worldlines can be greater than 45 degrees from vertical.  That is, their recession velocities can be greater than the speed of light.
This does not contradict special relativity because the motion is not in the observer's inertial frame.  No observer ever overtakes a light beam and all observers measure light locally to be travelling at $c$.

The second diagram is drawn using comoving distance so recession velocities have been removed.  Any slopes ($\dot{\chi}$) are due to peculiar velocities, $v_{\rm pec}=R\dot{\chi}$.  The peculiar velocity of light is constant, $\vpec=R\dot{\chi}=c$, but the peculiar velocity of light through comoving coordinates decreases as the Universe expands, $\dot{\chi}=c/R$.

\begin{figure}\bctr
\includegraphics[width=120mm]{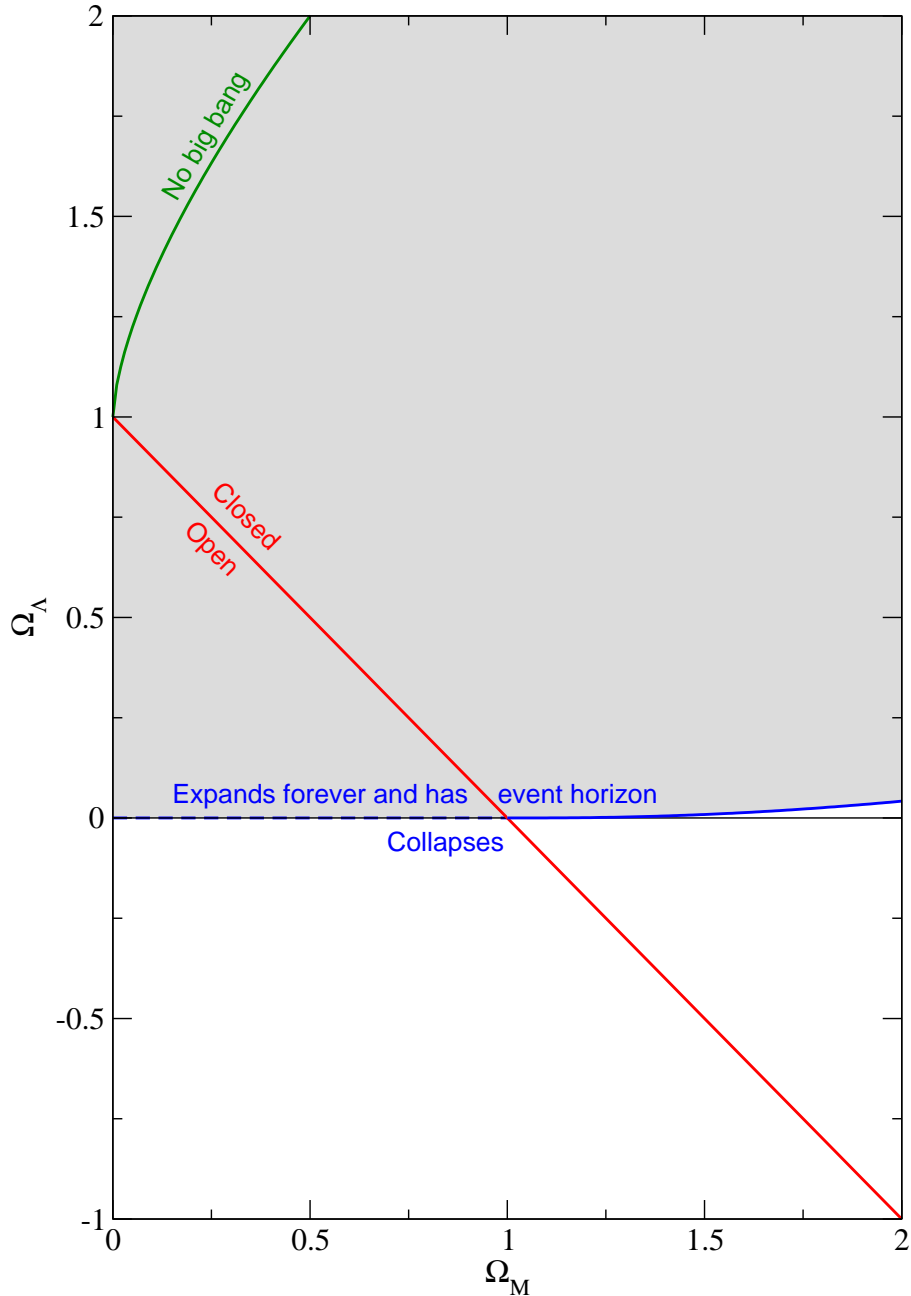} 
\caption[Which universes have event horizons?]{\small{The region in the $\om$--$\oll$ plane for which the Universe expands forever and has an event horizon ($q\rightarrow -1$ as $a\rightarrow \infty$) is shaded gray.  Open and flat universes with no cosmological constant ($0\le \om \le 1.0$, $\oll=0$) expand forever but have no event horizon.  Collapsing universes (non-shaded regions) can have an event horizon because light travels a finite distance in the finite age of the universe.  These differ from the event horizons in universes that expand forever because in the latter case light can only travel a finite distance even given infinite time.  }}
\label{fig:omol}
\ectr\end{figure}

\begin{figure} \bctr
\includegraphics[width=140mm]{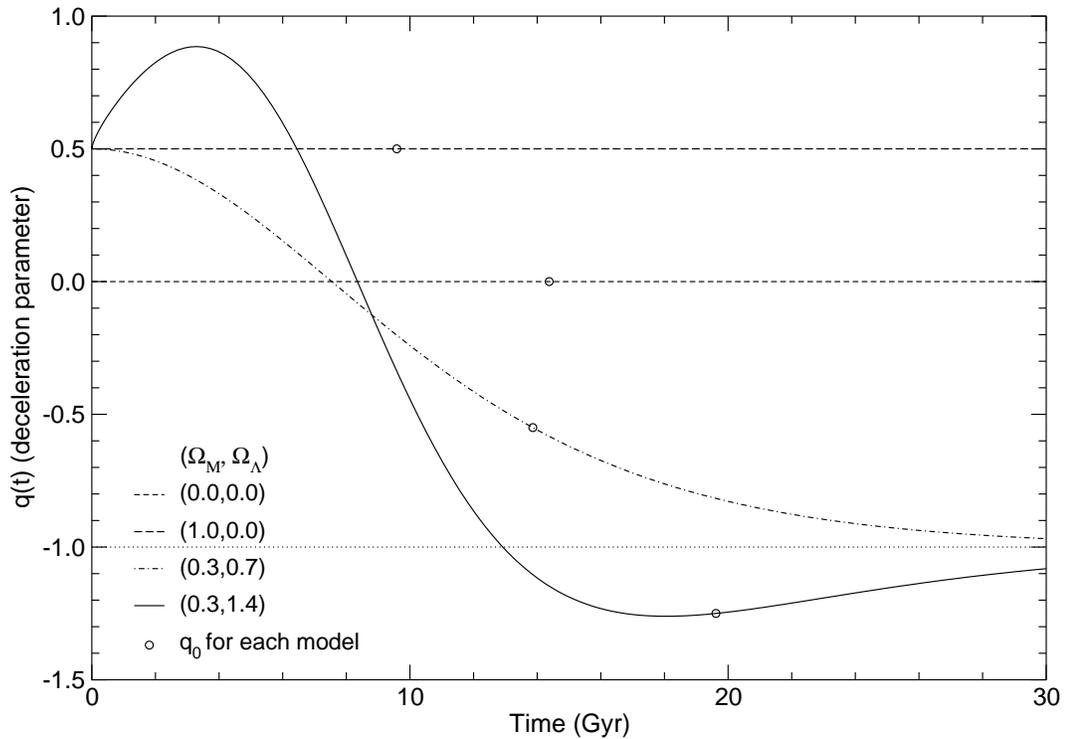}
\caption[The deceleration parameter versus time]{{\small The deceleration parameter is plotted against time for four model universes.  Both the $\omol=(0,0)$ and $\omol=(1,0)$ models have constant deceleration parameters, but in general the deceleration parameter of a universe changes with time.  The $\omol=(0.3,0.7)$ universe initially has a positive $q$ (decelerates) but tends towards $q=-1$ from above as $a\rightarrow \infty$.  This behaviour is typical of {\em flat} and {\em open} ($\om+\oll\le 0$) eternally expanding universes with a positive cosmological constant.  The $\omol=(0.3,1.4)$ model tends toward $q=-1$ from below as $a\rightarrow\infty$.  This behaviour is typical of {\em closed} ($\om+\oll> 0$) eternally expanding universes with a positive cosmological constant.   The proper distance to the event horizon decreases while $q<-1$.   During inflation and in a pure de Sitter model, $(\Omega_M,\Omega_{\Lambda})=(0,1)$, the deceleration parameter stays constant at $q=-1$.  The present day deceleration parameter for each model is marked with a circle.}}
\label{fig:q-t}
\ectr \end{figure}

The cosmological {\em event horizon} separates events we are able to see at some time, from events we will never be able to see.  
At any particular time the event horizon forms a sphere around us beyond which events are forever inaccessible.
An event horizon exists if light can only travel a finite distance during the lifetime of the universe.  This can occur 
 if the universe has a finite age, or if the universe accelerates such that light can travel only a finite 
distance given infinite time.  This second criterion is satisfied for all eternally expanding universes with
 a cosmological constant, so most observationally viable cosmological models have event horizons (see Fig.~\ref{fig:omol}).  The acceleration history of a universe is recorded by the deceleration parameter $q=-\ddot{R}R/\dot{R}^2$, where negative $q$ corresponds to acceleration (see Fig.~\ref{fig:q-t}).  In the $\omol=(0.3,0.7)$ model shown in Fig.~\ref{fig:dist} galaxies we currently observe at redshift $z\sim1.8$ are just passing over our event horizon.  Thus these galaxies are the most distant objects from which we will ever receive information about the present day.
The comoving distance to the event horizon (Eq.~\ref{eq:eventhorizont}) always decreases since the distance light can travel in the time remaining before the end of the universe (at $t\rightarrow \infty$), always decreases.  (Event horizons can also exist in universes that collapse to a big crunch, but we do not discuss those here.)
Like objects falling across the event horizon of a black hole, objects crossing our cosmological event horizon appear time dilated (from our perspective) as they approach the horizon.  
However, unlike an observer crossing the event horizon of a black hole, the galaxy crossing our cosmological event horizon does not see all distant clocks speeding up.  
This asymmetry in the analogy reflects the fact that the location of the cosmological event horizon is observer dependent.  


The cosmological {\em particle horizon} separates comoving distances (particles) we can currently see from those we cannot currently see.  At any particular time the particle horizon forms a sphere around us beyond which we cannot {\em yet} see.  
The comoving distance to the particle horizon always increases since the distance light has travelled since the beginning of the Universe always increases.  The particle horizon can be larger than the event horizon because  although we cannot see {\em events} that occur beyond our event horizon, we can still see many  galaxies that are beyond our current event horizon by light they emitted long ago.  That is, we can see some galaxies as they were when they were young, but we will never be able to watch them grow up.  In the comoving distance diagrams, it is clear that at $R_0\chi_0 \sim 47\,Glyr $($\sim 3ct_0$) lies the most distant comoving object that we can currently observe (i.e. intersects our past light cone at $t \sim 0$). This is the current distance to the particle horizon and is sometimes referred to as the size of the observable Universe.  Note that no photons have actually travelled $47\,Glyr$, our particle horizon is at $47\,Glyr$ simply because the objects that emitted those photons in the early Universe have moved that far away as the Universe expanded.  When we use the recent limits on the cosmological parameters from the WMAP project~\citep{bennett03} we can constrain the current distance to the particle horizon to be $47.3^{+5.4}_{-5.0}\,Glyr$. Note that all galaxies become increasingly redshifted as we watch them approach the cosmological {\em event} horizon ($z\rightarrow\infty$ as $t\rightarrow\infty$).  However, objects with $z=\infty$ now (and at any $t<\infty$) lie on our {\em particle} horizon.

The {\em Hubble sphere} is the surface on which comoving objects are receding at the speed of light.  It is a sphere around us with radius $D=c/H$.  It is not an horizon of any kind since both objects and light can cross it in both directions.  Note that objects that recede at the speed of light do not have an infinite redshift (Eq.~\ref{eq:chiz} for $\chi = c/\dot{R}$).  

Our {\em past light cone} traces the events in the Universe that we can currently see.  It is shaped like a cone in the conformal diagram (bottom diagram) however in proper distance (top diagram) our past light cone is shaped like a teardrop.  Starting at $t=0$, the outward curving part of the teardrop shows photons which were initially outside our Hubble sphere but eventually managed to reach us.  This shows that objects receding faster than the speed of light are observable, as we discuss in Sect.~\ref{sect:notobserve}.  During the outward curving part of the light cone these photons were receding from us, even as they propagated towards us at what local observers would have measured to be the speed of light.  The turning point between receding photons and approaching photons occurs when the Hubble sphere expands beyond our past light cone leaving the past-light-cone photons in a subluminally receding region of space. 
Our past light cone (Eq.~\ref{eq:lightconet}) approaches the event horizon (Eq.~\ref{eq:eventhorizont}) as $t_0 \rightarrow \infty$.

In conformal coordinates it is straightforward to determine causality because light, and thus all the horizons, follow straight lines with $|{\rm slope}|=1$ (Eq.~\ref{eq:conformalmetric}).  
The conformal transformation transforms an infinite proper time to a finite conformal time.  This diagram is therefore complete on the time axis.  
The horizontal axis could extend further because the comoving distance can be infinite in extent, but we will never see any objects at any time from beyond the maximum comoving distance of the particle and event horizons ($\sim 62\, Glyr$ in the $\Lambda$CDM model shown in Fig.~\ref{fig:dist}).  This is not to be confused with a Penrose diagram, which is drawn using ``null coordinates''.  Like a spacetime diagram in conformal coordinates, Penrose diagrams are useful to determine causality because light follows straight lines with $|{\rm slope}|=1$.  However in a Penrose diagram surfaces of constant time and constant distance are not straight lines.

One final feature we need to discuss is the decay of peculiar velocity in FRW universes.  This is a standard result recognised soon after the expansion of the Universe was discovered and clear derivations can be found in the recent works \citet[][Sect.~15.3]{peacock99} and \citet[][Sect.~6.2(c)]{padmanabhan96}.
A non-relativistic ``test galaxy'' with initial peculiar velocity $v_{\rm pec,0}$ will later find itself with a reduced peculiar velocity according to $\vpec = v_{\rm pec,0}/a$.  
Often peculiar velocity decay is explained in the following manner: since peculiar velocities are measured with respect to the local comoving frame, they decrease as the test galaxy catches up to galaxies that were initially receding from it.  
This description is pedagogically useful, but not entirely correct, because not all peculiar velocities decay at the same $1/a$ rate.  Photons for example, are not slowed down as they travel.  They maintain a peculiar velocity of $c$ in all comoving frames.  However, their momentum decays as their wavelength is redshifted.  Since $\lambda \propto a$ the momentum of photons, $p=hc/\lambda$, decays as $1/a$.    An analogous process can be said to apply to  massive particles.  Their de Broglie wavelength is given by $\lambda_{\rm de Broglie}=h/p$.  Applying $\lambda_{\rm de Broglie} \propto a$ results in momentum decreasing as $1/a$.  For non-relativistic particles, $p=mv$, so $v\propto 1/a$.  However, the rate of decay decreases as particles become more relativistic.  
A derivation of this effect is given in Appendix~\ref{sect:math} and some implications are discussed in Sections.~\ref{sect:nostretch} and~\ref{sect:rel-sol}.

\clearemptydoublepage

\chapter{Expanding confusion}\label{chap:misconcept}
\vspace{-5mm}
In this chapter we resolve conflicting views in the literature regarding the general relativistic (GR) description of the expanding universe, and provide observational evidence against the special relativistic interpretation of recession velocities that is the basis of many of the misconceptions.  In Section~\ref{sect:misconceptions} we clarify common misconceptions about superluminal recession velocities and horizons.  Firstly, we show that recession velocities can exceed the speed of light and that inflationary expansion and the current expansion both have superluminal and non-superluminal regions.   Secondly, we show that we can {\em observe} galaxies that are receding faster than the speed of light, contrary to special relativistic calculations in which a velocity of $c$ corresponds to an infinite redshift. This is a point that even the most knowledgable researchers on this subject frequently misrepresent. We also develop a more informative way to depict the particle horizon on spacetime diagrams.  Examples of misconceptions occurring in the literature are given in Appendix~\ref{sect:quotes}. 

In Section~\ref{sect:data} we provide an explicit observational test demonstrating that special relativistic concepts applied to the expanding Universe are in conflict with observations.
In particular, using data taken by \cite{perlmutter99} we show the SR interpretation of cosmological redshifts is inconsistent with the supernovae magnitude-redshift relation at the $\sim 23 \sigma$ level.    We discuss the relevance of distance and velocity in cosmology in Section~\ref{sect:misc_discussion}.

This chapter is based on the work published in~\cite{davis03luminal}.

\section{Clarifying Misconceptions}\label{sect:misconceptions}
For more than half a century the redshifts of galaxies have been almost universally accepted to be a result of the expansion of the Universe.   The expansion has become fundamental to our understanding of the cosmos.
However, this interpretation leads to several concepts that are widely misunderstood.  
Since the expansion of the Universe is the basis of the big bang model, these misunderstandings are fundamental.  
Not only popular science books written by astrophysicists, astrophysics textbooks but also professional astronomical literature addressing the expansion of the Universe, contains misleading, or easily misinterpreted, statements concerning recession velocities, horizons and the ``observable universe''.

Probably the most common misconceptions surround the expansion of the Universe at distances beyond which Hubble's law predicts recession velocities faster than the speed of light~\inciteFirst{1--8}, 
despite efforts to clarify the issue 
\citep{murdoch77,silverman86,stuckey92,ellis93,harrison93,kiang97,harrison00,davis00,gudmundsson02,kiang01,davis03luminal}. 
Misconceptions include misleading comments suggesting we cannot observe galaxies that are receding faster than light~\incite{9--13}. 
and related, but more subtle, confusions surrounding cosmological event horizons~\incite{14--15}. 
The concept of the expansion of the Universe is so fundamental to our understanding of cosmology and the misconceptions so abundant that it is important to clarify these issues and make the connection with observational tests as explicit as possible.

\subsection[Recession velocities can be superluminal]{Misconception \#1: Recession velocities cannot exceed the speed of light}

A common misconception is that the expansion of the Universe cannot be faster than the speed of light. 
Since Hubble's law predicts superluminal recession at large distances ($D>c/H$) it is sometimes stated that Hubble's law needs special relativistic corrections when the recession velocity approaches the speed of light~\incite{6--7}. 
However, it is well-accepted that general relativity, not special relativity, is necessary to describe cosmological observations.  Supernovae surveys calculating cosmological parameters, galaxy-redshift surveys and cosmic microwave background anisotropy tests, all use general relativity to explain their observations.  When observables are calculated using special relativity, contradictions with observations quickly arise (Section~\ref{sect:sr}).  
Moreover, we know there is no contradiction with special relativity when faster than light motion occurs {\em in a non-inertial reference frame}.  General relativity was derived to be able to predict motion when global inertial frames were not available \citep[][Ch.~1]{rindler77}.   
Galaxies that are receding from us superluminally are at rest locally (when their peculiar velocity, $v_{\rm pec}=0$) and motion in their local inertial frames remains well described by special relativity.
They are in no sense catching up with photons ($v_{\rm pec}=c$).  Rather, the galaxies and the photons (that are directed away from us) are both receding from us at recession velocities greater than the speed of light. 

In special relativity, redshifts arise directly from velocities.  It was this idea that led Hubble in 1929 to convert the redshifts of the ``nebulae'' he observed into velocities, and predict the expansion of the Universe with the linear velocity-distance law that now bears his name.  
The general relativistic interpretation of the expansion interprets cosmological redshifts as an indication of velocity since the proper distance between comoving objects increases. 
However, the velocity is due to the rate of expansion of space, not movement through space, and therefore cannot be calculated with the special relativistic Doppler shift formula.  Hubble \& Humason's calculation of velocity therefore should not be given special relativistic corrections at high redshift, contrary to their suggestion~\incite{16}. 

The general relativistic and special relativistic relations between velocity and cosmological redshift are 
\citep[e.g.][]{davis00}: 
\bea
{\rm GR}&\;\; v_{\rm rec}(t,z)  =& \frac{c}{R_0}\;\dot{R}(t)\int^{z}_{0}\frac{dz^\prime}{H(z^\prime)},  \label{eq:vGR}\\
                               & & \nonumber\\ 
{\rm SR} &\;\;\;\;\vp (z)       =& c \: \frac{(1+ z)^2-1}{(1 + z)^2+1}. \label{eq:vSR} 
\eea
These velocities are with respect to the comoving observer who observes the receding object to have redshift, $z$.
The GR description is written explicitly as a function of time because when we observe an object with redshift, $z$, we must specify the epoch at which we wish to calculate its recession velocity.  For example, setting $t= t_0$ yields the recession velocity today of the object that emitted the observed photons at $t_{\rm em}$. Setting $t=t_{\rm em}$ yields the recession velocity at the time the photons were emitted (see Eqs.~\ref{eq:vchi} \&~\ref{eq:chiz}).  The changing recession velocity of a comoving object is reflected in the changing slope of its worldline in the top panel of Fig.~\ref{fig:dist}.
There is no such time dependence in the SR relation.

\begin{figure}
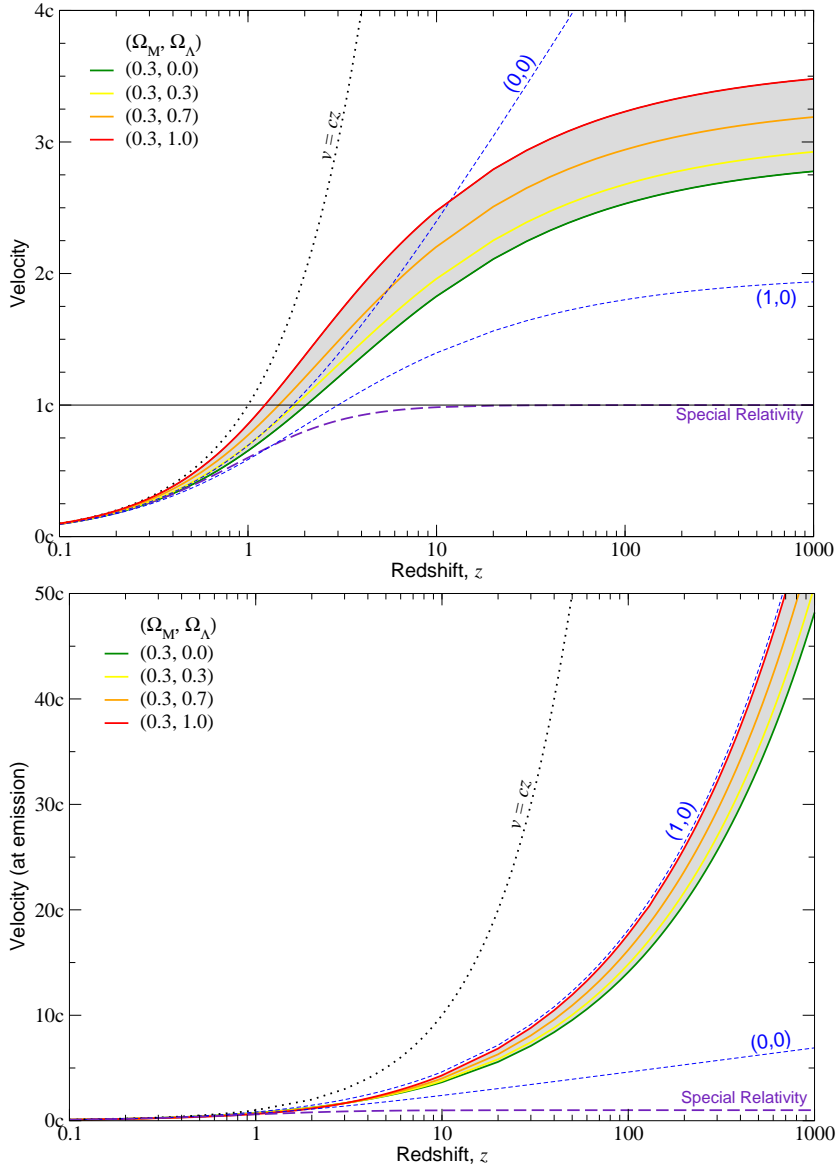
\bctr
\includegraphics[width=110mm]{./figs/zv-new2colour.eps}
\includegraphics[width=110mm]{./figs/zv-new-em-colour.eps}
\caption[Velocity as a function of redshift.]{{\small Velocity as a function of redshift under various assumptions.   The linear approximation, $v=cz$, is the low redshift approximation of both the GR and SR results.  The SR result is calculated using Eq.~\ref{eq:vSR} while the GR result uses Eq.~\ref{eq:vGR}.  The upper panel shows recession velocities at the time of observation (present day), i.e., uses $\dot{R}(t)=\dot{R}_0$.  The lower panel shows recession velocities at the time of emission, i.e., uses $\dot{R}(t)=\dot{R}(t_{\rm em})$.
The region shaded gray shows a range of Friedmann-Robertson-Walker (FRW) models as labeled in the legend.  These include the observationally favoured cosmological model $\omol=(0.3,0.7)$.  The recession velocity of all galaxies with $z\gsim 1.5$ currently exceeds the speed of light in all viable cosmological models.  Observations now routinely probe regions that are receding faster than the speed of light.}}
\label{fig:vz}\ectr
\end{figure}

Despite the fact that special relativity incorrectly describes cosmological redshifts it has been used for decades to convert cosmological redshifts into velocity because the special relativistic Doppler shift formula (Eq.~\ref{eq:vSR}), shares the same low redshift approximation, $v=cz$, as Hubble's Law (Fig.~\ref{fig:vz}).   
It has only been in the last decade that routine observations have been deep enough that the distinction has become significant. 
Figure~\ref{fig:vz} shows a snapshot of the GR velocity-redshift relation for various models as well as the SR velocity-redshift relation and their common low redshift approximation, $v=cz$.  Present day recession velocities exceed the speed of light in all viable cosmological models for objects with redshifts greater than $z\sim 1.5$.
At higher redshifts special relativistic ``corrections'' can be more incorrect than the simple linear approximation (Fig.~\ref{fig:mag-z}).

Some of the most common misleading applications of relativity arise from the misconception that nothing can recede faster than the speed of light.  These include texts asking students to calculate the velocity of a high redshift receding galaxy using the special relativistic Doppler shift equation~\incite{17--21}, 
as well as the comment that galaxies recede from us at speeds ``approaching the speed of light''~\incite{4--5, 8}, 
 or that quasars recede at a certain percentage of the speed of light\footnote{Redshifts are usually converted into velocities using $v=cz$, which is a good approximation for $z\lsim 0.3$ (see Fig.~\ref{fig:vz}) but inappropriate for today's high redshift measurements.  When a ``correction'' is made for high redshifts, the formula used is almost invariably the inappropriate special relativistic Doppler shift equation (Eq.~\ref{eq:vSR}).}~\incite{3, 18--21}. 

Although velocities of distant galaxies are in principle observable, the set of synchronized comoving observers required to measure proper distance 
(\citeauthor{weinberg72}, 1972, p.~415; \citeauthor{rindler77}, 1977, p.~218) 
 is not practical.  Instead, more direct observables such as the redshifts of standard candles can be used to observationally rule out the special relativistic interpretation of cosmological redshifts (Section~\ref{sect:data}).

\vspace{1cm}
\subsection[Inflation is misnamed ``superluminal expansion.'']{Misconception \#2: Inflation results in superluminal expansion but the normal expansion of the universe does not}
\label{sect:inflationissuperluminal}

Inflation is sometimes described as ``superluminal expansion''~\incite{22--23}. 
This is misleading because it implies that non-inflationary expansion is not superluminal.  However, {\em any} expansion described by Hubble's law has superluminal recession velocities for sufficiently distant objects.  Even during inflation, objects within the Hubble sphere ($D<c/H$) recede at less than  the speed of light, while objects beyond the Hubble sphere ($D>c/H$) recede faster than the speed of light.  This is identical to the situation during non-inflationary expansion, except the Hubble constant during inflation was much larger than subsequent values. Thus the distance to the Hubble sphere was much smaller.  
During inflation the proper distance to the Hubble sphere stays 
constant and is coincident with the event horizon -- this is also identical to the asymptotic behaviour of an eternally expanding universe with a cosmological constant $\oll>0$ (Fig. \ref{fig:dist}, top panel).  

The oft-mentioned concept of structures ``leaving the horizon'' during the inflationary period refers to  structures once smaller than the Hubble sphere becoming larger than the Hubble sphere.  If the exponentially expanding regime, $R=R_0e^{Ht}$, were extended to the end of time, the Hubble sphere would be the event horizon.  However, in the context of inflation the Hubble sphere is not a true event horizon because structures that have crossed the horizon can ``re-enter the horizon'' after inflation stops.  The horizon they ``re-enter'' is the revised event horizon determined by how far light can travel in the FRW universe that remains after inflation.  This revised event horizon is larger than the event horizon that would have existed if inflation had continued forever.

It would be more appropriate to describe inflation as superluminal expansion if all distances down to the Planck length, $l_{\rm pl}\sim 10^{-35}m$, were receding faster than the speed of light.  Solving $D_{\rm H}=c/H = l_{\rm pl}$ gives $H = 10^{43}s^{-1}$ (inverse Planck time) which is equivalent to $H= 10^{62} \,km\,s^{-1}Mpc^{-1}$.   If Hubble's constant during inflation exceeded this value it would justify describing inflation as ``superluminal expansion''. 


\subsection[We can observe galaxies that are receding superluminally]{Misconception \#3: Galaxies with recession velocities exceeding the speed of light exist but we cannot see them}
\label{sect:notobserve}

Amongst those who acknowledge that recession velocities can exceed the speed of light, the claim is sometimes made that objects with recession velocities faster than the speed of light are not observable~\incite{9--13}. 
We have seen that the speed of photons propagating towards us (the slope of our past light cone in the upper panel of Fig.~\ref{fig:dist}) is not constant, but is rather $v_{\rm rec}-c$.  Therefore all photons beyond the Hubble sphere, even those photons propagating in our direction ($\vpec=-c$), have a total velocity away from us.   How is it then that light from beyond the Hubble sphere can ever reach us?  Although the photons are in the superluminal region and therefore recede from us (in proper distance), the Hubble sphere also recedes.  In decelerating universes $H$ decreases as $\dot{R}$ decreases (causing the Hubble sphere to recede). In accelerating universes $H$ also tends to decrease since $\dot{R}$ increases more slowly than $R$.   As long as the Hubble sphere recedes faster than the photons immediately outside it, $\dot{D}_{\rm H}>v_{\rm rec}-c$, the photons end up in a subluminal region and approach us\footnote{The behaviour of the Hubble sphere is model dependent.    The Hubble sphere recedes as long as the deceleration parameter $q=-\ddot{R}R/\dot{R}^2>-1$.  In some closed eternally accelerating universes (specifically $\om +\oll>1$ and $\oll>0$) the deceleration parameter can be less than minus one in which case we see faster-than-exponential expansion and some subluminally expanding regions can be beyond the event horizon (light that was initially in subluminal regions can end up in superluminal regions and never reach us).  Exponential expansion, such as that found in inflation, has $q=-1$.  Therefore the Hubble sphere is at a constant proper distance and coincident with the event horizon.  This is also the late time asymptotic behaviour of eternally expanding FRW models with $\oll>0$ (see Fig.~\ref{fig:dist}, upper panel).} (the photons we are referring to are those with $\vpec=-c$).  
Thus photons near the Hubble sphere that are receding slowly are overtaken by the more rapidly receding Hubble sphere\footnote{The myth that superluminally receding galaxies are beyond our view, may have propagated through some historical preconceptions.  Firstly, objects on our event horizon {\em do} have infinite redshift, tempting us to apply our SR knowledge that infinite redshift corresponds to a velocity of $c$.  Secondly, the once popular steady state theory predicts exponential expansion, for which the Hubble sphere and event horizon {\em are} coincident.}.


Our teardrop shaped past light cone in the top panel of Fig.~\ref{fig:dist} shows that any photons we now observe that were emitted in the first $\sim$ five billion years were emitted in regions that were receding superluminally, $v_{\rm rec}>c$.  Thus their total velocity was away from us.  Only when the Hubble sphere expands past these photons do they move into the region of subluminal recession and approach us. 
The most distant objects that we can see now were outside the Hubble sphere when their comoving coordinates intersected our past light cone. Thus, they were receding superluminally when they emitted the photons we see now. Since their worldlines have always been beyond the Hubble sphere these objects were, are, and always have been, receding from us faster than the speed of light. 

An example of an object that has always been receding faster than the speed of light is the object with redshift $z=3$ in the middle (comoving) panel of Fig.~\ref{fig:dist}.  On the same diagram the object with redshift $z=1$ 
is initially beyond the Hubble sphere, but as the Universe decelerates the $z=1$ galaxy finds itself within the Hubble sphere and it is currently receding subluminally.  At around $t=7.4\,Gyr$ the Universe began to accelerate and the Hubble sphere began to contract (in comoving coordinates).  At about $19\,Gyr$ the $z=1$ galaxy will again find itself outside the Hubble sphere.  Note that we label the galaxy $z=1$ because that is its current redshift.  However, the redshift we observe that galaxy to have will evolve over time.

Evaluating Eq.~\ref{eq:vGR} for the observationally favoured $\omol=(0.3,0.7)$ universe shows that all galaxies beyond a redshift of $z=1.46$ are currently receding faster than the speed of light (Fig.~\ref{fig:vz}). 
Hundreds of galaxies with $z > 1.46$ have been observed.
The highest spectroscopic redshift observed in the Hubble deep field is $z=6.68$ 
\citep{chen99} 
and the Sloan digital sky survey has identified four galaxies at $z>6$ 
\citep{fan03}.  
All of these galaxies were, are and always will be receding superluminally, and yet we see them. The particle horizon, not the Hubble sphere, marks the size of our observable Universe because we cannot have received light from, or sent light to, anything beyond the particle horizon\footnote{The current distance to our particle horizon and its velocity are difficult to determine due to the unknown duration of inflation.  The particle horizon depicted in Fig.~\ref{fig:dist} assumes no inflation.}.  Our effective particle horizon is the cosmic microwave background (CMB), at redshift $z\sim 1100$, because we cannot see beyond the surface of last scattering.  Although the last scattering surface is not at any fixed comoving coordinate, the current recession velocity of the points from which the CMB was emitted is $3.2c$ (Fig.~\ref{fig:vz}).  At the time of emission their speed was $58.1c$, assuming $\omol=(0.3,0.7)$.  Thus we routinely observe objects that are receding faster than the speed of light and the Hubble sphere is not an horizon\footnote{Except in the special cases when the expansion is exponential, $R=R_0e^{Ht}$, such as the de Sitter universe ($\om=0,\oll>0$), during inflation or in the asymptotic limit of eternally expanding FRW universes.}.

\subsection[Particle horizons on spacetime diagrams]{Ambiguity: The depiction of particle horizons on spacetime diagrams}
\label{sect:particlehorizon}

\begin{figure}
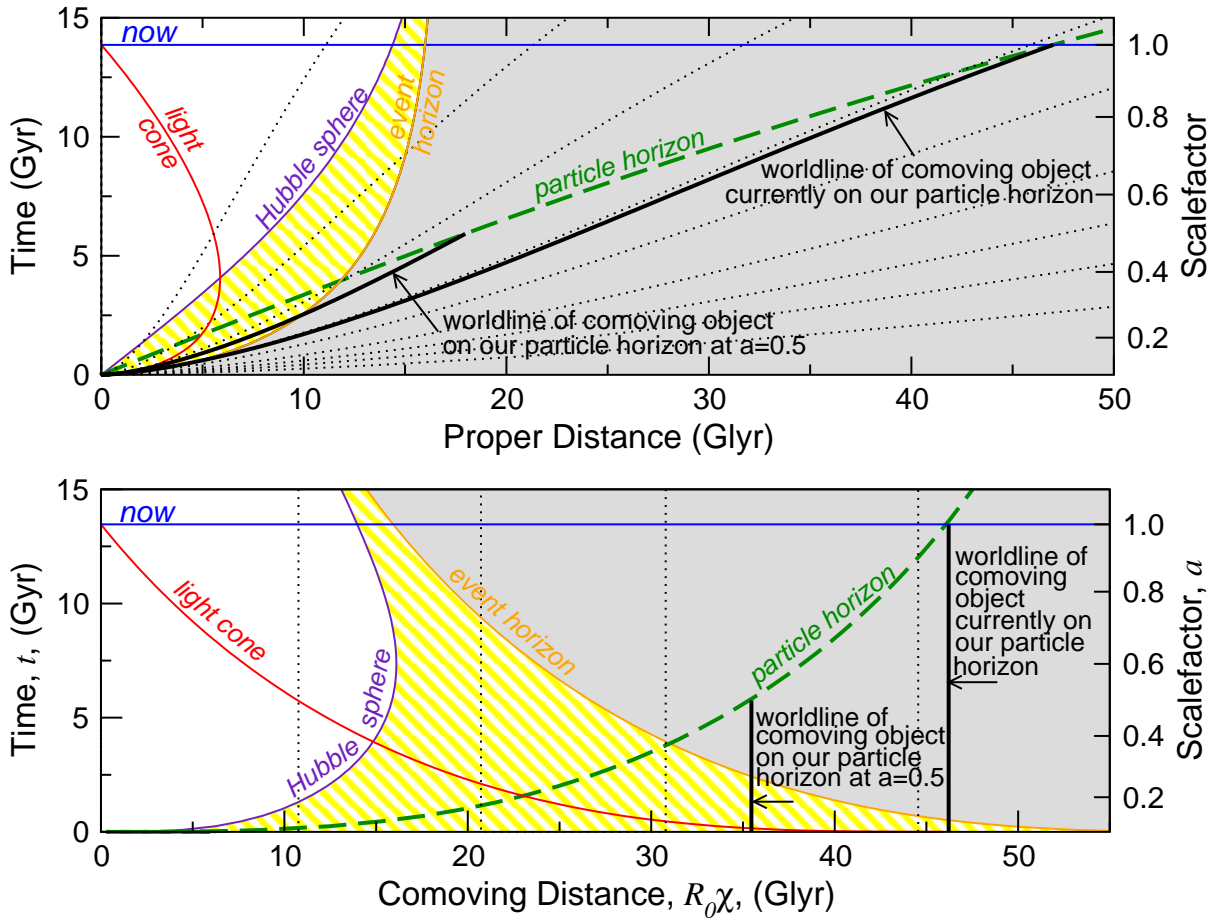
 \bctr
\includegraphics[width=160mm]{./figs/0-dist0307-colour2-particlehorizon2.eps}\vspace{2mm}
\includegraphics[width=160mm]{./figs/0-comov0307-colour2-particlehorizon-H70.eps}
\caption[Depicting the particle horizon on spacetime diagrams]{{\small The traditional depiction of the particle horizon on spacetime diagrams is the worldline of the object currently on our particle horizon (thick solid line).  All the information in this depiction is contained in a single point, the current distance to the particle horizon. An alternative way to plot the particle horizon is to plot the distance to the particle horizon as a function of time (thick dashed line and Fig.~\ref{fig:dist}).  This alleviates the need to draw a new worldline when we need to determine the particle horizon at another time (for example the worldline of the object on our particle horizon when the scalefactor $a=0.5$).}}
\label{fig:ph}
\ectr \end{figure}

Here we identify an inconvenient feature of the most common depiction of the particle horizon on spacetime diagrams and provide a useful alternative (Fig.~\ref{fig:ph}).  The particle horizon at any particular time is a sphere around us whose radius equals the distance to the most distant object we can see. 
The particle horizon has traditionally been depicted as the {\em worldline} or comoving coordinate of the most distant particle that we can currently see 
\citep{rindler56,ellis93}.  
The only information this gives is contained in a single point: the current radius of the particle horizon, and this indicates the current radius of the observable Universe.   The rest of the worldline can be misleading as it does not represent a boundary between events we can see and events we cannot see, nor does it represent the radius of the particle horizon at different times.  An alternative way to represent the particle horizon is to plot the radius of the particle horizon as a function of time 
\citep{kiang91}.
The particle horizon at any particular time defines a unique distance which appears as a single point on a spacetime diagram.  Connecting the points gives the radius of the particle horizon vs time.  It is this time dependent series of particle horizons that we plot in Fig.~\ref{fig:dist}.  
(\cite{rindler56}
 calls this the boundary of our creation light cone -- an outgoing light cone starting at the big bang.)  Drawn this way, one can read from the spacetime diagram the radius of the particle horizon at any time.  There is no need to draw another worldline.  

Specifically, what we plot as the particle horizon is $\chi_{\rm ph}(t)$ from Eq.~\ref{eq:chipht} rather than the traditional $\chi_{\rm ph}(t_0)$.  To calculate the distance to the particle horizon at an arbitrary time $t$ it is not sufficient to multiply $\chi_{\rm ph}(t_0)$ by $R(t)$ since the comoving distance to the particle horizon also changes with time.  

The particle horizon is sometimes distinguished from the event horizon by describing the particle horizon as a ``barrier in space'' and the event horizon as a ``barrier in spacetime''.  
This is not a useful distinction because both the particle horizon and event horizon are surfaces in spacetime -- they both form a sphere around us whose radius varies with time.   
When viewed in comoving coordinates the particle horizon and event horizon are mirror images of each other (symmetry about $z\sim 10$ in the middle and lower panels of Fig.~\ref{fig:dist}). 
The traditional depiction of the particle horizon would appear as a straight vertical line in comoving coordinates, i.e.,~the comoving coordinate of the present day particle horizon (Fig.~\ref{fig:ph}, lower panel).

The proper distance to the particle horizon is {\em not} $D_{\rm PH}=ct_0$.  Rather, it is the proper distance to the most distant object we can observe, and is therefore related to how much the universe has expanded, i.e.~how far away the emitting object has become, since the beginning of time.  In general this is $\sim 3 c t_0$. The relationship between the particle horizon and light travel time arises because the comoving coordinate of the most distant object we can see {\em is} determined by the {\em comoving} distance light has travelled during the lifetime of the Universe (Eq.~\ref{eq:chipht}).

\section[Observational evidence for the GR interpretation of cosmological redshifts]{Observational evidence for the GR interpretation of cosmological redshifts}\label{sect:data}\label{sect:sr}

\subsection{Duration-redshift relation for Type Ia Supernovae}

Many misconceptions arise from the idea that recession velocities are limited by SR to less than the speed of light so in Section~\ref{sect:mag-z} we present an analysis of supernovae observations yielding evidence against the SR interpretation of cosmological redshifts.
But first we would like to present an observational test that {\em can not} distinguish between special relativistic and general relativistic expansion of the Universe.  

General relativistic cosmology predicts that events occurring on a receding emitter will appear time dilated by a factor,
\beq \gamma_{\rm GR}(z) = 1+z. \eeq
A process that takes $\Delta t_0$ as measured by the emitter appears to take $\Delta t = \gamma_{\rm GR}\Delta t_0$ as measured by the observer when the light emitted by that process reaches them.
Wilson (1939) 
suggested measuring this cosmological time dilation to test whether the expansion of the Universe was the cause of cosmological redshifts. 
Type Ia supernovae (SNe Ia) lightcurves provide convenient standard clocks with which to test cosmological time dilation.   Recent evidence from supernovae includes Leibundgut \etal~(1996) 
who gave evidence for $1+z$ time dilation using a single high-$z$ 
supernova and Riess \etal~(1997) 
who showed  $1+z$ time dilation for a single SN Ia  at the 96.4\% confidence level using the time variation of spectral features.  Goldhaber \etal~(1997) 
show five data points of lightcurve width consistent with $1+z$ broadening and extend this analysis in Goldhaber \etal~(2001) 
to rule out any theory that predicts zero time dilation (for example ``tired light'' scenarios (see Wright, 2001)), 
at a confidence level of $18\sigma$. 
All of these tests show that $\gamma=(1+z)$ time dilation is preferred over models that predict {\em no} time dilation.   

We want to know whether the same observational test can show that GR time dilation is preferred over SR time dilation as the explanation for cosmological redshifts.  When we talk about SR expansion of the universe we are assuming that we have an inertial frame that extends to infinity (impossible in the GR picture) and that the expansion involves objects moving through this inertial frame.  The time dilation factor in SR is,
\bea \gamma_{\rm SR}(z) &=& (1-v_{\rm pec}^2/c^2)^{-1/2},\\
               &=&  \frac{1}{2}(1+z+\frac{1}{1+z})\approx 1 + z^2/2.\label{eq:gammaSRz}\eea
This time dilation factor relates the proper time in the moving emitter's inertial frame ($\Delta t_0$) to the proper time in the observer's inertial frame ($\Delta t_1$).  To measure this time dilation the observer has to set up a set of synchronized clocks (each at rest in the observer's inertial frame) and take readings of the emitter's proper time as the emitter moves past each synchronized clock.  The readings show that the emitter's clock is time dilated such that $\Delta t_1 = \gamma_{\rm SR} \Delta t_0$. 

We do not have this set of synchronized clocks at our disposal when we measure time dilation of supernovae and therefore Eq.~\ref{eq:gammaSRz} is not the time dilation we observe. 
For the observed time dilation of supernovae we have to take into account an extra time dilation factor that occurs because the distance to the emitter (and thus the distance light has to propagate to reach us) is increasing.   In the time $\Delta t_1$ the emitter moves a distance $v\Delta t_1$ away from us.  The total proper time we observe ($\Delta t$) is $\Delta t_1$ plus an extra factor describing how long light takes to traverse this extra distance ($v\Delta t_1/c$),
\beq \Delta t = \Delta t_1 (1+v/c). \eeq
The relationship between proper time at the emitter and proper time at the observer is thus,
\bea \Delta t &=& \Delta t_0 \gamma_{\rm SR} (1+v/c), \\
              &=& \Delta t_0 \sqrt{\frac{1+v/c}{1-v/c}},\\
              &=& \Delta t_0 (1+z).\eea
This is identical to the GR time dilation equation.  Therefore using time dilation to distinguish between GR and SR expansion is impossible.

Leibundgut \etal~(1996), 
Riess \etal~(1997) 
and Goldhaber \etal~(1997, 2001) 
do provide excellent evidence that expansion is a good explanation for cosmological redshifts.  What they can not show is that GR is a better description of the expansion than SR.  Nevertheless, other observational tests provide strong evidence against the SR interpretation of cosmological redshifts, and we demonstrate one such test in the next section.

\begin{figure}[h]\bctr
\includegraphics[width=84mm]{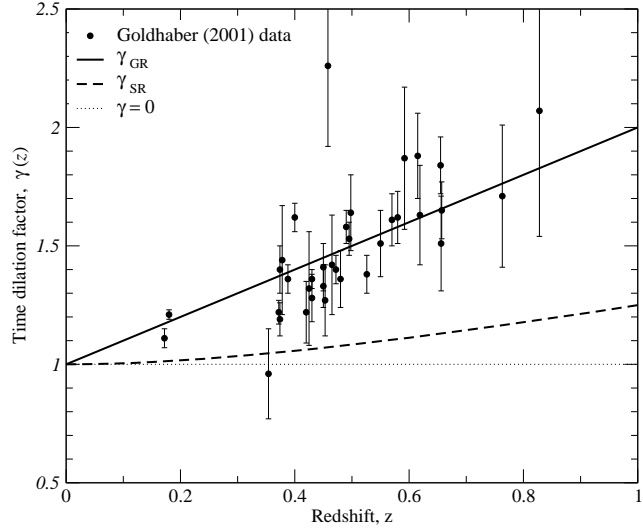}
\caption[Supernovae time dilation factor vs redshift.]{\small Supernovae time dilation factor vs redshift.  The solid line is the time dilation factor predicted by both general relativity and special relativity.  The thick dashed line is the special relativistic time dilation factor that a set of synchronized clocks spread throughout our inertial frame would observe, without taking into account the changing distance light has to travel to reach us.  Once the change in the emitter's distance is taken into account SR predicts the same time dilation effect as GR, $\gamma = (1+z)$.  The thin dotted line represents any theory that predicts no time dilation (e.g. tired light). The 35 data points are from Goldhaber \etal~(2001).    
They rule out no time dilation at a confidence level of $18\sigma$.  
}
\label{fig:Goldhaber}
\ectr\end{figure}


\begin{figure} \bctr
\includegraphics[width=86mm]{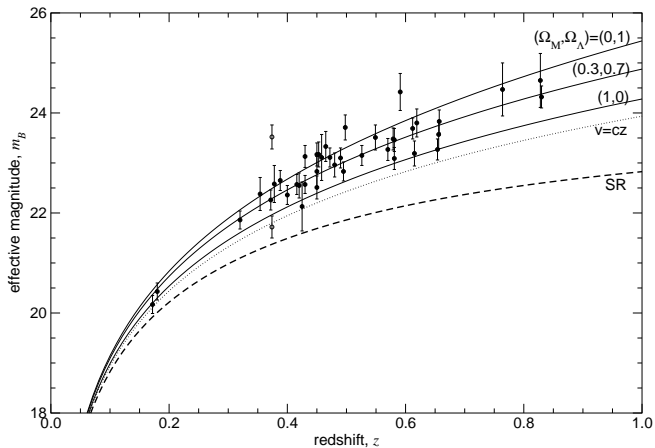}
\caption[Magnitude-redshift relation for SNIa]{\small Magnitude-redshift relation for several models with data taken from Perlmutter et al. 1999 [Fig.~2(a)].  The SR prediction has been added (as described in text), as has the prediction assuming a linear $v=cz$ relationship.  The interpretation of the cosmological redshift as an SR Doppler effect is  ruled out at more than $23\sigma$ compared with the $\Lambda$CDM concordance model. 
The linear $v=cz$ model is a better approximation than SR, but is still ruled out at $12\sigma$.  
}
\label{fig:mag-z}
\ectr \end{figure}

\subsection{Magnitude-redshift relationship for SNe Ia}\label{sect:mag-z}
Another observational confirmation of the GR interpretation that {\em is} able to rule out the SR interpretation is the curve in the magnitude-redshift relation.  SNe Ia are being used as standard candles to fit the magnitude-redshift relation out to redshifts close to one 
\citep{perlmutter99,reiss98}.  
Recent measurements are accurate enough to put restrictions on the cosmological parameters $\omol$.  We perform a simple analysis of the supernovae magnitude-redshift data to show that it also strongly excludes an SR interpretation of cosmological redshifts (Fig.~\ref{fig:mag-z}).

Figure~\ref{fig:mag-z} shows the theoretical curves for several GR models accompanied by the observed SNe Ia data from 
\cite{perlmutter99} 
 [their Fig.~2(a)].  The conversion between luminosity distance, $D_L$ (Eq.~\ref{eq:luminosity}), and effective magnitude in the B-band given in 
\cite{perlmutter99}, 
is $m_{\rm B}(z)=5\log H_0 D_L + M_{\rm B}$ where $M_{\rm B}$ is the absolute magnitude in the $B$-band at the maximum of the light curve. They marginalize over $M_{\rm B}$ in their statistical analyses.  We have taken $M_{\rm B} =-3.45$ which closely approximates their plotted curves.  

We superpose the curve deduced by interpreting Hubble's law special relativistically.  One of the strongest arguments against using SR to interpret cosmological redshifts is the difficulty in interpreting observational features such as magnitude.    We calculate $D(z)$ special relativistically by assuming the velocity in $v=HD$ is related to redshift via Eq.~\ref{eq:vSR}, so, 
\beq D(z)=\frac{c}{H}\frac{(1+z)^2-1}{(1+z)^2+1}.\eeq  
Special relativity does not provide a technique for incorporating acceleration into our calculations for the expansion of the Universe, so the best we can do is assume that the recession velocity, and thus Hubble's constant, are approximately the same at the time of emission as they are now\footnote{There are several complications that this analysis does not address.  (1) SR could be manipulated to give an evolving Hubble's constant and (2) SR could be manipulated to give a non-trivial relationship between luminosity distance, $D_L$, and proper distance, $D$.  However, it is not clear how one would justify these ad hoc corrections.}.  We then convert $D(z)$ to $D_L(z)$ using Eq.~\ref{eq:luminosity}, so $D_L(z)=D(z)(1+z)$.  This version of luminosity distance has been used to calculate $m(z)$ for the SR case in Fig.~\ref{fig:mag-z}. 

Special relativity fails this observational test dramatically being $23\sigma$ from the general relativistic $\Lambda$CDM model $\omol=(0.3,0.7)$.  
We also include the result of assuming $v=cz$.  Equating this to Hubble's law gives, $D_L(z)=cz(1+z)/H$.  For this observational test the linear prediction is closer to the GR prediction (and to the data) than SR is.  Nevertheless the linear result lies $12\sigma$ from the $\Lambda$CDM concordance result.



\subsection{Cosmological redshift evolution}\label{sect:futuretests-z}
\begin{figure}[!t] \bctr
\includegraphics[width=84mm]{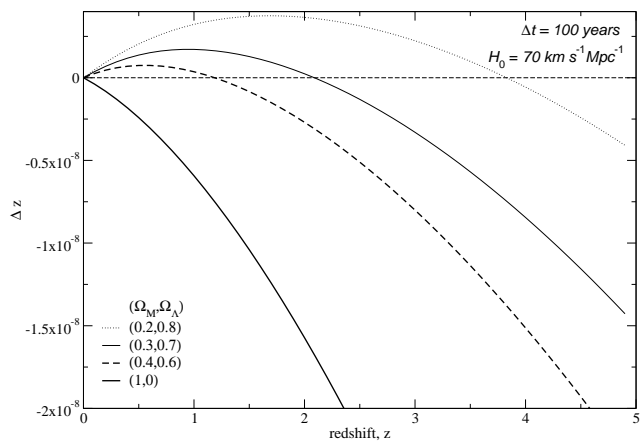}
\caption[Redshift evolution for four cosmological models]{\small The change in the redshift  of a comoving object as predicted by FRW cosmology for various cosmological models.  The horizontal axis represents the initial redshifts.  The timescale taken for the change is 100 years.  The changes predicted are too small for current instrumentation to detect.}
\label{fig:dz-z}
\ectr \end{figure}

Current instrumentation is not accurate enough to perform some other observational tests of GR.  For example \cite{sandage62} showed that the evolution in redshift of distant galaxies due to the acceleration or deceleration of the universe is a direct way to measure the cosmological parameters.  The change in redshift with time is given by,
\bea
1+z 	&=& \frac{R_0}{R_{\rm e}} = \frac{\Delta t_0}{\Delta t_{\rm e}}\\
\frac{dz}{dt_0} &=& \frac{R_{\rm e}\frac{dR_0}{dt_0}-R_0\frac{dR_{\rm e}}{dt_0}}{R_{\rm e}^2}\\
	&=& \frac{\dot{R}_0}{R_{\rm e}} - \frac{R_0}{R_{\rm e}^2}  \frac{dR_{\rm e}}{dt_{\rm e}} \frac{dt_{\rm e}}{dt_0}\\
 	&=& \frac{R_0}{R_{\rm e}}\left(\frac{\dot{R}_0}{R_0}-\frac{\dot{R}_{\rm e}}{R_{\rm e}}\frac{1}{(1+z)}\right)\\
	&=& H_0 (1+z) - H_{\rm e},\label{eq:dz}\eea
\citep[c.f.][Eq.~3]{loeb98} where $H_{\rm e}=\dot{R}_{\rm e}/R_{\rm e}$ is Hubble's constant at the time of emission,
\beq H_{\rm e} = \frac{\dot{R}_{\rm e}}{R_{\rm e}} = H_0\left[1+\om z + \oll\left(\frac{1}{(1+z)^2}-1\right)\right]^{1/2}.\eeq
  Unfortunately the magnitude of the redshift variation is small over human timescales.  \cite{ebert75}, \cite{lake81}, \cite{loeb98} and references therein each reconfirmed that the technology of the day did not yet provide precise enough redshifts to make such an observation viable.  Figure~\ref{fig:dz-z} shows the expected change in redshift due to cosmological acceleration or deceleration is only $\Delta z \sim 10^{-8}$ over 100 years.   Current Keck/HIRES spectra 
can measure quasar absorption line redshifts to an accuracy of $\Delta z \sim 10^{-5}$~\citep{outram99}.  Observations of many sources' redshifts, treated statistically, could improve this limit, but not enough.  This observational test must wait for future technology.

\section{Discussion}\label{sect:misc_discussion}

Recession velocities of individual galaxies are of limited use in observational cosmology because they are not directly observable.  For this reason some of the physics community considers recession velocities meaningless and would like to see the issue swept under the rug~\incite{24--26}. 
It is arguable that we should refrain from quoting our observables in terms of velocity or distance, and stick to the observable, redshift.  This avoids any complications with superluminal recession and avoids any confusion between the variety of observationally-motivated definitions of distance commonly used in cosmology\footnote{\vspace{-5mm}\bea \mbox{Proper Distance} & D  = &R \chi\\
   \mbox{Luminosity Distance}& D_L= &R S_k(\chi)(1+z)\label{eq:luminosity}\\
   \mbox{Angular Diameter Distance}& D_{\theta}= &R S_k(\chi)(1+z)^{-1}\label{eq:angdiam}\eea }.  

However, redshift is not the only observable that indicates distance and velocity.  The host of low redshift distance measures and the multitude of available evidence for the Big Bang model all suggest that higher redshift galaxies are more distant from us and receding faster than lower redshift galaxies.  Moreover, we cannot currently sweep distance and velocity under the rug if we want to explain the cosmological redshift itself.  Expansion has no meaning without velocity and distance.  If recession velocity were meaningless we could not refer to an ``expanding Universe'' and would have to restrict ourselves to some operational description such as ``fainter objects have larger redshifts''.  However, using general relativity the relationship between cosmological redshift and recession velocity is straightforward.  Data such as the time-dilation effect seen in SNe Ia light curves provide independent evidence that cosmological redshifts are due to the expansion of the Universe.  Since the expansion of the Universe is so fundamental to our modern world view, we consider the concepts of distance and velocity pertinent to an understanding of our Universe.

When distances are large enough that light has taken a substantial fraction of the age of the Universe to reach us there are more observationally convenient distance measures than proper distance, such as luminosity distance (Eq.~\ref{eq:luminosity}) and angular-diameter distance (Eq.~\ref{eq:angdiam}).  The most convenient distance measure depends on the method of observation.  Nevertheless, all distance measures can be converted between each other, and so collectively define a unique concept. In this chapter and for the rest of this thesis we take proper distance to be the fundamental {\em radial} distance measure.    Proper distance is the spatial geodesic measured along a hypersurface of constant cosmic time (as defined in the Robertson-Walker metric).  It is the distance measured along a line of sight by a series of infinitesimal comoving rulers at a particular time, $t$ (\citeauthor{weinberg72}, 1972, p.~415; \citeauthor{rindler77}, 1977, p.~218).  Both luminosity and angular diameter distances are calculated from observables involving distance perpendicular to the line of sight and so contain the angular coefficient $S_k(\chi)$.  They parametrize radial distances but are not geodesic distances along the three dimensional spatial manifold\footnote{Note also that the standard definition of angular size distance is purported to be the physical size of an object, divided by the angle it subtends on the sky.  The physical size used in this equation is not actually a length along a spatial geodesic, but rather along a line of constant $\chi$ \citep{liske00}.  The correction is negligible for the small angles usually measured in astronomy.}.  They are therefore not relevant for the calculation of recession velocity\footnote{Murdoch, H.~S. 1977, ``[McVittie] regards as equally valid other definitions of distance such as luminosity distance and distance by apparent size.  But while these are extremely useful concepts, they are really only definitions of observational convenience which extrapolate results such as the inverse square law beyond their range of validity in an expanding universe''}~\citep{murdoch77}. 
Nevertheless, if they were used, our results would be similar.  Only angular size distance can avoid superluminal velocities~\citep{murdoch77} because $D_\theta=0$ for both $z=0$ and $z\rightarrow \infty$.  Even then the rate of change of angular size distance does not approach $c$ for $z\rightarrow \infty$. 

Throughout this thesis we use proper time, $t$, as the temporal measure.  This is the time that appears in the RW metric and the Friedmann equations. This is a convenient time measure because it is the proper time of comoving 
observers.  Moreover, the homogeneity of the Universe is dependent on this choice of time coordinate --- if any other time coordinate were chosen (that is not a trivial multiple of $t$) the density of the Universe would be distance dependent.  
Time can be defined differently, for example to make the SR Doppler shift formula  (Eq.~\ref{eq:vSR}) correctly calculate recession velocities from observed redshifts \citep{page93}.  
However, to do this we would have to sacrifice the homogeneity of the universe and the synchronous proper time of comoving objects (Chapter~\ref{chap:coord}).

\section{Conclusion}
We have clarified some common misconceptions surrounding the expansion of the Universe, and shown with numerous references how abundant these misconceptions are.  
Superluminal recession is a feature of all expanding cosmological models that are homogeneous and isotropic and therefore obey Hubble's law.  This does not contradict special relativity because the superluminal motion does not occur in any observer's inertial frame.  All observers measure light locally to be travelling at $c$ and nothing ever overtakes a photon.  Inflation is often called ``superluminal recession'' but even during inflation objects with $D<c/H$ recede subluminally while objects with $D>c/H$ recede superluminally.  Precisely the same relationship holds for non-inflationary expansion.  We showed that the Hubble sphere is not an horizon --- we routinely observe galaxies that have, and always have had, superluminal recession velocities.  All galaxies at redshifts greater than $z\sim 1.45$ today are receding superluminally in the $\Lambda$CDM concordance model.  We have also provided a more informative way of depicting the particle horizon on a spacetime diagram than the traditional worldline method.  
An abundance of observational evidence supports the general relativistic big bang model of the universe.  The observed duration of supernovae light curves \citep{goldhaber01} shows that cosmological redshifts are well explained by the expansion of the Universe, but does not distinguish between GR and SR expansion.   Using magnitude-redshift data from supernovae \citep{perlmutter99} we were able to rule out the SR interpretation of cosmological redshifts at the $\sim23\sigma$ level.   
These observations provide strong evidence that 
the general relativistic interpretation of the cosmological redshifts is preferred over tired light and special relativistic interpretations.  The general relativistic description of the expansion of the Universe agrees with observations, and does not need any modifications for $\vrec > c$. 


\clearemptydoublepage

\chapter{The effect of the expansion of space on non-comoving systems}\label{chap:chain}


In Chapter~\ref{chap:misconcept} we addressed misconceptions surrounding the expansion of the Universe.  In this chapter we address the question of what effect the expansion of the Universe has on local systems that are not expanding with the Hubble flow.

Debate persists over what spatial scales participate in the expansion of the Universe \citep{munley95,shi98,tipler99,chiueh02,dumin02}, and the effect of the expansion of the Universe on local systems is a topic of current research~\citep{lahav91,anderson95,cooperstock98,hamilton01,baker02}.   A persistent confusion is that galaxies set up at rest with respect to us and then released will start to recede as they pick up the Hubble flow.  This is similar to the assumption that, without a force to hold them together, galaxies (or even our bodies) would be stretched as the Universe expands.
In this chapter we clarify the nature of the expansion of the Universe, by looking at the effect of the expansion on objects that {\em are not} receding with the Hubble flow.  This is an extension of previous discussions \citep[e.g.][]{silverman86,stuckey92,stuckey92b,ellis93,tipler96b,munley95}.

To clarify the influence of the expansion of the universe we consider the `tethered galaxy' problem \citep{harrison95,peacock01}.  We set up a distant galaxy at a constant distance from us and then allow it to move freely.  The essence of the question is, once it has been removed from the Hubble flow and then let go, what effect, if any, does the expansion of the Universe have on its movement?
In the next section we derive and illustrate solutions to the tethered galaxy problem for arbitrary values of the density of the universe $\om$ and the cosmological constant $\oll$. 
We show that the untethered galaxy's behaviour depends upon the model universe used.  In all cases the untethered galaxy rejoins the Hubble flow -- {\em but} the untethered galaxy does not always start to recede from us.  In decelerating universes the untethered galaxy, initially at rest, falls through our position and joins the Hubble flow on the opposite side of the sky.   This does not argue against the concept of expanding space \citep{peacock99,peacock01}, but highlights the common false assumption that there is a force or drag associated with the expansion of space.  We show that an object that is not participating in the expansion does rejoin the Hubble flow in all eternally expanding universes, but does not feel any force causing it to rejoin the Hubble flow.  This qualitative result extends to all objects with a peculiar velocity.
 Our calculations agree with and generalize the results obtained by \citet{peacock01} however we also point out an interesting interpretational difference.
In Section~\ref{sect:rel-sol} we extend the analysis to relativistic peculiar velocities.

The cosmological redshift is important because it is the most readily observable evidence of the expansion of the Universe.  In Section~\ref{sect:redshift} we point out a consequence of the fact that the cosmological redshift is not a special relativistic Doppler shift; we derive the counter-intuitive result that our tethered galaxy, constrained to have zero total velocity, does not have zero redshift. To our knowledge this is the first explicit derivation of this counter-intuitive behaviour.  The approaching jet of some active galactic nuclei (AGN) provide examples of receding blueshifted objects (Sect.~\ref{sect:obs-cons}).  
We show that a total velocity of zero does not result in zero redshift even in the empty universe case.  This is particularly surprising because we would expect an {\em empty} FRW universe to be well described by special relativity in flat Minkowski spacetime.  In Chapter~\ref{chap:coord} we examine the empty universe case and use it to demonstrate the relationship between special relativity and FRW cosmology.

This chapter is primarily based on the work published in~\cite{davis03chain}.

\section{The tethered galaxy problem}\label{sect:nostretch}
\label{sect:problem}

\begin{figure}
\bctr
\includegraphics[width=140mm]{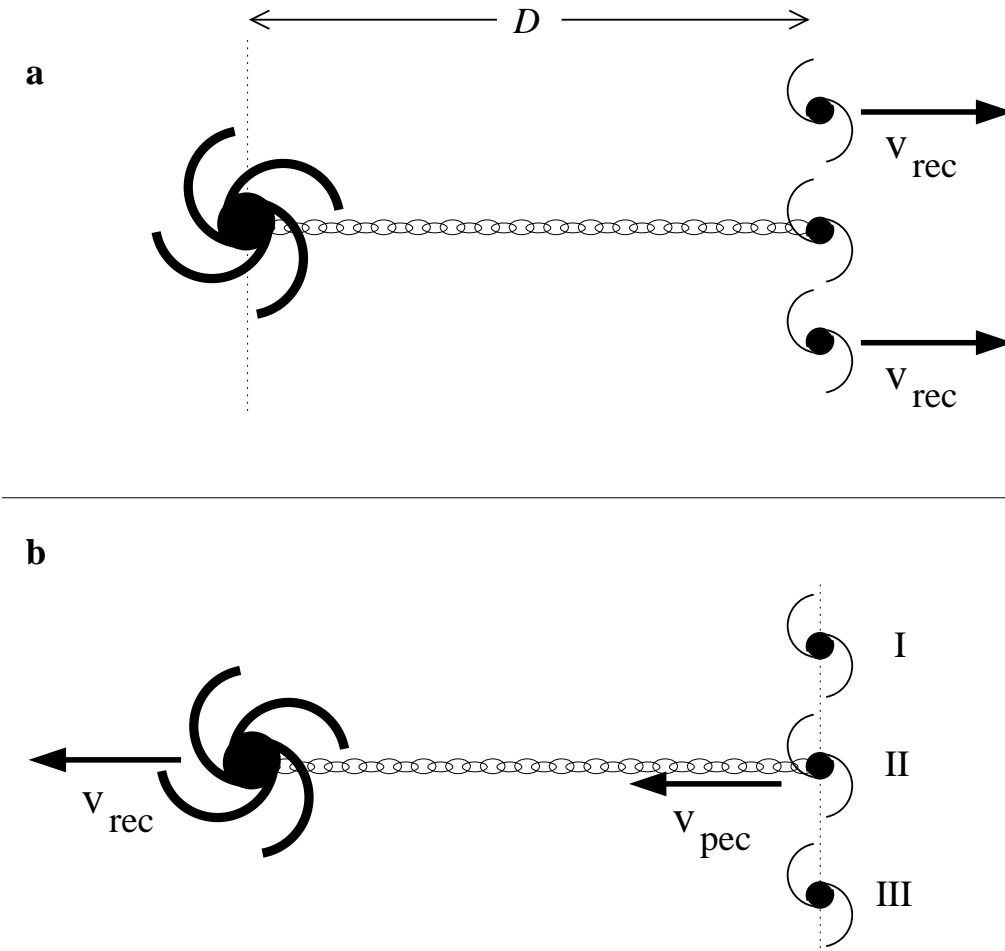}
\caption[The tethered galaxy]{\small (a) A small distant galaxy (considered to be a massless
test particle) is tethered to an observer in a large galaxy. The
proper distance to the small galaxy, $D$, remains fixed; the small
galaxy does not share the recession velocity of the other galaxies
at the same distance. The tethered galaxy problem is ``What path
does the small galaxy follow when we unhook the tether?'' (b)
Drawn from the perspective of the local comoving frame (out of
which the test galaxy was boosted), the test galaxy has a peculiar
velocity equal to the recession velocity of the large galaxy.
Thus, the tethered galaxy problem can be reduced to ``How far does
an object, with an initial peculiar velocity, travel in an
expanding universe?''}
\label{fig:problem}
\ectr
\end{figure}
Figure~\ref{fig:problem} illustrates the tethered galaxy problem.  
  Suppose we separate a small test galaxy from the Hubble flow by tethering it to an observer's galaxy such that the proper distance between them remains constant.  We neglect all practical considerations of such a tether since we
can think of the tethered galaxy as one that has received a peculiar velocity boost towards the observer that exactly matches its recession velocity.  We then remove the tether (or turn off the boosting rocket).
This satisfies the initial condition of constant proper distance, $\dot{D}_0=0$, and the idea of tethering is incidental. 
For simplicity we will refer to this as the untethered or test galaxy.  
Note that this is an artificial setup; we have had to arrange for the galaxy to be moved out of the Hubble flow in order to apply this zero total velocity condition.  Thus it is not necessarily a primordial condition, merely an initial condition we have arranged for our experiment.  Nevertheless, the discussion can be generalized to any object that has obtained a peculiar velocity and in Sect~\ref{sect:obs-cons} we describe a similar situation that is found to occur naturally.  

Recall that total velocity is $v_{\rm tot} = \dot{D} =\dot{R}\chi + R\dot{\chi}= v_{\rm rec} + v_{\rm pec}$.
We define ``approach'' and ``recede'' as $\dot{D}<0$ and $\dot{D}>0$ respectively.  The motion of this test galaxy reveals the effect the expansion of the Universe has on local dynamics.
To enable us to isolate the effect of the expansion of the Universe we  assume that the galaxies have negligible mass.    
By construction the tethered galaxy at an initial time $t_0$ has zero total velocity,  $\dot{D}_0 = 0$.  In other words, its initial peculiar velocity exactly cancels its initial recession velocity,
\bea
v_{\rm pec_0}  &=& - v_{\rm rec_0,\label{eq:initial1}}\\
R_0\dot{\chi}_0 &=& -\dot{R}_0\chi_0.\label{eq:initial}
\eea
With this initial condition established we untether the galaxy
and let it coast freely.  The question is then: Does the test galaxy approach, recede or stay at the same distance?

The momentum with respect to the local comoving frame, $p$, decays as $1/a$ \citep{weinberg72,misner73,peebles93,padmanabhan96}, see Section.~\ref{sect:pecdecay}. 
This scale factor dependent decrease in momentum is an important basis for many of the results that follow. 
For non-relativistic velocities  $p=mv_{\rm pec}$.  This means,
\bea v_{\rm pec} &=& \frac{v_{\rm pec_0}}{a}, \label{eq:Rinv}\\
R\dot{\chi}      &=& \frac{-\dot{R}_0\chi_0}{a}, \label{eq:Rchidot} \\
\chi   &=& \;\;\chi_o\left[1-{H}_0\int_{\rm t_o}^{t}\frac{dt}{a^{2}}\right],     \label{eq:chisolution}\\ 
D   &=& R\,\chi_o\left[1-{H}_0\int_{\rm t_o}^{t}\frac{dt}{a^{2}}\right].     \label{eq:Dsolution}
\eea
(For the relativistic solution see Section~\ref{sect:rel-sol}.)
The integral in Eqs.~(\ref{eq:chisolution}) \& (\ref{eq:Dsolution})  can be performed numerically 
by using $dt=da/\dot{a}$ where we obtain $\dot{a}$
directly from the Friedmann equation, 
\beq 
\dot{a}=\frac{da}{dt}=H_0\left[1+\om\left(\frac{1}{a}-1\right)+\oll(a^2-1)\right]^{1/2} . \label{eq:frieda}
\eeq 
The normalized matter density $\om=8\pi G\rho_0/3H_0^2$ and cosmological constant 
$\oll=\Lambda/3H_0^2$ are constants calculated at the present day.   
The scale factor $a(t)$ is derived by integrating the Friedmann equation \citep{felten86}.  In \cite{davis03chain} the constants look slightly different, because we attributed to $\chi$ the dimensions of distance.  The formalism used here is more explicit and hopefully clearer.  We define both $\chi$ and $a=R/R_0$ to be dimensionless and use $R$ and $R_0$ to explicitly track the dimension of distance. 

\begin{figure}\bctr
\includegraphics[width=140mm]{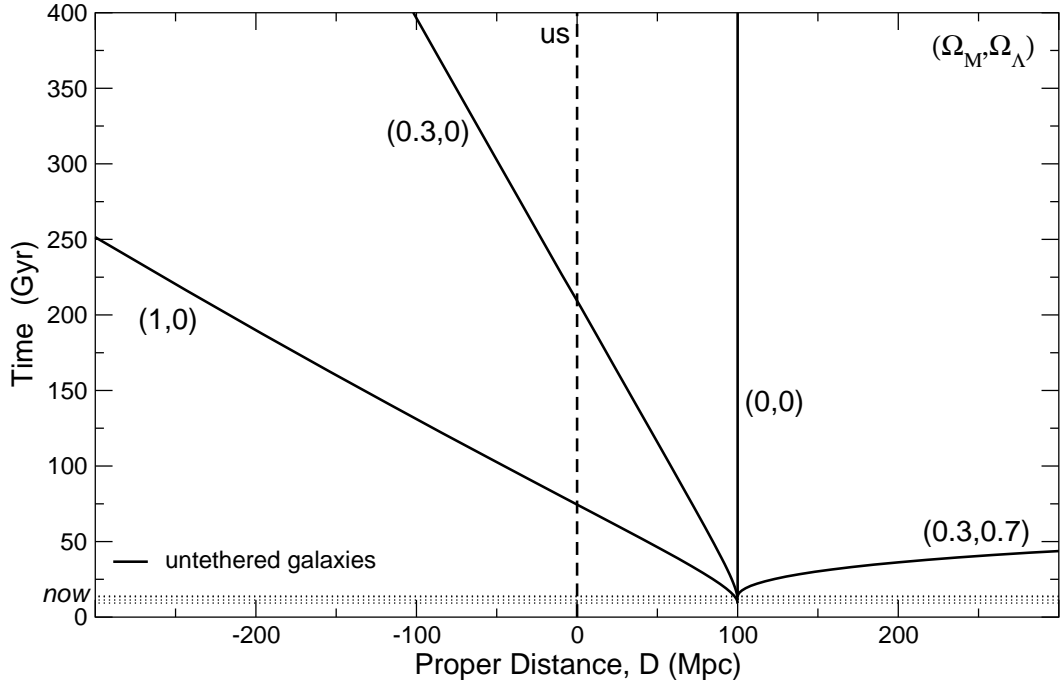}
\caption[Solutions to the tethered galaxy problem (proper distance)]{\small Solutions to the tethered galaxy problem (Eq.~\ref{eq:Dsolution}). For four cosmological models we
untether a galaxy at a distance of $D_0=100$\,Mpc with an initial
peculiar velocity equal to its recession velocity (total initial
velocity is zero) and plot its path. In each case the peculiar
velocity decays as $1/a$. Its final position depends on the model.
In the $\omol=(0.3,0.7)$ accelerating universe, the untethered
galaxy recedes from us as it joins the Hubble flow, while in the
decelerating examples, $\omol=(1,0)\ \mbox{and}\ (0.3,0)$, the
untethered galaxy approaches us, passes through our position and
joins the Hubble flow in the opposite side of the sky. In the
$\omol=(0,0)$ model the galaxy experiences no acceleration and
stays at a constant proper distance as it joins the Hubble flow (Eq.~\ref{eq:q}). In Sec.~\ref{sect:redshift} and
Fig.~\ref{fig:z0vpecvrec} we derive and illustrate the
counter-intuitive result that such a galaxy will be blueshifted.
We are the comoving galaxy represented by the thick dashed line
labeled ``us.'' There is a range of values labeled ``now,''
because the current age of the Universe is different in each
model.} \label{fig:Dsolution}\ectr
\end{figure}

\begin{figure}\bctr 
\includegraphics[width=140mm]{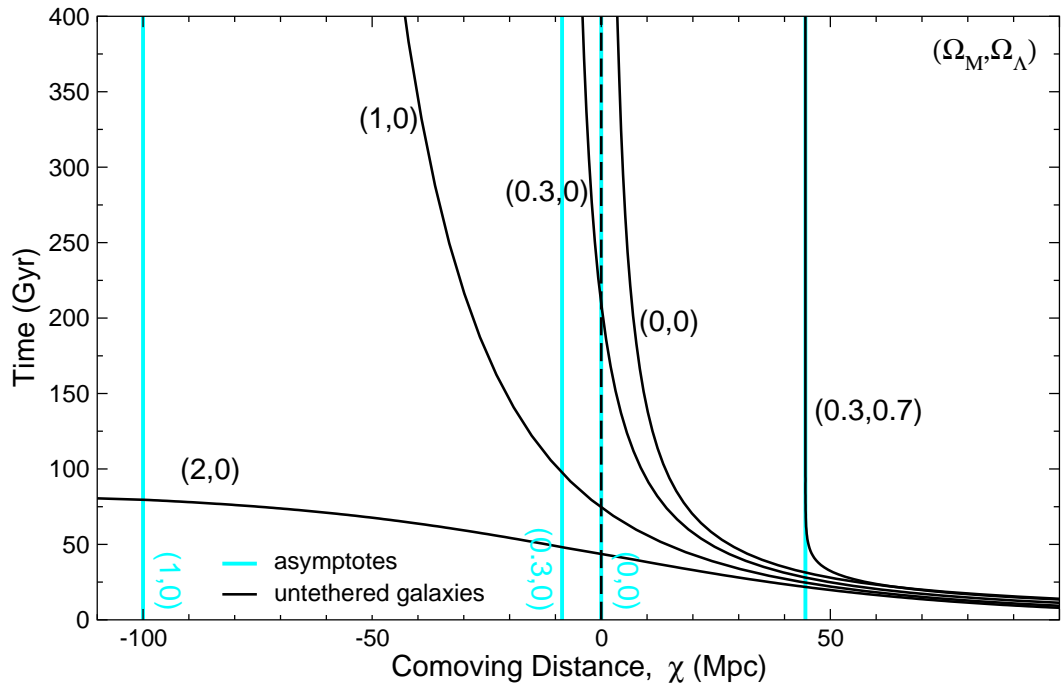}
\caption[Solutions to the tethered galaxy problem (comoving distance)]{\small Solutions to the tethered galaxy problem in comoving
coordinates (Eq.~\ref{eq:chisolution}) for five cosmological
models. In all the models the comoving coordinate of the
untethered galaxy decreases (our initial condition specified a
negative peculiar velocity). In models that do not recollapse the
untethered galaxy coasts and approaches an asymptote as it joins
the Hubble flow. The rate of increase of the scale factor
determines how quickly an object with a peculiar velocity joins
the Hubble flow. In the accelerating universe $\omol = (0.3,0.7)$,
the perturbed galaxy joins the Hubble flow more quickly than in
the decelerating universes $(1,0)$ and $(0.3,0)$, with the $(0,0)$
universe in between. The $\omol = (2,0)$ model is the only model
shown that recollapses. In the recollapsing phase of this model
the galaxy's peculiar velocity increases as $a$ decreases and the
galaxy does not join the Hubble flow  (Eq.~\ref{eq:joinflow}). In
the (0,0) model the proper distance, $D$, to the untethered galaxy is
constant, and therefore its comoving distance $\chi=D/R$ tends
toward zero (our position) as $R$ tends toward infinity. The
different models have different starting points in time because
the current age of the Universe is different in each model.}
\label{fig:chisolution}\ectr
\end{figure}

Equation~(\ref{eq:Dsolution}) provides the general solution to the tethered galaxy problem.  
Figure~\ref{fig:Dsolution} shows this solution for four different models.  In the currently 
favored, $\omol=(0.3,0.7)$, model the untethered galaxy recedes.  In the empty, $\omol=(0,0)$ 
universe,  it stays at the same distance while in the previously favored 
Einstein-de Sitter model, $\omol=(1,0)$, 
and the $\omol=(0.3,0)$ model, it approaches.  
The different behaviours in each model ultimately stem from the different compositions of the universes, since the composition dictates the acceleration.  When the cosmological constant is large enough to cause the expansion of the universe to accelerate, the test galaxy will also accelerate away.  When the attractive force of gravity dominates, decelerating the expansion, the test galaxy approaches.  This may not seem surprising, but it is surprising when you have a preconceived notion that the expansion is a ``stretching of space'' and therefore should be dragging all points in the universe apart.  We consider stretching of space a useful concept, but warn that we should not follow the analogy too closely. 

General solutions in comoving coordinates of the tethered galaxy problem are given by
Equation~\ref{eq:chisolution} and are plotted in Fig.~\ref{fig:chisolution}
for the same four models shown in Fig.~\ref{fig:Dsolution},
as well as for a recollapsing model, $\omol=(2,0)$.  

\subsection{Expansion makes galaxies join the Hubble flow}
As demonstrated in Fig.~\ref{fig:chisolution}, the untethered galaxy asymptotically joins the Hubble flow in every cosmological model that expands forever.  When we think of the Hubble flow we automatically think of galaxies receding from us.  So it is natural to assume that as an object starts to join the Hubble flow, it starts to recede.
However, Fig.~\ref{fig:Dsolution} shows that whether the untethered galaxy joins the Hubble flow by approaching or receding from us is a different, model dependent issue. 
The untethered galaxy asymptotically joins the Hubble flow for all cosmological models that expand forever since,
\bea
\dot{D}          &=&v_{\rm rec} + v_{\rm pec}\\
                 &=&v_{\rm rec} + v_{\rm pec_o}/a. \label{eq:joinflow}
\eea
As $a \rightarrow \infty$ we have $\dot{D} =v_{\rm rec} = HD$; pure Hubble flow.  Note that this is entirely due to the expansion of the universe ($a$ increasing).  

We further see that the expansion does not effect dynamics since when we calculate the acceleration of the comoving galaxy, all terms in $\dot{R}$ (or $\dot{a}$) cancel out,
\bea \ddot{D}&=&\dot{v}_{\rm rec} -\frac{v_{\rm pec_0}}{a}\frac{\dot{a}}{a},\\
             &=&\dot{v}_{\rm rec} - v_{\rm pec}\;\frac{\dot{R}}{R},\\
            &=&(\ddot{R}\chi + \dot{R}\dot{\chi})  - \dot{R}\dot{\chi}, \label{eq:balance}\\
             &=&\ddot{R}\chi,   \label{eq:onlyacceleration}\\
             &=&-q\;H^2\,D,   \label{eq:q}
\eea
where the deceleration parameter 
$q(t)=-\ddot{R}R/\dot{R}^{2}$.
Notice that the second term in Eq.~(\ref{eq:balance}) owes its existence to $\dot{\chi} \neq 0$
(which is only true if $v_{\rm pec} \neq 0$) and here represents the galaxy moving to lower comoving coordinates.  The resulting reduction in recession velocity is exactly canceled by
the third term which is the decay of peculiar velocity.
Thus all terms in $\dot{R}$ cancel and we conclude that the expansion, $\dot{R} > 0$, 
does not cause acceleration, $\ddot{D} > 0$.   Thus, the expansion does not cause the untethered galaxy to recede (or to approach) but does result in the untethered galaxy joining the Hubble flow ($v_{\rm pec}\rightarrow 0$).

An alternative way to obtain Eq.~(\ref{eq:q}) is to differentiate Hubble's Law, $\dot{D}=HD$.
This method ignores $v_{\rm pec}$ and therefore does not include the explicit cancellation of the two terms in Eq.~(\ref{eq:balance}) of the more general calculation.  The fact that the results are the same emphasizes that the acceleration of the test galaxy is the same as that of comoving galaxies and there is no additional acceleration on our test galaxy pulling it into the Hubble flow. 

\begin{figure*}
\bctr \includegraphics[width=150mm]{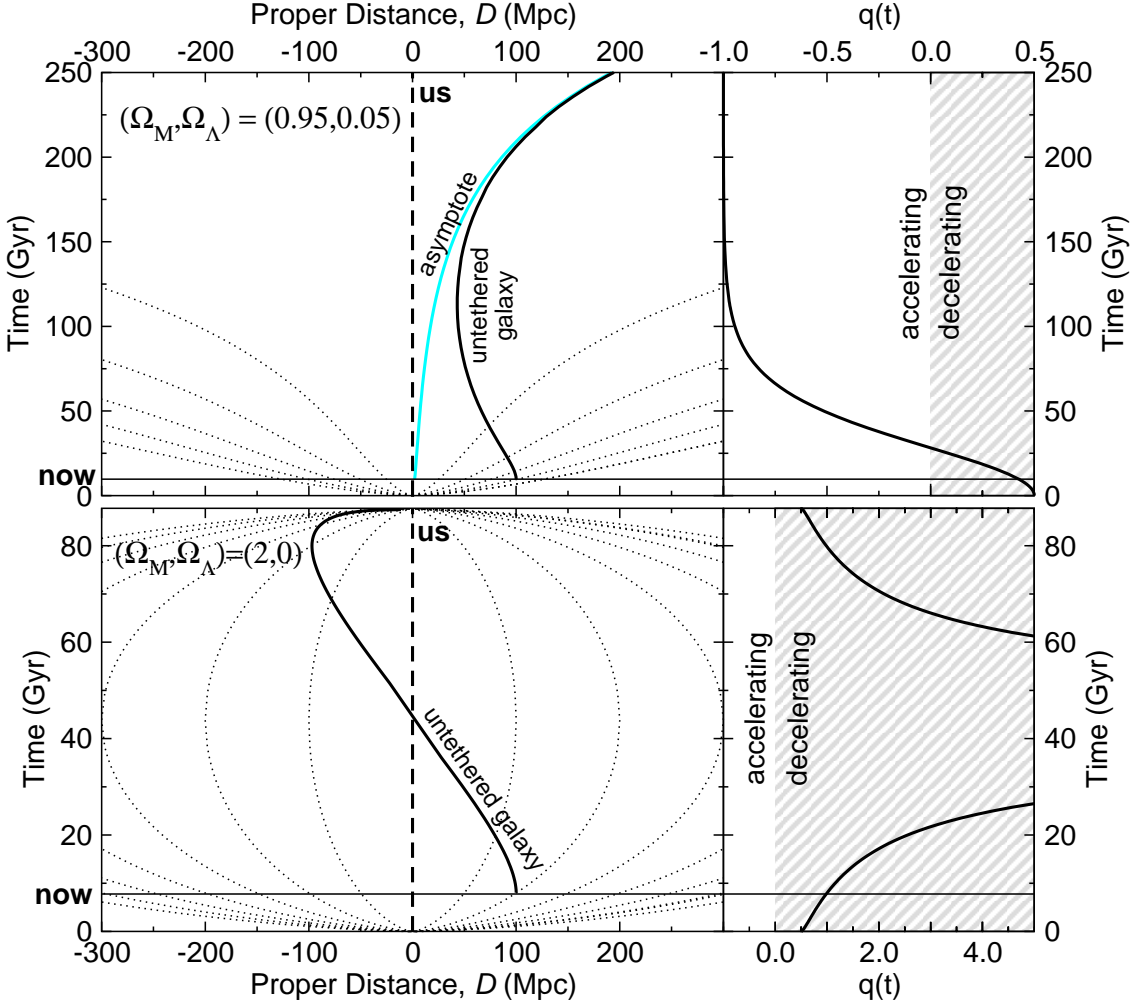}
\caption[Tethered galaxy (a) as $q$ changes sign and (b) in collapsing universe]{\small Upper panels: The deceleration parameter $q(t)$
determines the acceleration of the untethered galaxy
(Eq.~\ref{eq:q}) and can change sign. This particular model
shows the effect of $q$ (right panel) on the position of the
untethered galaxy (left panel). Initially $q > 0$ and the proper
distance to the untethered galaxy decreases (as in an $\omol =
(1,0)$ universe), but $q$ subsequently evolves and becomes
negative, reflecting the fact that the cosmological constant
begins to dominate the dynamics of the Universe. With $q < 0$, the
acceleration $\ddot{D}$ changes sign. This makes the approaching
galaxy slow down, stop, and eventually recede. The dotted lines
are fixed comoving coordinates. Lower panels: The $\omol=(2,0)$
universe expands and then recollapses ($\dot{a}$ changes sign),
and the peculiar velocity increases and approaches $c$ as
$R\rightarrow0$ (Eq.~\ref{eq:relativistic}).}
\label{fig:weirdexamples}\ectr
\end{figure*} 

\subsection[Acceleration of the expansion causes $\Delta v_{\rm tot}$]{Acceleration of the expansion makes the untethered galaxy approach or recede}
Since we start with the initial condition $\dot{D}_0=0$, whether the galaxy approaches or recedes from us is determined by whether it is accelerated towards us ($\ddot{D} < 0$) or away from us ($\ddot{D} > 0$).
Equation~(\ref{eq:q}) shows that, in an expanding universe,  whether the galaxy approaches us or recedes from us does not depend on the velocity of the 
Hubble flow (since $H>0$), or the distance of the untethered galaxy (since $D>0$), but on the sign of $q$.
When the universe accelerates ($q < 0$) the galaxy recedes from us. 
When the universe decelerates ($q > 0$) the galaxy approaches us. 
Finally, when $q = 0$ the proper distance stays the same 
as the galaxy joins the Hubble flow.  
Thus the expansion does not `drag' the untethered galaxy away from us, even though the untethered galaxy does end up joining the Hubble flow.  Only the {\em acceleration of the expansion} can result in a change in distance between us and the untethered galaxy.  We have shown that the direction of that change is not always outwards.

Notice that in Eq.~(\ref{eq:q}), $q= q(t) =q(a(t))$ is a function of 
scale factor,
\beq 
q(a)=\left(\frac{\om}{2a}-\oll a^2\right)\left[1+\om(\frac{1}{a}-1)+\oll(a^2-1)\right]^{-1},
\eeq 
which for $a(t_0)=1$ becomes the current deceleration parameter $q_0 = \om/2 -\oll$.  
Thus, for example, an $\omol=(0.66,0.33)$ model has $q_0=0$, but $q$ decreases with time; therefore the untethered galaxy recedes. 
The upper panels of Figure~\ref{fig:weirdexamples} show how a changing deceleration parameter affects the untethered galaxy. There is a time-lag between the onset of acceleration ($q < 0$) and the galaxy beginning to recede ($v_{\rm tot} > 0$) as is usual when accelerations and velocities are in different directions.

The example of an expanding universe in which an untethered galaxy approaches us exposes the common fallacy that ``expanding space'' is in some sense trying to drag all pairs of points apart.
The fact that in an $\omol=(1,0)$ universe the untethered galaxy, initially at rest, falls through our position and joins the Hubble flow on the other side of us does not argue against the idea of the expansion of space \citep{peacock99,peacock01}.  It does however highlight the common false assumption of a force or drag associated with the expansion of space.  
We have shown that an object with a peculiar velocity does rejoin the Hubble flow in eternally expanding universes but {\em does not feel any force} causing it to rejoin the Hubble flow.  This qualitative result extends to all objects with a peculiar velocity.

\begin{figure}
\bctr
\includegraphics[width=140mm]{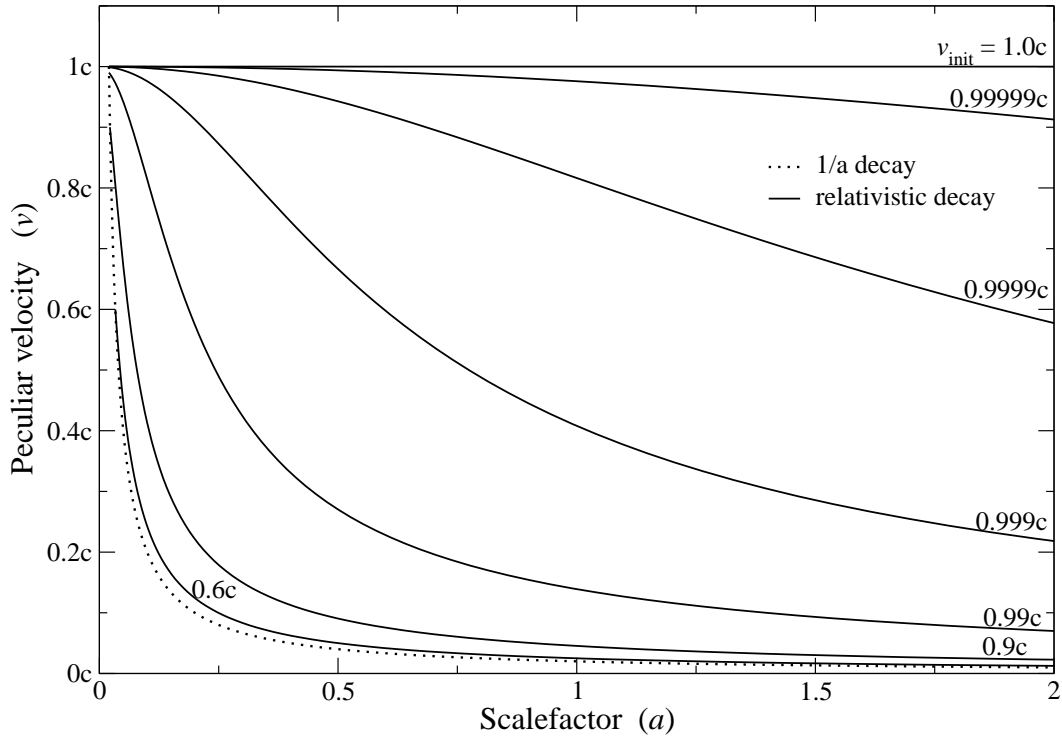}
\caption[Relativistic peculiar velocity decay]{\small The decay of velocity as the Universe expands is shown for relativistic particles when (1) peculiar velocity is assumed to decay as 1/a (dotted line) and (2) when peculiar velocity is treated special relativistically (solid lines).  The initial scalefactor is arbitrarily set to $a=0.02$.   Each line is labeled with the initial velocity except the non-relativistic ($1/a$) decay which begins with $v_{\rm init}=c$.   Relativistic velocities decay more slowly than they would if 1/a decay was assumed.  For all but the relativistic $v_{\rm init}=c$ case the peculiar velocities tend towards zero, so the particles tend towards comoving with the Hubble flow.  The extreme case of $v_{\rm init}=c$ does not decay in the relativistic case, reflecting the fact that photon velocity does not decay.   (The initial scalefactor for the $v=0.9c$ and $v=0.6c$ lines are $a=0.11$ and $a=0.167$ respectively so their starting points lie on the $1/a$ decay curve for easy comparison.)  This graph shows the continuum in behaviour between $1/a$ velocity decay for massive particles, and no velocity decay for photons.}
\label{fig:vrel}
\ectr
\end{figure}

\section{Relativistic peculiar velocity decay}\label{sect:rel-sol}

When a universe collapses, the scale
factor $a$ decreases. Thus
$v_{\rm pec}\propto 1/a$ means that the
peculiar velocity increases with time.   In collapsing
universes, untethered galaxies do not ``join the Hubble flow,'' they diverge from the Hubble flow.
This behavior is shown for the $\omol=(2,0)$ model in
Fig.~\ref{fig:chisolution}. Collapsing universes require the
relativistic formula for peculiar velocity decay to
avoid the infinite peculiar velocities that result from $v_{\rm
pec}\propto 1/a$ as $a\rightarrow0$.  The special
relativistic formula for momentum is $p=\gamma mv_{\rm pec}$, where
$\gamma=(1-v_{\rm pec}^2/c^2)^{-1/2}$.  Since momentum decays as
$1/a$ ($p=p_0a_0/a$), we obtain, 
\begin{equation}
v_{\rm pec}= \frac{\gamma_0 v_{\rm pec_0}}{\sqrt{a^2+\gamma_0^2
v_{\rm pec_0}^2/c^2}} . \label{eq:relativistic}
\end{equation}
Therefore, as $a\rightarrow0$, $v_{\rm pec}\rightarrow c$.
Equation~(\ref{eq:relativistic}) was used to produce the lower
panels of Fig.~\ref{fig:weirdexamples}.

The relativistic formula for momentum should also be used in eternally expanding universes if relativistic velocities are set as the initial condition in Eq.~\ref{eq:initial1}.
Using Eq.~\ref{eq:relativistic} in Eq.~\ref{eq:joinflow} results in a residual dependence on $\dot{R}$ in Eq.~\ref{eq:onlyacceleration}.  The residual is negligible for $v<\hspace{-1.5mm}<c$, and becomes negligible for $v\sim c$ as $a\rightarrow \infty$.   
Note that Eq.~\ref{eq:vpeczpec} is relativistic and therefore the results of Section~\ref{sect:redshift} hold for $v_{\rm pec}\sim c$.

Using this relativistic formula for peculiar velocity decay also removes an apparent discontinuity.  It may seem strange that momentum decaying as $1/a$ means the peculiar velocities of massive objects decay until the objects are comoving, and yet the peculiar velocities of photons always stay at $c$.  It seems that photons are getting some velocity boost that massive particles miss out on.  
However, when we treat the momentum decay relativistically we find that the velocity decay gets slower and slower as the particles become more relativistic.  In the limit of an initial peculiar velocity of $c$, Eq.~\ref{eq:relativistic} shows that peculiar velocities do not decay at all\footnote{\beq \lim_{v_{\rm pec,0}\rightarrow c}v_{\rm pec} = \lim_{v_{\rm pec,0}\rightarrow c}\frac{v_{\rm pec,0}}{\sqrt{(a^2/\gamma_0^2) + (v_{\rm pec,0}^2/c^2})} = \frac{c}{\sqrt{(a^2/\infty) + 1}} = c\eeq}.  Therefore this formula, which we illustrate in Fig.~\ref{fig:vrel}, shows the continuum in behaviour between non-relativistic particles and photons.

\section{Zero velocity corresponds to non-zero redshift}\label{sect:redshift}

\begin{figure*}\bctr
\includegraphics[width=150mm]{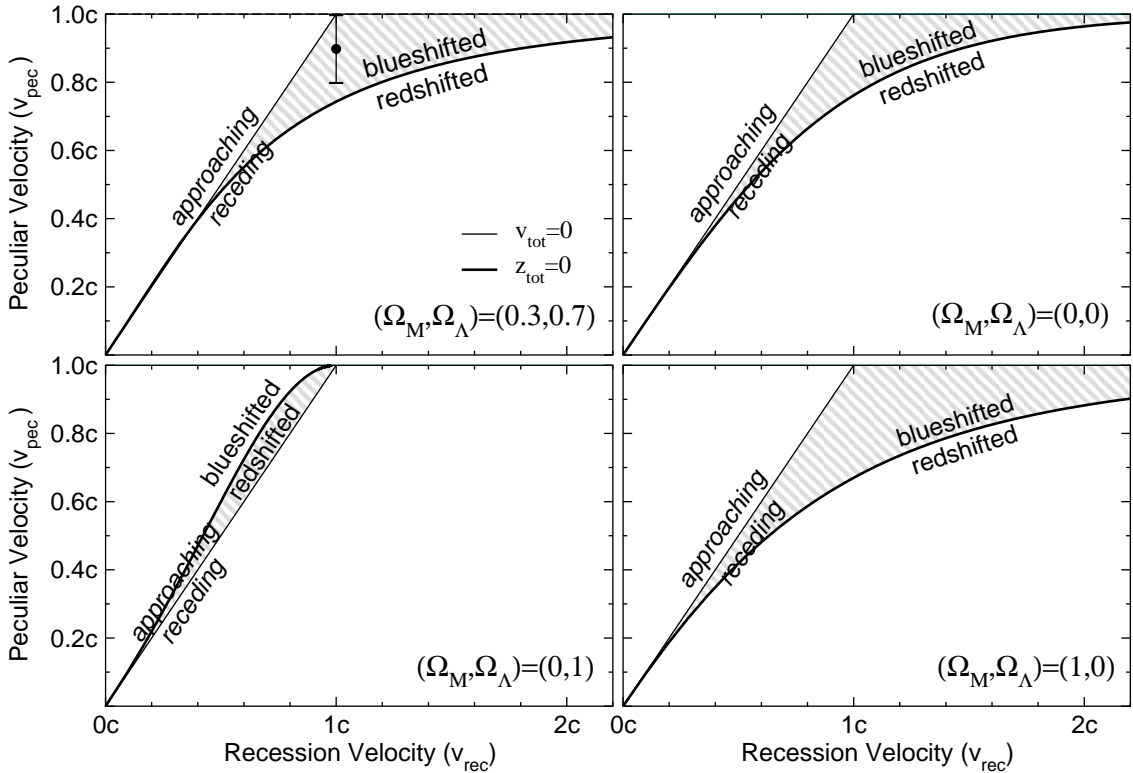}
\caption[Receding blueshifted and approaching redshifted galaxies]{\small{These graphs show the combination of recession velocity and peculiar velocity that result in a redshift of zero, for four cosmological models. The purpose of these graphs is to display the counter-intuitive result that in an expanding universe a redshift of zero does not correspond to zero total velocity ($\dot{D}=0$).  Gray striped areas show the surprising situations where receding galaxies appear blueshifted or approaching galaxies appear redshifted. Other models (e.g. $\omol=(0.05,0.95)$, Fig.~\ref{fig:weirdexamples}, top panel) can have both approaching redshifted and receding blueshifted regions simultaneously. Recession velocities are calculated at the time of emission; the results are qualitatively the same when recession velocities are calculated at the time of observation. Thus galaxies that were receding at emission and are still receding, can be blueshifted.  Notice that in each panel for low velocities (nearby galaxies) the $z_{\rm tot} = 0$ line asymptotes to the $v_{\rm tot} = 0$ line.  See Section~\ref{sect:obs-cons} for a discussion of the AGN jet data point in the upper left panel.
}}
\label{fig:z0vpecvrec} 
\ectr\end{figure*}

In the context of special
relativity (Minkowski space), objects at rest with respect to an
observer have zero redshift. However, in an expanding universe
special relativistic concepts do not generally apply. ``At rest''
is defined to be ``at constant proper distance'' ($v_{\rm
tot}=\dot{D}=0$), so our untethered galaxy with $\dot{D}_0 = 0$
satisfies the condition for being at rest. Will it therefore have
zero redshift?  Do objects with $v_{\rm tot}=0$ have 
$z_{\rm tot} = 0$? Although radial recession and peculiar velocities add
vectorially, their corresponding redshift components
combine\footnote{Light is emitted by the tethered galaxy. Let
$\lambda_{\rm observed}$ be the wavelength we observe,
$\lambda_{\rm emitted}$ be the wavelength measured in the comoving
frame of the emitter (the frame with respect to which it has a
peculiar velocity $v_{\rm pec}$) and $\lambda_{\rm rest}$ be the
wavelength of light in the rest frame of the emitter. Then
$1+z_{\rm tot} = \frac{\lambda_{\rm observed}}{\lambda_{\rm rest}}
= \frac{\lambda_{\rm observed}}{\lambda_{\rm
emitted}}\frac{\lambda_{\rm emitted}}{\lambda_{\rm rest}} =
(1+z_{\rm rec})(1+z_{\rm pec})$.} 
as $(1+z_{\rm tot}) = (1+z_{\rm
rec})(1+z_{\rm pec})$ \citep{kiang01}. The condition that $z_{\rm tot}=0$ gives,
\begin{equation}
(1+z_{\rm pec}) = \frac{1}{(1+z_{\rm rec})}. \label{eq:pecrec}
\end{equation}
The special relativistic relation between peculiar velocity and
Doppler redshift is,
\begin{equation}
v_{\rm pec}(z_{\rm pec}) = c \biggl[\frac{(1+z_{\rm pec})^{2} -
1}{(1+z_{\rm pec})^{2} + 1}\biggr] , \label{eq:vpeczpec} 
\end{equation}
while the general relativistic relation between recession velocity
(at emission\footnote{To calculate the current recession velocity
(as opposed to the recession velocity at the time of emission)
replace $z_{\rm rec}$ with $z=0$ in Eq.~(\ref{eq:vreczrec}) (except in the upper
limit of the integral).}) and cosmological redshift
is~\cite[][Eq.~13]{harrison93},
\begin{equation}
 v_{\rm rec}(z_{\rm rec}) = c\; \frac{H(z_{\rm rec})}{1+z_{\rm rec}} \int_0^{z_{\rm rec}}\frac{dz}{H(z)}, \label{eq:vreczrec} 
\end{equation}
where $H(z_{\rm rec})=H(t_{\rm em})$ is Hubble's constant at the
time of emission (Eq.~\ref{eq:H}).
In Fig.~\ref{fig:z0vpecvrec} 
we plot the $v_{\rm tot} = 0$ and the
$z_{\rm tot} = 0$ solutions to show they are not coincident. To obtain the
$z_{\rm tot} = 0$ curve, we do the following: For a given $v_{\rm
rec}$ we use Eq.~(\ref{eq:vreczrec}) to calculate $z_{\rm rec}$
(for a particular cosmological model). Equation~(\ref{eq:pecrec})
then gives us a corresponding $z_{\rm pec}$ and we can solve for
$v_{\rm pec}$ using Eq.~(\ref{eq:vpeczpec}). The result is the
combination of peculiar velocity and recession velocity required
to give a total redshift of zero.  The fact that the $z_{\rm tot}
= 0$ curves are different from the $v_{\rm tot} = 0$ line in all
models shows that $z_{\rm tot} = 0$ is not equivalent to $v_{\rm
tot} = 0$. When $v_{\rm tot}=0$ the blueshift due to the approaching peculiar velocity 
does not balance the redshift due to the expansion of the Universe.

That the $z_{\rm tot} = 0$ line is not the same as the $v_{\rm
tot} = 0$ line even in the $q=0$, $\omol = (0,0)$ model (upper
right Fig.~\ref{fig:z0vpecvrec}) 
is particularly surprising
because we might expect an empty expanding FRW universe to be
well-described by special relativity in flat Minkowski spacetime.
Zero velocity approximately corresponds to zero redshift only for 
$v_{\rm rec} \lsim 0.3 c$ or $z_{\rm rec} \lsim 0.3$, even in 
the $\omol=(0,0)$ model.  
Therefore an empty FRW universe does not trivially reduce to Minkowski space.
We discuss this in depth in Chapter~\ref{chap:coord}.  
We find that in the $\omol = (0,0)$ model, a galaxy with zero total 
velocity ($\dot{D} = 0$) will be blueshifted.  An analytical derivation of
the solution for the empty universe is given in
Section~\ref{sect:analytic-empty}.

The fact that approaching galaxies can be redshifted and receding
galaxies can be blueshifted is an interesting illustration of the
fact that cosmological redshifts are not Doppler shifts. The
expectation that when
$v_{\rm tot}=0$, $z_{\rm tot}=0$, comes from special relativity
and does not apply to galaxies in the general relativistic
description of an expanding universe, even an empty one.


\section{Observational consequences}\label{sect:obs-cons}

\begin{figure}\bctr
\includegraphics[width=110mm]{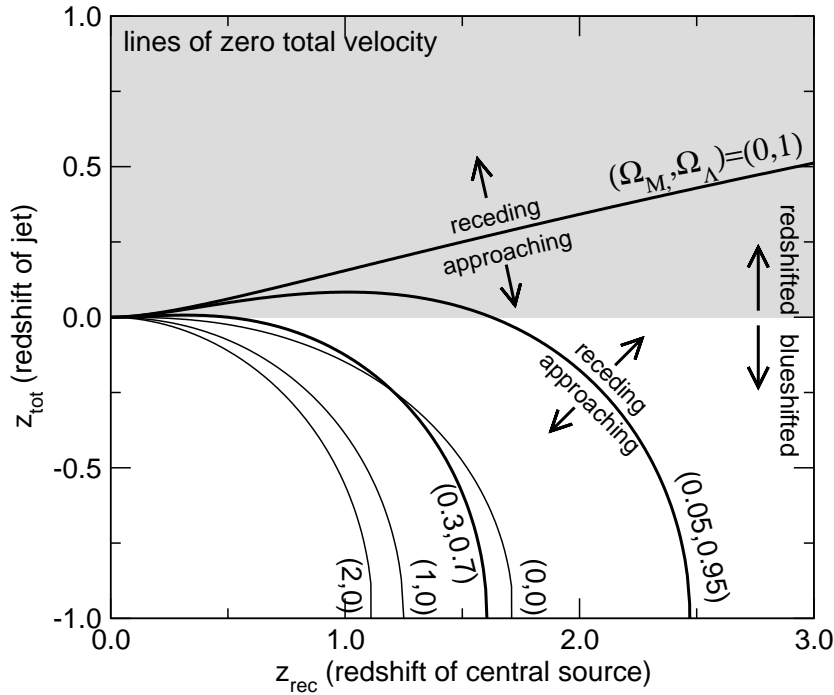}
\caption[Receding/approaching vs redshifted/blueshifted using AGN observables]{\small{This graph expresses the same information as Fig.~\ref{fig:z0vpecvrec} but in terms of observables.  An AGN with the central source of redshift $z_{\rm rec}$, is assumed to be comoving.  The observed redshift of a knot in a jet, $z_{\rm tot}$, is the total redshift resulting from the special relativistic Doppler shift, due to its peculiar velocity, combined with the cosmological redshift. The $z_{\rm tot}=0$ boundary separates the redshifted region (upper) from the blueshifted region (lower). The curves correspond to a total velocity of zero ($\dot{D}=0$) for different models, $\omol$, as labeled.  
The regions representing receding objects and approaching objects are indicated for the $\omol=(0.05,0.95)$ and $\omol=(0,1)$ models as examples (recession or approach {\it at emission} is plotted).  In contrast with expectations based on special relativity, receding objects are not necessarily redshifted, nor are blueshifted objects necessarily approaching us.
}}
\label{fig:v0ztotzrec} 
\ectr\end{figure}

The result for the tethered galaxy can be applied to the related
case of active galactic nuclei outflows. Some compact
extragalactic radio sources at high redshift are seen to have
bipolar outflows of relativistic jets of plasma. Jets directed
toward us (and in particular the occasional `knots' in them) are
analogs of a tethered (or boosted) galaxy. These knots have
peculiar velocities in our direction, but their recession
velocities are in the opposite direction and can be larger. Thus
the proper distance between us and the knot can be increasing.
They are receding from us (in the sense that $\dot{D} > 0$), yet,
as we have shown here, the radiation from the knot can be
blueshifted.  In Fig.~\ref{fig:v0ztotzrec} the zero-total-velocity condition is plotted in terms of the observable redshifts of a central-source and jet system. 

We can predict which radio sources have receding blueshifted jets.
The radio source 1146+531, for example, has a redshift $z_{\rm
rec}=1.629\pm0.005$~\citep{vermeulen95}. In an $\omol=(0.3,0.7)$
universe, its recession velocity at the time of emission was
$v_{\rm rec} \approx c$. Therefore the relativistic jet ($v_{\rm
pec}<c$) it emits in our direction was (and is) receding from us
and yet, if the parsec scale jet has a peculiar velocity within
the typical estimated range $0.8 \lsim v_{\rm pec}/c \lsim 0.99$,
it will be blueshifted. This example is the point plotted in the
upper left panel of Fig.~\ref{fig:z0vpecvrec}.

Collapsing universes also provide the possibility of
approaching-redshifted objects, but without involving peculiar
velocities. In the collapsing phase all galaxies are approaching
us. However, if the galaxy is distant enough, it may have been
receding for the majority of the time its light took to propagate
to us. In this case the galaxy appears redshifted even though it
may be approaching at the time of observation. 
This well known example
differs from the active galactic nuclei jet example because the
active galactic nuclei jet may appear blueshifted even though the
jet {\em never approaches us}.  Therefore approaching redshifted objects in a collapsing universe are not examples of the effect we discuss here.

\section{Summary}\label{sect:chainsummary}

We have pointed out and interpreted some counter-intuitive results
of the general relativistic description of our Universe. We have
shown that the unaccelerated expansion of the Universe has no
effect on whether an untethered galaxy approaches or recedes from
us. In a decelerating universe the galaxy approaches us, while in
an accelerating universe the galaxy recedes from us. The
expansion, however, {\em is} responsible for the galaxy joining
the Hubble flow, and we have shown that this happens whether the
untethered galaxy approaches or recedes from us.

The expansion of the Universe is a natural feature of general
relativity that also allows us to unambiguously convert observed
redshifts into proper distances and recession velocities, and to
unambiguously define approach and recede. We have used this
foundation to predict the existence of receding blueshifted and
approaching redshifted objects. To our knowledge
this is the first explicit derivation of this counter-intuitive
behavior.

Concepts such as ``recede'' or ``approach'' and quantities such as
$\dot{D}$ are of limited use in observational cosmology because
all our observations come to us via the backward pointing null
cone. This limitation will remain the case until a very patient
observer organizes a synchronized set of comoving observers to
measure proper distance~\citep{weinberg72,rindler77}.  However, the
issue we are addressing --- the relationship between observed
redshifts and expansion --- is a conceptual one and is closely
related to the important conceptual distinction between the
theoretical and empirical Hubble laws~\citep{harrison93}.

\clearemptydoublepage

\begin{flushright}
\begin{minipage}{5cm}
Time is defined so that motion looks simple.
\begin{flushright}
{\it Misner, Thorne and Wheeler, 1970}
\end{flushright}
\end{minipage}
\end{flushright}

\chapter{The empty universe}\label{chap:coord}
\vspace{-1cm}
We have shown in Sect.~\ref{sect:redshift} that the SR velocity-redshift relation does not hold even in the limit of an empty FRW universe.  Yet the empty universe is the one case in which our SR concepts should apply.  
Moreover, we have seen in Chapter~\ref{chap:misconcept} that many misconceptions about the expansion of the Universe arise from the misapplication of special relativistic concepts.
  If we are to understand the expansion of the Universe, and in particular the concept of ``stretching space'', it is important to understand the relationship between the FRW concept of recession velocities and velocities as they appear in special relativity.  

An alternative description of the empty FRW universe is the Milne universe.  Edward Milne proposed what is now known as the Milne universe as an alternative to expanding space \citep{milne35}.  His theory was named kinematic relativity.  The Milne universe is an expanding universe, but one that is based on special relativity and occurs in flat Minkowski spacetime\footnote{Minkowski spacetime is simply the flat spacetime of special relativity.}.  In the Milne universe the big bang is an explosion, particles emanate from the origin at $t=0$ with all possible speeds less than the speed of light.  These particles move outwards through pre-existing Minkowski space.  For a more detailed description of the Milne universe we refer the reader to \citeauthor{rindler77} (1977) Sect.~9.4 and \citeauthor{ellis88book} (1988) Sect.~4.8.

The Milne universe is unable to explain the acceleration and deceleration of expansion that occurs in a non-empty universe.  However, for an empty universe the Milne model is as appropriate as the FRW description.  A simple coordinate transformation (\citeauthor{rindler77}, 1977, Sect.~9.4; \citeauthor{peacock99}, 1999, Sect.~3.3) converts the Robertson-Walker metric in the empty universe case to the Minkowski metric of special relativity.  This makes the Milne universe an excellent testing ground to demonstrate how SR concepts fit into the FRW picture.  

In Sect.~\ref{sect:conversion} we review the well known transformation between FRW and Minowski space in the empty universe case.  We then use this transformation in Sect.~\ref{sect:analytic-empty} to show how the SR Doppler shift is related to the cosmological redshift in FRW models.

\section{Coordinate change from FRW to Minkowski space}\label{sect:conversion}

Here we show the transformation between the Minkowski metric and the Robsertson-Walker metric in the empty universe case.  This derivation follows a similar one by \citet[Sect.~3.3]{peacock99}.  In the empty universe $R=R_0H_0 t$ (from integrating Eq.~\ref{eq:dotR}).  So the Robertson-Walker metric for an empty universe is,
\beq  ds^2 = -c^2dt^2 + R_0^2H_0^2t^2[d\chi^2+\sinh^2(\chi)d\psi^2], \eeq
with symbols as defined in Appendix~\ref{sect:math}.  Radial distance ($d\psi=0$) along a hypersurface of constant time ($dt=0$) is given by $D=R_0H_0t\chi$, so recession velocity is $v_{\rm rec}=R_0H_0\chi$.  Minkowski space has the metric,
\beq ds^2 = -c^2dT^2 + dl^2 + l^2d\psi^2. \eeq
Fundamental particles in the Milne universe emanate from the origin of Minkowski space at $T=0$ and travel with a constant velocity $v_{\rm M}=l/T$.  (To qualify as an empty universe these fundamental particles must be massless test particles.)  Fundamental observers ride on fundamental particles.  These fundamental observers each have an inertial frame that extends throughout Minkowski space and their coordinates can be converted to any other fundamental observers frame using Lorentz transformations.  Therefore, using Lorentz transformations we can write the metric from the point of view of any fundamental observer, 
\beq ds^2 = -c^2(dT^\prime)^2 + (dl^\prime)^2 + l^2d\psi^2, \label{eq:movingmetric}\eeq
where this observer's coordinates are time dilated and length contracted (along the radial direction) with respect to the stationary observer at the origin, 
\bea dT&=&\gamma_{\rm M} dT^\prime,\label{eq:timedil}\\
     dl&=&\gamma_{\rm M} dl^\prime,\label{eq:lengthcon} \eea
with $\gamma_{\rm M}=(1-v_{\rm M}^2/c^2)^{-1/2}$.
The infinitesimal distances perpendicular to the radial axis of motion  are not length contracted, so the angular part of the metric, $l^2d\psi^2$ remains unchanged.
Introducing a velocity parameter $\chi^\prime$ such that $v_{\rm M} = c\tanh \chi^\prime$ allows us to simplify, $\gamma_{\rm M} = \cosh \chi^\prime$.  This in turn gives $l=cT^\prime \sinh \chi^\prime$, which when differentiated gives,
\beq dl = cT^\prime \cosh \chi^\prime d\chi^\prime + c \sinh \chi^\prime dT^\prime. \eeq
Along a surface of constant time, $dT^\prime=0$, we have $dl=cT^\prime \cosh \chi^\prime d\chi^\prime$ thus, 
\beq dl^\prime = cT^\prime d\chi^\prime. \eeq
Substituting these back into the metric Eq.~\ref{eq:movingmetric} gives, 
\beq ds^2 = -c^2dT^{\prime\,2} + c^2T^{\prime\,2}[d\chi^{\prime\, 2} + \sinh^2(\chi^\prime)\,d\psi^2]. \eeq
This has the form of the RW metric with $T^\prime=t$, $\chi^\prime=\chi$ and $R_0H_0=c$.   

\begin{figure}
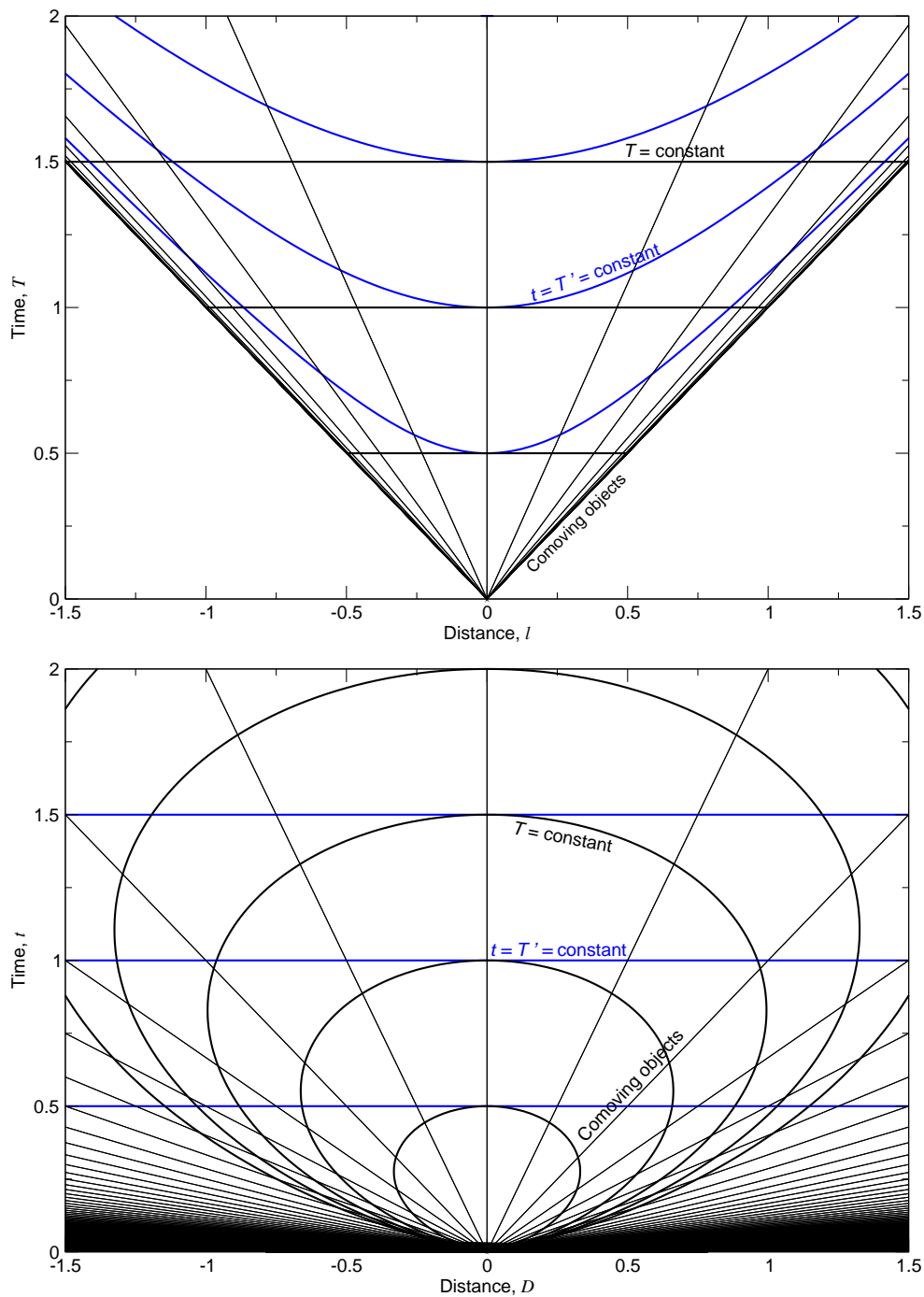
\bctr
\includegraphics[width=130mm]{./figs/T-l.eps}\vspace{2mm}
\includegraphics[width=130mm]{./figs/t-D.eps}
\caption[Subsection of Minkwoski space transformed into FRW coordinates]{\small Spacetime diagrams for the Milne universe (upper panel) and the empty FRW universe (lower panel).  The same comoving observers are shown in both cases.  The same surfaces of constant $t=T^\prime$ and constant $T$ are also shown in both cases.  In the Milne universe density increases along a surface of constant $T$.  In the FRW universe density is constant along a surface of constant $t$.   The Milne universe only fills up a quarter of Minkowski space, but when converted to FRW coordinates the space is complete.}
\label{fig:T-r-t-D}
\ectr\end{figure}

This transformation itself is quite revealing.  It shows that time, $t$, in the Robertson-Walker metric is the proper time of fundamental observers, $T^\prime$.  In Fig.~\ref{fig:T-r-t-D} we show two spacetime diagrams.  The first is drawn in Minkowski space, from the point of view of the stationary observer at the origin whose coordinates are $l$ and $T$.  The second is drawn in FRW space, using the coordinates $D$ and $t$.  On these spacetime diagrams we show the trajectories of fundamental (comoving) observers.  Fundamental observers in Minkowski space have constant $\chi$ ($v_{\rm M}=c\tanh \chi =$ constant) and therefore translate to comoving observers in FRW space.

When the coordinate transformation is made from FRW to Minkowski spacetime homogeneity is not maintained.  On the upper spacetime diagram of Fig.~\ref{fig:T-r-t-D} it is evident that density in the Milne universe is not constant along a surface of constant $T$ (unevenly spaced comoving observers).  That is, in any observer's inertial frame density increases with distance from the origin.  This is one of the main differences between Minkowski and FRW coordinates.  The Milne universe, as measured in any observer's inertial frame, is {\em not homogeneous}.  It is, however, isotropic.  

The Milne universe can be made to satisfy the cosmological principle -- homogeneity and isotropy -- under certain conditions \citep[for a concise but thorough summary see][]{rindler77}.  Firstly, every fundamental observer believes that they are the central observer, so the Milne universe has no unique centre\footnote{The Milne universe does originate from a point in the pre-existing Minkowski spacetime, but all fundamental observers believe they are the centre of the expansion, so the centre is a frame dependent feature and no unique centre can be defined}.  
Moreover, if you consider the proper time of fundamental (comoving) observers, $T^\prime$, as your constant time surface, $dT^\prime=0$, then the Milne universe {\em is} homogeneous.  That is, fundamental observers all measure the same density at the same proper time.  This is exactly the choice made in FRW coordinates.  The time coordinate, $t$, is chosen to be the proper time of comoving observers.  When this choice is made the universe is homogeneous along a surface of constant $t$ (equally spaced comoving observers in Fig.~\ref{fig:T-r-t-D}, lower panel).

Close to $l=0$ and $D=0$ the two spacetime diagrams of Fig.~\ref{fig:T-r-t-D} are very similar.  That is why SR is a good approximation to FRW in an infinitesimal region surrounding any point.  However, far from that point the deviation of SR from the GR description becomes dramatic.  This is why recession velocities in FRW space 
do not obey the special relativistic Doppler shift equation, even in the empty universe.  
  In the next section we extend this to demonstrate how recession velocities and the SR Doppler shift equation {\em are} related.

\section{Cosmological redshift vs SR Doppler shift}\label{sect:analytic-empty}
In Section~\ref{sect:redshift} we showed that a galaxy with zero total velocity does not have zero redshift even in the empty universe case.  This demonstrates that cosmological redshifts are not SR Doppler shifts.  In this section we go the other way and show how cosmological redshifts and SR Doppler shifts {\em are} linked.

First we need to calculate  recession velocity as a function of redshift in the empty FRW universe.  
For an empty expanding universe, $H(z)=H_0(1+z)$, so Eq.~(\ref{eq:vreczrec}) becomes,
\beq v_{\rm rec}=c H_0 \! \int_0^{z_{\rm rec}}\frac{dz}{H_0(1+z)}. \eeq
Velocity as a function of redshift in the empty FRW universe is therefore\footnote{This allows us to find the analytic solution for the combination of recession and peculiar velocity that would give a redshift of zero.
If we substitute $1+z_{\rm rec}=e^{v_{\rm rec}/c}$ into Eq.~(\ref{eq:pecrec}) followed by Eq.~(\ref{eq:vpeczpec}), we find,
\begin{equation}
v_{\rm pec}=c\left[\frac{e^{-2v_{\rm rec}/c}-1}{e^{-2v_{\rm rec}/c}+1}\right].\label{eq:vpec00} 
\end{equation}
Equation~(\ref{eq:vpec00}) shows that only in the limit of small recession velocity does $v_{\rm pec}= -v_{\rm rec}$ give a total redshift of zero.  Equation~\ref{eq:vpec00} generates the thick black $z=0$ line in Fig.~\ref{fig:z0vpecvrec}, upper right panel.},
\beq v_{\rm rec} = c \ln (1+z_{\rm rec}).\label{eq:vrecclnz}\eeq
Now we are in a position to show how the SR redshift equation fits into the GR picture.  We observe a distant object to have redshift $z$.  
How can we use the special relativistic Doppler shift equation to accurately calculate velocity from this redshift?
We showed in Sect.~\ref{sect:conversion} that $\vrec = R_0H_0\chi = c\chi$, therefore Eq.~\ref{eq:vrecclnz} yields $\chi=\ln (1+\zrec)$. Using $v_{\rm M} = c\tanh{\chi}$ and an inverse hyperbolic identity we find, 
\bea \chi &=& \tanh^{-1}(v_{\rm M}/c), \\
\ln (1+z)&=&\frac{1}{2}\ln\left(\frac{1+v_{\rm M}/c}{1-v_{\rm M}/c}\right),\\
     1+z  &=& \sqrt{\frac{1+v_{\rm M}/c}{1-v_{\rm M}/c}}.\eea
Thus we find the redshift of an object in an empty universe can be described by the special relativistic Doppler shift equation -- {\em only} if the velocity used in the equation is the velocity in Minkowski space, $v_{\rm M}=l/T$.  This velocity {\em does not} obey $v_{\rm M}=HD$.

\cite{peacock99} claims that using the special relativistic Doppler formula to calculate recession velocity from large cosmological redshifts, although generally incorrect, is appropriate in the case of an empty universe\footnote{Peacock, J.~A. (1999), 
 p.~87, ``For small redshifts, the interpretation of the redshift as a Doppler shift ($z=v/c$) is quite clear.  What is not so clear is what to do when the redshift becomes large.  A common but incorrect approach is to use the special-relativistic Doppler formula and write
\beq 1+z = \sqrt{\frac{1+v/c}{1-v/c}}.\eeq
This would be appropriate in the case of a model with $\Omega=0$ (see below), but is wrong in general.'' }.
We maintain it is not appropriate, even in the empty FRW universe.  However, if we have made the change in coordinate systems as outlined in Section~\ref{sect:conversion} we can use the special relativistic formula to calculate $v_{\rm M}$.  This is {\em not} the velocity that appears in Hubble's law.

For completeness we note that there is another way to relate the SR Doppler shift, which holds locally, to the cosmological redshift.   To calculate the cosmological redshift one can integrate over a series of infinitesimal SR Doppler shifts between one observer and the object whose redshift they are measuring.   The derivation of this effect can be found in \cite{padmanabhan96}, Sect.~6.2(a), and we summarize it in Appendix~\ref{sect:infinitesimal}.

\clearemptydoublepage

\part{Entropy of Event Horizons}

\chapter{Testing the generalized second law of thermodynamics}\label{chap:gsl}

According to the generalized second law of thermodynamics (GSL) entropy never decreases when all event horizons are attributed with an entropy proportional to their area.  
In this chapter we test the GSL as it pertains to cosmological event horizons.  We investigate the change in entropy when dust, radiation and black holes cross a cosmological event horizon.  If the total entropy decreases in any of these cases then the total entropy of the Universe violates the GSL.    We provide analytic derivations of the entropy variation in these scenarios for small departures from de Sitter space and use numerical calculations to generalize for flat, open and closed Friedmann-Robertson-Walker (FRW) universes.  In most cases the loss of entropy from within the cosmological horizon is more than balanced by an increase in cosmological event horizon entropy, maintaining the validity of the generalized second law of thermodynamics.  However, an intriguing set of open universe models show an apparent entropy decrease when black holes disappear over the cosmological event horizon.  We anticipate that this apparent violation of the generalized second law is due to insufficient models of black hole event horizons in evolving backgrounds and will disappear when solutions are available for black holes embedded in arbitrary spacetimes.

This Chapter is based on work published in \cite{davies_davis02} and \cite{davis03event}.

\section{Horizon entropy}\label{sect:horizonentropy}

A significant advance in physical theory was made by Bekenstein with the suggestion \citep{bekenstein73} 
that the area of the event horizon of a black hole is a measure of its entropy. This hinted at a deep link between information, gravitation and quantum mechanics that remains tantalizingly unresolved today. Bekenstein's claim was bolstered by Hawking's application of quantum field theory to black holes \citep{hawking76}, 
from which he deduced that these objects emit thermal radiation with a characteristic temperature,
\beq T_{\rm b} = \frac{1}{8\pi m_{\rm b}}, \eeq
for a Schwarzschild hole, where $m_{\rm b}$ is the mass of the black hole, and we use units $G = \hbar = c = k = 1$.  Hawking's calculation enabled the entropy of a black hole $S_{\rm b}$ to be determined precisely as,
\bea S_{\rm b} &=& 16\pi m_{\rm b}^2, \\
            &=& \frac{A_{\rm b}}{4}, \label{eq:Sbh}\eea
where $A_{\rm b}$ is the event horizon area.  Eq.~\ref{eq:Sbh} also applies to spinning and charged black holes. It was then possible to formulate a generalized second law of thermodynamics (GSL),
\beq \dot{S}_{\rm env} + \dot{S}_{\rm b} \ge 0, \eeq
where $S_{\rm env}$ is the entropy of the environment exterior to the black hole and an overdot represents differentiation with respect to proper time, $t$. Thus when a black hole evaporates by Hawking radiation its horizon area shrinks, so its entropy decreases, but the environment gains at least as much entropy from the emitted heat radiation \citep{hawking75}. 
Conversely, if a black hole is immersed in heat radiation at a higher temperature, radiation will flow into the black hole and be lost. The corresponding entropy reduction in the environment is offset by the fact that the black hole gains mass and increases in area and entropy.

\cite{gibbons77} 
conjectured that event horizon area, including cosmological event horizons, might quite generally have associated entropy.  A prominent example is de Sitter space, a stationary spacetime which possesses a cosmological event horizon at a fixed distance, $r_{\rm c}=(3/\Lambda)^{1/2}$, from the observer (to conform with the notation of \cite{davis03event} and \cite{davies_davis02} we use $r$ in this chapter to represent proper distance rather than $D$ as was used in Part I). It was known \citep[see e.g.][Sect.~5.4]{birrell_davies82} 
that a particle detector at rest in de Sitter space responds to a de Sitter-invariant quantum vacuum state as if it were a bath of thermal radiation with temperature,
\beq T_{\rm deS} = \frac{1}{2\pi\;(3/\Lambda)^{1/2}}. \eeq
It thus seemed plausible that the GSL could be extended to de Sitter space. Subsequent work by \cite{davies84}, 
and \cite{davies_ford_page86} 
supported this conclusion. There were, however, some problems. Although the de Sitter horizon has thermal properties, the stress-energy-momentum tensor of the de Sitter vacuum state does not correspond to that of a bath of thermal radiation\footnote{The stress-energy-momentum tensor for thermal radiation is $T_{\mu}\,^{\nu} = {\rm diag}(1,-1/3,-1/3,-1/3)\, \rho_{\rm r}$, where $\rho_{\rm r}$ is the radiation energy density.} 
(unlike for the black hole case).  It is given instead by \citep{dowker76}, 
\beq T_{\mu}\,^{\nu} = \frac{\Lambda^4}{8640\pi^2}\; \delta_{\mu}\,^{\nu},\eeq
($T_{\mu}\,^{\nu}$ is used here for the stress-energy-momentum tensor, elsewhere $T$ is used for temperature). This  corresponds to the stress-energy-momentum tensor associated with a cosmological constant, and so merely renormalizes $\Lambda$.   Secondly, there is no asymptotically flat external spacetime region for de Sitter space, which precludes assigning a mass parameter to the de Sitter horizon. This makes it hard to interpret trading in energy and entropy, as is conventional in thermodynamic considerations, between de Sitter space and an environment. A final problem is that in the black hole case Bekenstein attributed the entropy of the hole to its total hidden information content, which is readily evaluated. For a cosmological horizon, which may conceal a spatially infinite domain lying beyond, the total hidden entropy would seem to be ill-defined. 

The foregoing concerns are amplified in the case of more general cosmological horizons that are non-stationary and do not even have an associated well-defined temperature (a comoving particle detector sees a bath of radiation with a non-thermal spectrum). 
 For the most general derivations of horizon entropy to date see the recent pioneering work of 
\cite{padmanabhan02b,padmanabhan02c,padmanabhan02a}.

 We consider the general class of Friedmann-Robertson-Walker (FRW) models with scalefactor $R(t)$.
One may define a conformal vacuum state adapted to the conformally flat geometry\footnote{All FRW models are conformally flat.  Redefining the time coordinate to be $d\tau=dt/R(t)$ allows us to write the Robertson-Walker metric as, $ ds^2 =  R(t)^2[-c^2d\tau^2 +d\chi^2+S_k^2(\chi)d\psi^2].$ The part of the metric inside the square brackets is flat Minkowski spacetime.} of these spaces, and consider the response of a quantum particle detector \citep[Sect.~3.3]{birrell_davies82} 
to such a state. The response will generally be non-zero, but the perceived spectrum will not be thermal. This raises the question: just how far can one extend the GSL to event horizons? Could it apply even to non-stationary cosmological models in spite of the absence of a clear thermal association? And if the GSL cannot be thus extended, what are the criteria that determine the limits of its application? 

We consider these questions to be of significance to attempts to link information, gravitation and thermodynamics \citep{padmanabhan03}, and to recent discussions about the total information content of the universe \citep{lloyd02}. 
They may also assist in attempts to formulate a concept of gravitational entropy \citep{padmanabhan02b,padmanabhan02c,padmanabhan02a}
, and to clarify the status of the holographic principle \citep{susskind95,bousso02}. 

In this chapter we explore the range of validity of the GSL.  We assume cosmological event horizons do have entropy proportional to their area, as \cite{gibbons77} 
proposed. The total entropy of a universe is then given by the entropy of the cosmological event horizon plus the entropy of the matter and radiation it encloses.  In Sect.~\ref{sect:dustFRW} and Sect.~\ref{sect:radFRW} we assess the loss of entropy as matter and radiation disappear over the cosmological event horizon and show that the loss of entropy is more than balanced by the increase in the horizon area.  We then consider in Sect.~\ref{sect:bhFRW} the case of an FRW universe filled with a uniform non-relativistic gas of small black holes.  This enables a direct entropic comparison to be made between black hole and cosmological event horizon area.  As the black holes stream across the cosmological horizon, black hole horizon area is lost, but the cosmological horizon area increases.  We may thus assess the relative entropic `worth' of competing horizon areas.

\section{Dust crossing the cosmological event horizon}

\label{sect:dustFRW}
The simplest case to consider is the classic homogeneous, isotropic FRW universe filled with pressureless dust.  The dust in this model is assumed to be comoving.  The dust is therefore in the most ordered state possible and has zero entropy which allows us to restrict our thermodynamic considerations to the cosmological event horizon alone.  
\begin{figure}\bctr
\includegraphics[width=70mm]{./figs/tXehrehAehVeh5-half1.eps} 
\includegraphics[width=76mm]{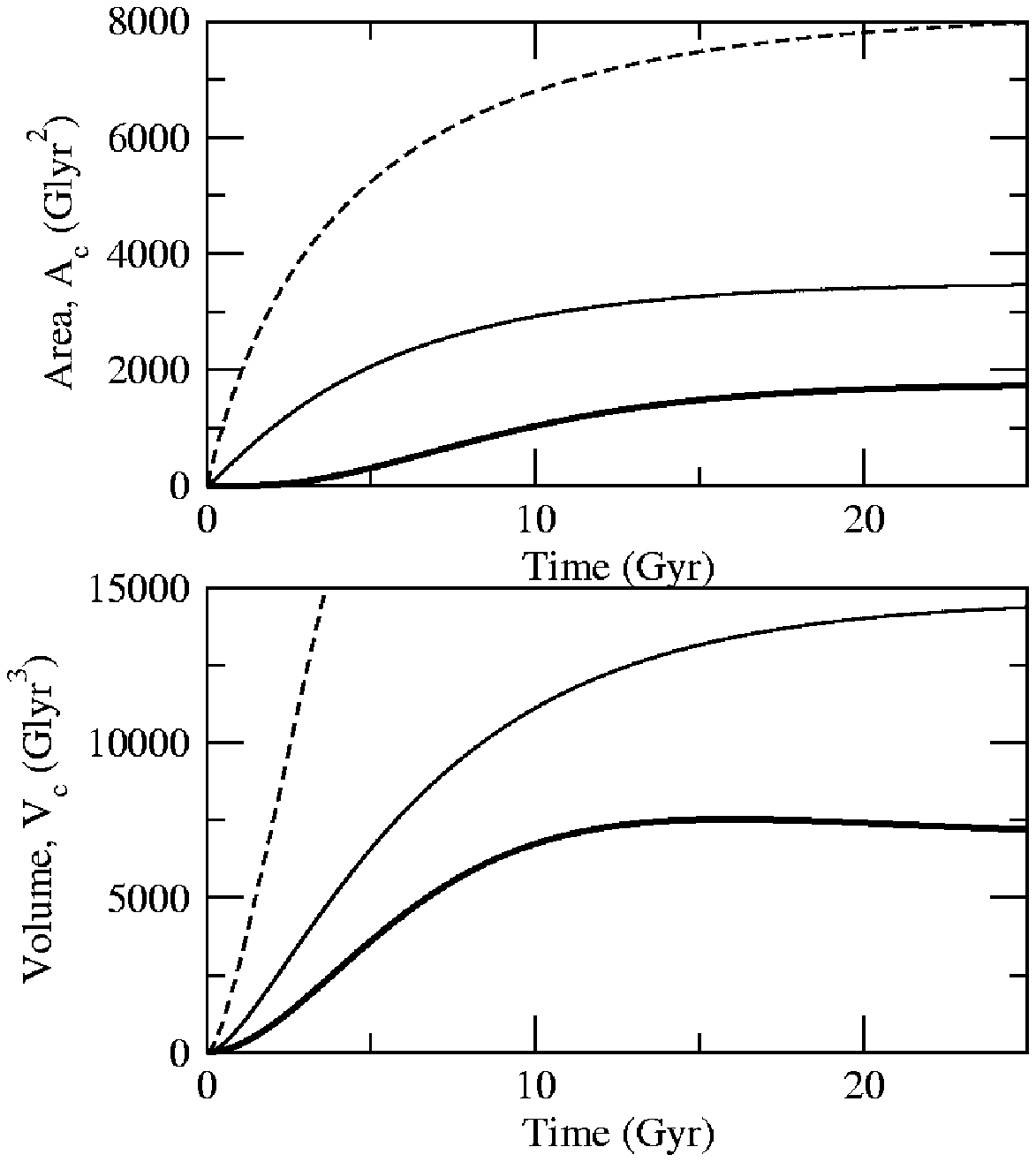} 
\caption[The event horizon in $\omol=(0.3,0.3), (0.3,0.7)$ and $(0.3,1.4)$]{\small{The comoving distance, proper distance, area and volume of the cosmological event horizon are shown for three different cosmological models.  The models' matter (energy) density and cosmological constant $\omol$ is given in the legend.  The dimensionless comoving distance is not shown for the $\omol=(0.3,0.7)$ case since $R_0$ is undefined in this model.  Note that although the radius and volume within the cosmological event horizon both decrease for periods in the $\omol=(0.3,1.4)$ universe, the area always increases.}}
\label{fig:tXehrehAehVeh}
\ectr\end{figure}
\begin{figure}\bctr
\includegraphics[width=100mm]{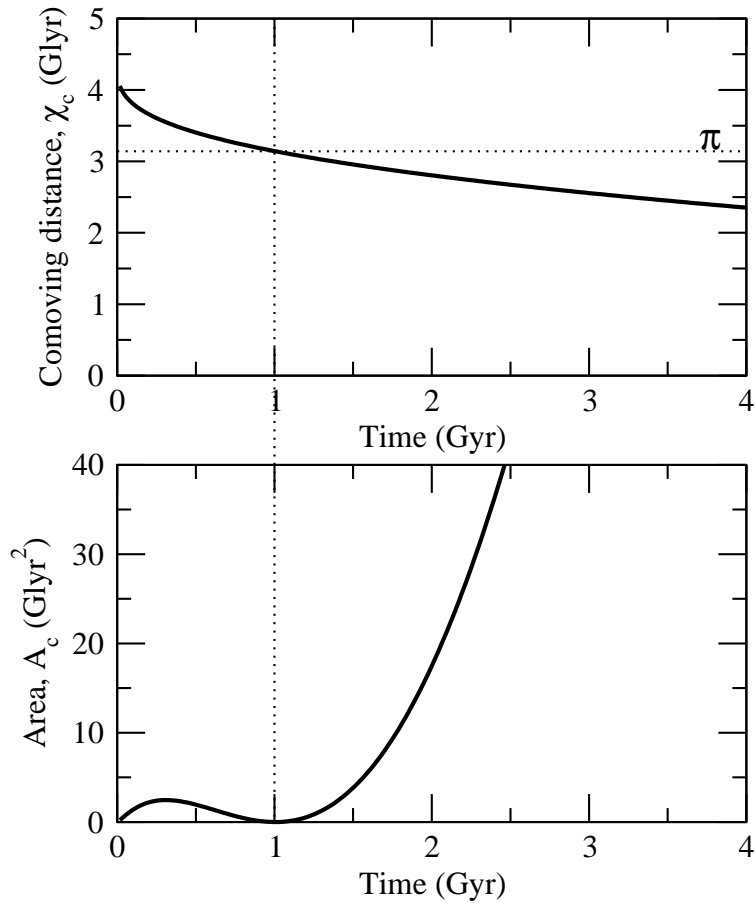}
\caption[Event horizon emerges out of the antipode in $\omol=(0.3,1.4)$]{\small{This is a close-up of the region near the origin of Fig.~\ref{fig:tXehrehAehVeh} for the $\omol=(0.3,1.4)$, $k=+1$ case, which appears to show a curious rise and fall in event horizon area at early times.  However, this is an artefact of the finite spatial size of closed FRW universes.  When the comoving distance to the event horizon exceeds $\pi$, as it does for $t\lsim 1\,Gyr$ in this example, it is possible for an observer to see past the antipode.   In this example the event horizon appears out of the antipode at $\sim 1.0\,Gyr$.
}}
\label{fig:tXehrehAehVeh-zoom}
\ectr\end{figure}

The time dependence of the scalefactor, $R(t)$, is given by the Friedmann equations,
\bea \dot{\rho} &=& -3H(\rho + p),\label{eq:Fried1}\\
3H^2&=&8\pi\rho + \Lambda - 3k/R^2,\label{eq:Fried2}\eea 
where $\rho$ and $p$ are the density and pressure of the cosmological fluid respectively.  We assume the present day Hubble's constant $H_0=70\,kms^{-1}Mpc^{-1}$ for all quantitative calculations.  
Eternally expanding models possess event horizons if light can not travel more than a finite distance in an infinite time,
\beq \chi_c(t) = \int_t^\infty \frac{dt^\prime}{a(t^\prime)} < \infty. \label{eq:chi}\eeq
The integral in Eq.~\ref{eq:chi} represents the comoving distance to the cosmological event horizon, $\chi_c$, at time $t$.  
The proper distance to the cosmological event horizon is then $r_{\rm c} = R(t)\chi_{\rm c}$.  The area of the cosmological horizon generalized to curved space is, 
\beq A_{\rm c}=4\pi R^2(t)S_k^2(\chi_{\rm c}),\label{eq:Ac}\eeq
which reduces to $A_{\rm c}=4\pi r_{\rm c}^2$ in flat space (again, $S_k(\chi)=\sin\chi, \chi, \sinh\chi$ for $k=1,0,-1$ respectively).  

\subsection{Cosmological event horizon area never decreases}

\cite{davies88} 
showed that the cosmological event horizon area of an eternally expanding FRW universe never decreases, assuming the dominant energy condition holds, $\rho+p\ge 0$. This is analogous to Hawking's area theorem for black holes \cite{hawking72}.  
In black holes the dominant energy condition is violated by quantum effects, allowing black holes to evaporate and shrink.  There is no known analogous shrinking in cosmological horizon area.

It is interesting to note that the area of the cosmological event horizon increases even in models in which the radius of the event horizon decreases.  Closed eternally expanding universes have a decreasing event horizon radius at late times, but the effect of curvature ($S_k(\chi)$ term) forces the area to increase nevertheless, for example the $\omol=(0.3,1.4)$ model in Fig.~\ref{fig:tXehrehAehVeh}.

\section{Radiation crossing the cosmological event horizon} \label{sect:radFRW}

To investigate the interplay of entropy exchange between the cosmological event horizon and an environment we consider an eternally-expanding FRW universe with a positive cosmological constant, filled with radiation of temperature $T(t)$ \citep[see][]{davis03event}.  Such a universe has an event horizon radius that tends toward the de Sitter value, $r_{\rm deS} = 1/H$, at late times.   Most $\Lambda>0$ universes tend toward de Sitter at late times except the few that have a large enough matter density to begin recollapse before they become cosmological constant dominated.  We include constants in this and subsequent sections to explicitly ensure environment and horizon entropy are being compared in the same units.  The entropy of the cosmological event horizon is, 
\beq S_{\rm c} = \left(\frac{k c^3}{\hbar G}\right) \frac{A_{\rm c}}{4}. \eeq
\begin{figure}\bctr
\includegraphics[width=100mm]{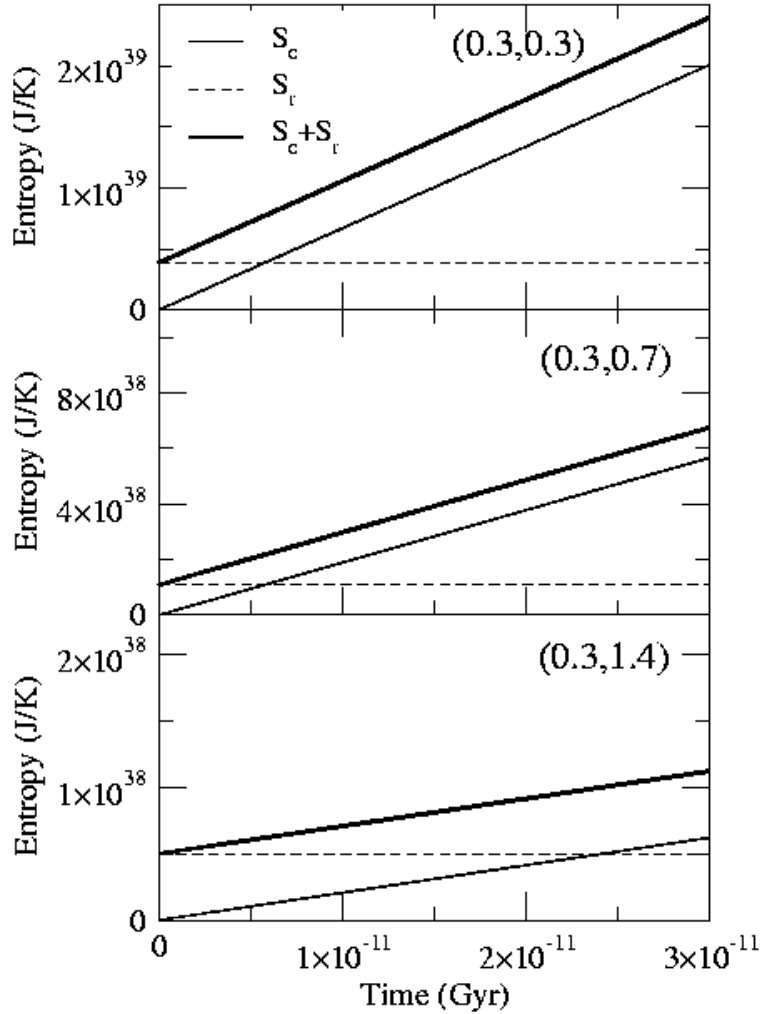}  
\caption[Radiation disappearing over the cosmological event horizon]{\small{This shows the radiation entropy $S_{\rm r}$ compared to the cosmological horizon entropy $S_{\rm c}$ in three radiation filled FRW universes.  Each graph is labeled with the model, $\omol$.  Only early times are shown because that is the only time that the radiation entropy is comparable to the horizon entropy.  The radiation entropy is not constant but decreases rapidly.  However, the decrease is orders of magnitude slower than the increase in cosmological event horizon entropy, so does not show up on this scale.  Total entropy $S_{\rm c} + S_{\rm r}$ never decreases so the GSL holds for these models.}}
\label{fig:Srad}
\ectr\end{figure}
\noindent Radiation energy density obeys $\rho_{\rm r} = \sigma T^4$ (where the radiation constant\footnote{The radiation constant is often denoted by $a$, but since we use $a$ for the scalefactor we denote the radiation constant by $\sigma$.  This is not to be confused with Stephan-Boltzmann's constant $\sigma_{\rm SB}$, which is related to the radiation constant by $\sigma = 4\sigma_{\rm SB}/c$.} $\sigma=\pi^2k^4/15c^3\hbar^3$) 
 while entropy density follows $s_{\rm r}=(4/3)\rho_{\rm r}\, T^{-1}$.  This means the total entropy  within an event horizon volume, $S_{\rm r}=s_{\rm r}V_{\rm c}$, is given by,
\beq S_{\rm r}= \frac{4}{3}\,\sigma^{1/4}\,\rho_{\rm r}^{3/4}\;V_{\rm c}.\label{eq:Srad}\eeq
The volume within a cosmological event horizon for closed, flat and open FRW universes is given by Eq.~\ref{eq:volume}.
We take $p = \rho_{\rm r}/3$ for radiation in the Friedmann equations (Eq.~\ref{eq:Fried1} and Eq.~\ref{eq:Fried2}).
The radiation density decays as $\rho_{\rm r} = \rho_0 a^{-4}$ (or $T\propto 1/a$) as the universe expands so the radiation entropy within a constant comoving volume ($V\propto a^3$) remains constant.  However, the radiation entropy within the cosmological horizon decreases as the comoving volume of the event horizon decreases ($\chi_c$ decreases in Eq.~\ref{eq:volume}) and radiation crosses the cosmological event horizon.

The evolution of the universe is dependent on the density of radiation, so the model universe we choose constrains the radiation density according to $\Omega_{\rm r} =  8\pi G \rho_0/3H_0^2$.
(The normalized radiation density, $\Omega_{\rm r}$, replaces $\om$ in Friedmann's equations with the difference that $\Omega_{\rm r}$ decays as $a^{-4}$.)  Allowing for this constraint we replace the dust of Sect.~\ref{sect:dustFRW} with radiation and calculate the loss of entropy over the cosmological event horizon as the universe evolves.  Although the radiation represents much more entropy than dust, in a realistic cosmological model this entropy is minuscule compared to that of the cosmological event horizon.  At the present day in a $(\Omega_{\rm r},\oll)=(0.3,0.7)$ radiation dominated FRW universe the radiation entropy would be 14 orders of magnitude smaller than the entropy of the horizon.  At early times the event horizon was tiny and the radiation was very hot -- it is only at early times that we could expect the radiation entropy to be significant enough to compete with the increase in event horizon area.  Figure~\ref{fig:Srad} shows some numerical solutions typical of a wide class of radiation-filled models.  In all cases we find that the total entropy increases with time ($\dot{S}_{\rm r}+\dot{S}_{\rm c}>0$) in conformity with the extended interpretation of the generalized second law of thermodynamics.  In the next section we show analytically that thermal radiation crossing the cosmological event horizon satisfies the GSL in the limit of small departures from de Sitter space as long as the radiation temperature is higher than the cosmological horizon temperature.  A rigorous analytical proof for the general FRW case, however, is lacking.

\subsection{Small departure from de Sitter space}\label{sect:rad-small-dep}
We consider a radiation-$\Lambda$ FRW cosmology at late times when it represents a small departure from de Sitter space \citep[see][]{davies_davis02}. 
The distance to the cosmological event horizon is approximately equal to the de Sitter horizon radius, $r_{\rm deS}=1/H$, corresponding to an entropy of $S_{\rm c} \approx \pi/H^2$.  Therefore the horizon entropy increases at a rate,
\bea \dot{S}_{\rm c} &\approx& -2\pi\dot{H}/H^3, \label{eq:sdotc1}\\
                     &=& 2\pi(16\pi\rho)/3H^3, \label{eq:sdotc}\eea
where we use $\dot{H}=-4\pi G(\rho+p)$ from Eq.~\ref{eq:Fried1} and Eq.~\ref{eq:Fried2} with $k=0$ and the pressure and density in this case are, $p_{\rm r}=\rho_{\rm r}/3$.
Fill the universe with thermal radiation at a temperature $T>T_{\rm deS}=H/2\pi$ so the typical wavelength of the radiation is less than the horizon distance.   While the radiation temperature remains greater than the temperature of the de Sitter horizon there is a net flow of energy lost across the horizon (radiation temperatures cooler than the temperature of the de Sitter horizon have a typical wavelength greater than the horizon distance and can therefore not be localized within the horizon). 
From Eq.~\ref{eq:Srad} this means the total radiation entropy within a horizon volume is $S_{\rm r} = (16\pi \sigma^{1/4}/9)\;\rho_{\rm r}^{3/4}H^{-3}$.  The loss of radiation as it passes across the cosmological event horizon occurs at the rate,
\beq \dot{S}_{\rm r} \approx \frac{-16\pi \sigma^{1/4}}{3} \; \rho_{\rm r}^{3/4}H^{-2},\label{eq:sdotr}\eeq
if $\dot{H}$ is small.  The GSL holds when $\dot{S}_{\rm c} + \dot{S}_{\rm r} \ge 0$.  Comparing Eq.~\ref{eq:sdotr} with Eq.~\ref{eq:sdotc} shows that the GSL holds when $\rho_{\rm r}^{1/4} > H/2\pi$, which is the condition that the radiation be hotter than the horizon temperature, as assumed.

\section{Black holes crossing the cosmological event horizon}\label{sect:bhFRW}
A way to directly compare the entropic worth of cosmological horizons and black hole horizons is to assess the change in entropy as black holes cross the cosmological horizon.  To this end we examine FRW universes containing a dilute pressureless gas of equal mass black holes.  We ignore the Hawking effect which would be negligible for black holes larger than solar mass over the timescales we address\footnote{Black hole evaporation time $\sim (m/m_{\rm solar})^3 \times 10^{66} yr$.}.  We also ignore interactions between black holes.
As the universe expands the density of the black hole gas decreases ($\rho_{\rm b}\propto a^{-3}$) and black holes disappear over the cosmological event horizon, resulting in a decrease in the black hole contribution to the total entropy within a horizon volume.  The area of the cosmological event horizon increases in turn\footnote{Cause and effect become confused when we try to assess cosmological event horizons in an analogous way to black holes.  The normal language used for cosmological event horizons would be to say that the matter density and cosmological constant of the universe determine the rate of expansion of the universe and thus determine the increase in distance to the event horizon.  Alternatively we can state that the loss of matter (energy) over the cosmological horizon results in the increase in distance to the event horizon.}.   To ascertain whether the GSL is threatened we ask: does the cosmological event horizon area increase enough to compensate for the loss of black hole entropy?

The area of the cosmological event horizon is easy to calculate in arbitrary (eternally expanding) FRW universes, as shown in Sect.~\ref{sect:dustFRW}.  Not so the event horizon area of black holes because the solutions require us to deal with an overdensity in an homogeneous but time-dependent background.  The Schwarzschild metric, 
\beq ds^2 = -\left(1-\frac{2Gm_{\rm b}}{rc^2}\right)c^2dt^2 + \left(1-\frac{2Gm_{\rm b}}{rc^2}\right)^{-1}dr^2 + r^2(d\theta^2 + \sin^2\theta d\phi^2),\eeq
applies for a black hole of mass $m_{\rm b}$ embedded in empty space.  The relationship between black hole mass and event horizon radius is $m_{\rm b}=r_{\rm b} c^2/2G$.  The Schwarzschild-de Sitter solution \citep{gibbons77} 
with metric,
\beq ds^2 = -\left(1-\frac{2Gm_{\rm b}}{rc^2}-\frac{\Lambda r^2}{3c^2}\right)c^2dt^2 + \left(1-\frac{2Gm_{\rm b}}{rc^2}-\frac{\Lambda r^2}{3c^2}\right)^{-1}dr^2 + r^2(d\theta^2 + \sin^2\theta d\phi^2),\eeq
applies for a black hole embedded in a de Sitter universe (a universe with zero mass density and a constant positive cosmological constant, $\Lambda$).  This solution should therefore be a better approximation than pure Schwarzschild at late times in a FRW universe with $\Lambda >0$.  The mass of a black hole in such a space is,
\beq m_{\rm b}= \frac{r c^2}{2G}\left(1-\frac{\Lambda r^2}{3c^2}\right). \label{eq:mbsdeS}\eeq
There are two positive real solutions for $r$.  The inner is identified with the black hole radius, $r_{\rm b}$.   The outer is identified with the cosmological event horizon radius, $r_{\rm c}$.   We approximate a black hole embedded in an arbitrary FRW universe using the Schwarzschild-de Sitter solution.  This will be a good approximation for eternally expanding FRW universes with $\Lambda>0$ at late times.  

We have the freedom to choose the mass of our black holes arbitrarily.  The number density of black holes is then constrained by the need to remain consistent with the matter density of the universe.  
Recall, the normalized matter density of the universe, $\om$, is related to the density by,
\beq      \rho_0=\frac{3H_0^2\,\om}{8\pi G}.\eeq
We assume that the black holes are the only contribution to the matter density of the universe, $\rho_0 = \rho_{\rm b_0}$.  Let $n_{\rm b_0}$ be the current number density of black holes. Then,
\bea \rho_{\rm b_0}&=& m_{\rm b}\;n_{\rm b_0,}\\
n_{\rm b_0}&=&\frac{3H_0^2\,\om}{8\pi\, G\,m_{\rm b}}.\eea
The black hole number density drops like $n_{\rm b}=n_{\rm b_0}a^{-3}$ as the universe expands.  For both Schwarzschild and Schwarzschild-de Sitter space the horizon surface area is given by\footnote{To calculate the area of a surface with $dr=0$ we integrate over the angular terms in the metric.  Therefore for both Schwarzschild and Schwarzschild-de Sitter horizons, $A_{\rm b}=\int_0^\pi \int_0^{2\pi}r^2\sin\theta\,{\rm d}\phi{\rm d}\theta=4\pi r^2$.} $A_{\rm b}=4\pi r_{\rm b}^2$.  However, in general FRW universes the surface area of a single black hole's event horizon will depend to some extent on the spacetime geometry of the cosmological model.  For very large black holes or very early epochs these corrections may be significant.  A full treatment of black hole solutions in time-dependent cosmological backgrounds is beyond the scope of this thesis.  As a first approximation, however, we may correct for the spacetime curvature of the embedding space by introducing the factor $S_k$ such that,
\beq A_{\rm b}=4\pi R^2(t)\,S_k^2(r_{\rm b}/R),\label{eq:Abhsingle}\eeq
 (c.f. Eq.~\ref{eq:Ac}).  This factor is chosen so that the areas of the black hole and cosmological horizons are equal when $r_{\rm b}=r_{\rm c}$. 
Thus the total surface area of all the black hole event horizons, $A_{\rm b,tot}$, is given from Eqs.~\ref{eq:mbsdeS}--\ref{eq:Abhsingle} and Eq.~\ref{eq:volume} by,
\beq 
A_{\rm b,tot}= A_{\rm b}\; n_{\rm b}\; V_{\rm c}.\label{eq:Abh}\eeq

\subsection{Small departures from de Sitter space}\label{sect:bh-small-dep}
Restricting for the moment to small perturbations about de Sitter space, we may proceed in the same fashion as the calculation in Eqs.~\ref{eq:sdotc1}--\ref{eq:sdotr} 
 above.  For a first approximation we stick to the Schwarzschild black hole with $r_{\rm b}=2m_{\rm b}$.  The total black hole area within the cosmological horizon is, 
\bea A_{\rm b,tot} &=& \; (4\pi r_{\rm b}^2) \;\left( \frac{\rho_{\rm b}}{m_{\rm b}}\right) \left(\frac{4}{3}\pi H^{-3}\right)\\ 
&=& \frac{32\pi^2 r_{\rm b}\rho_{\rm b}}{3\,H^3}.\eea
The rate of change of black hole area is therefore,
\beq \dot{A}_{\rm b,tot}= -\frac{32\pi^2 r_{\rm b} \rho_{\rm b}}{H^2}\label{eq:Atotbh} \eeq
where we have used $\dot{\rho}_{\rm b}=-3H(\rho_{\rm b}+p_{\rm b})$ from Eq.~\ref{eq:Fried1} with $p_{\rm b}=0$, and once again ignored the $\dot{H}$ term.  
Eq.~\ref{eq:sdotc1} gives the rate of change of the cosmological horizon entropy, $\dot{S}_{\rm c}$, with $\dot{H}=-4\pi (\rho+p)$ 
where $p_{\rm b}=0$ for black holes.  Using $\dot{A}_{\rm c}=4\dot{S}_{\rm c}$ gives,
\beq \dot{A}_{\rm c} = \frac{-32\pi^2\rho_{\rm b}}{H^3}.\label{eq:Acdotbh}\eeq   
For the GSL to hold we need,
\beq \dot{A}_{\rm b,tot} + \dot{A}_{\rm c} \ge 0.  \label{eq:Abhdot}\eeq
Using Eqs.~\ref{eq:Acdotbh} and~\ref{eq:Atotbh} this inequality becomes 
\bea r_{\rm b} &\lsim & 1/H,\\
                  r_{\rm b} &\lsim & r_{\rm c}.\label{eq:rblimitS}\eea 
Thus the GSL holds as long as the black holes are smaller than the cosmological event horizon.  

Converting to a Schwarzschild-de Sitter black hole slightly increases the size of the black hole and decreases the size of the cosmological event horizon.
Redoing the same analysis using the Schwarzschild-de Sitter mass-radius relationship (Eq.~\ref{eq:mbsdeS}) gives the total black hole area to be,
\beq A_{\rm b,tot} = \frac{32\pi^2 r_{\rm b}\rho_{\rm b}}{3\,H^3(1-H^2r_{\rm b}^2)}, \eeq
where we have used the fact that $r_{\rm c}=r_{\rm deS}=\sqrt{\Lambda/3}=1/H$ for a de Sitter cosmological horizon.  The rate of change of total black hole area is,
\beq \dot{A}_{\rm b,tot} = \frac{-32\pi^2 r_{\rm b}\rho_{\rm b}}{H^2(1-H^2r_{\rm b}^2)}, \eeq
assuming $\dot{H}$ is negligible.  The inequality Eq.~\ref{eq:Abhdot} reduces to,
\beq r_{\rm b} \;\lsim\;  r_{\rm c}\left(1-\frac{r_{\rm b}^2}{r_{\rm c}^2}\right),\label{eq:rblimitSdeS}\eeq
so for the GSL to hold the black holes have to be somewhat smaller than the de Sitter horizon (as the assumption of ``small departures from de Sitter'' ensures).  The limit on the right hand side of Eq.~\ref{eq:rblimitSdeS} is less than that of Eq.~\ref{eq:rblimitS} as we would expect, because in the Schwarzschild-de Sitter case the cosmological event horizon is smaller than the $r_{\rm c}=1/H$ which we used here.  

We have been unable to solve inequality~\ref{eq:Abhdot} exactly for all FRW universes.  Instead in the next section we provide numerical solutions for a broad range of black hole cosmological models including large departures from de Sitter space.

\subsection{Numerical extension to far-from-de-Sitter FRW models}

In this section we show the results of numerical calculations assessing the loss of black hole entropy across the cosmological event horizon in general FRW models that are large departures from de Sitter space.  We use the numerical calculations to find the comoving distance to the cosmological event horizon from which we can calculate both $A_{\rm c}$ (Eq.~\ref{eq:Ac}) and $V_{\rm c}$ (Eq.~\ref{eq:volume}), in turn allowing us to use Eq.~\ref{eq:Abh} for $A_{\rm b,tot}$.

We find that for cosmological models with realistic black hole sizes (up to, say, the size of a typical supermassive black hole) the increase in cosmological horizon area overwhelms the loss of black hole horizon area, in clear conformity with the extended GSL.  Greater interest, then, attaches to the case where the black holes are relatively large enough to represent a significant fraction of the total horizon area.  In a realistic case this would refer only to very early epochs, on the assumption that primordial black hole formation had taken place.  In what follows we concentrate on the case where the ratio of black hole horizon area to total horizon area is large.

\begin{figure*}\bctr
\includegraphics[width=49mm]{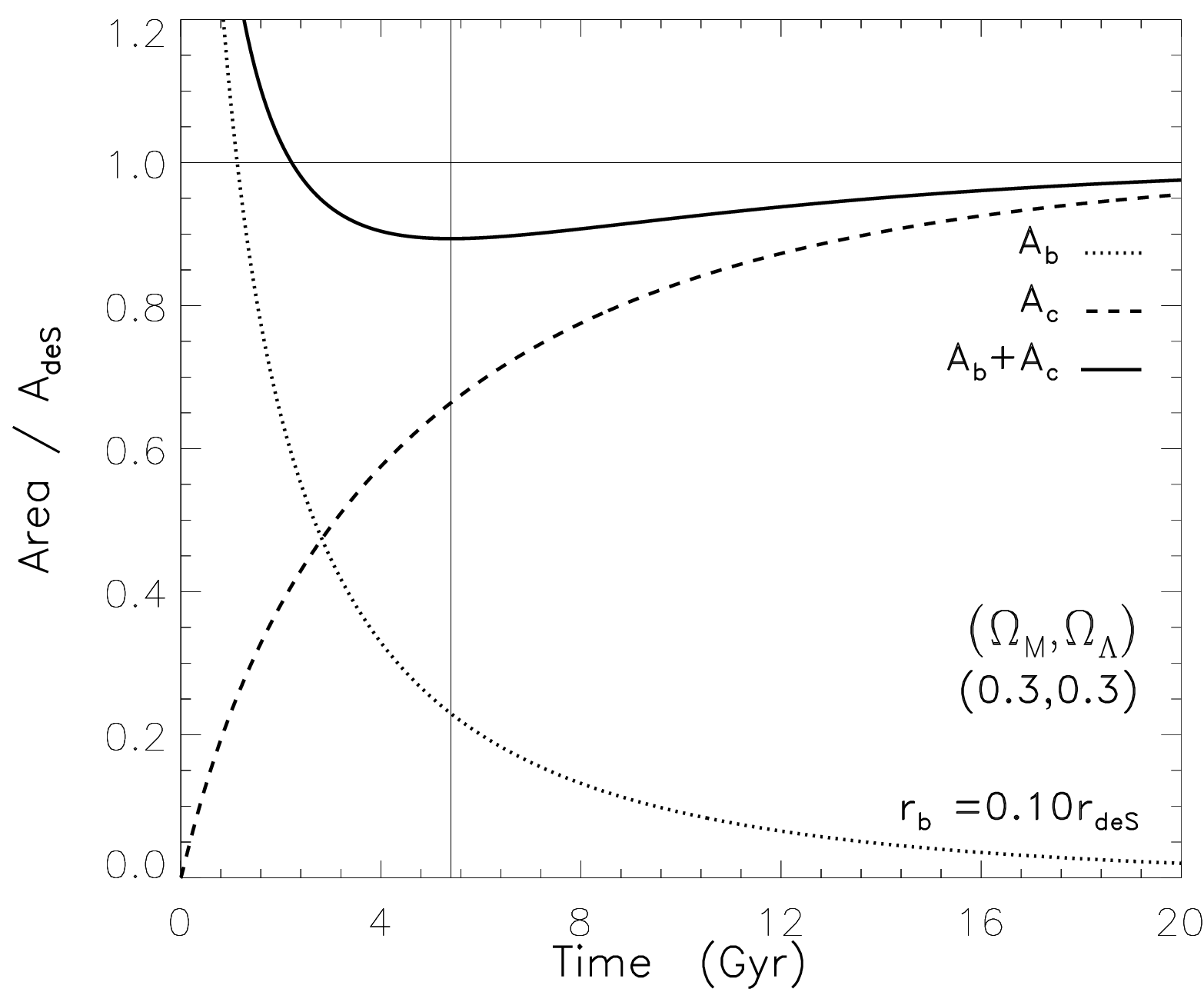}
\includegraphics[width=49mm]{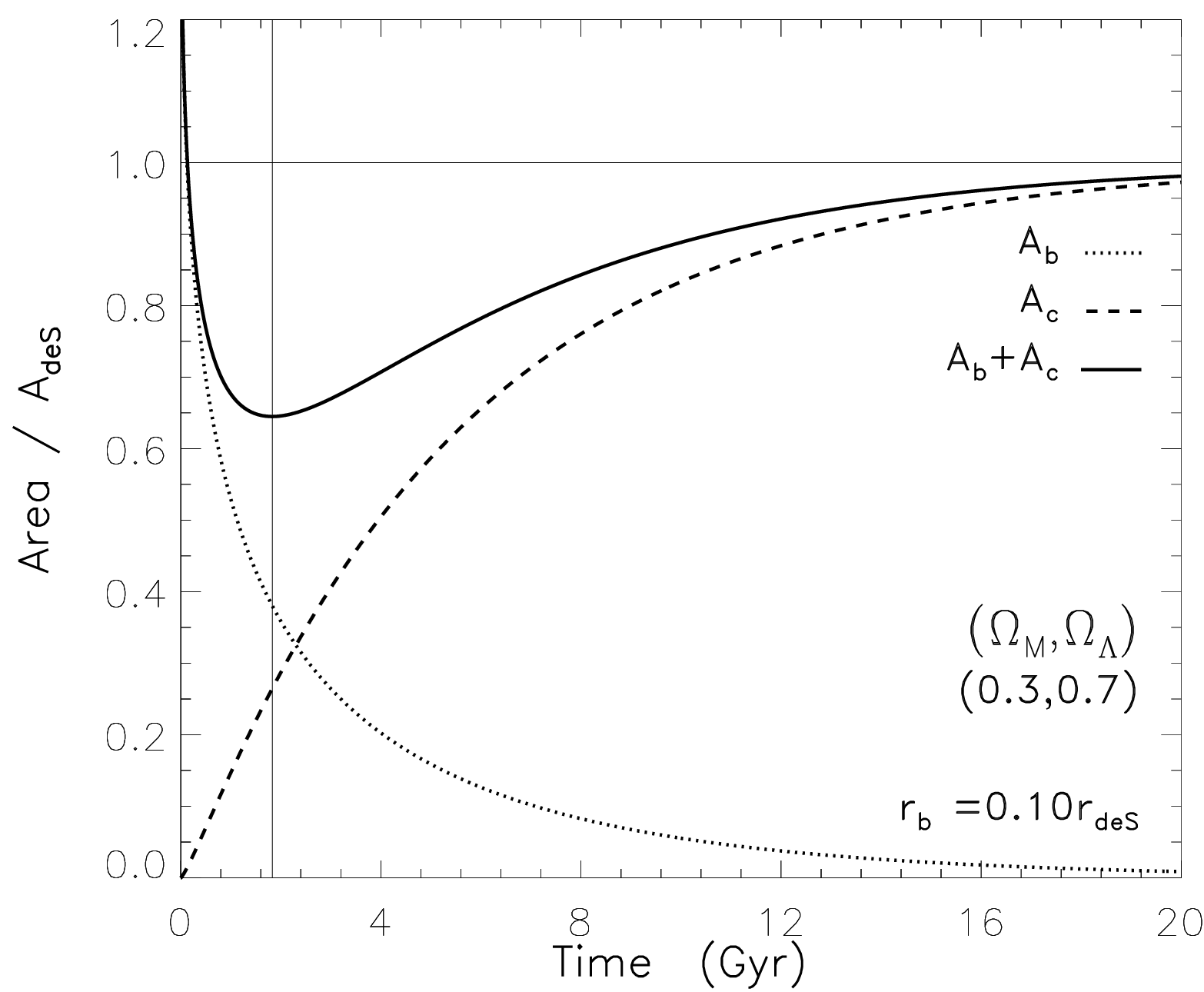}
\includegraphics[width=49mm]{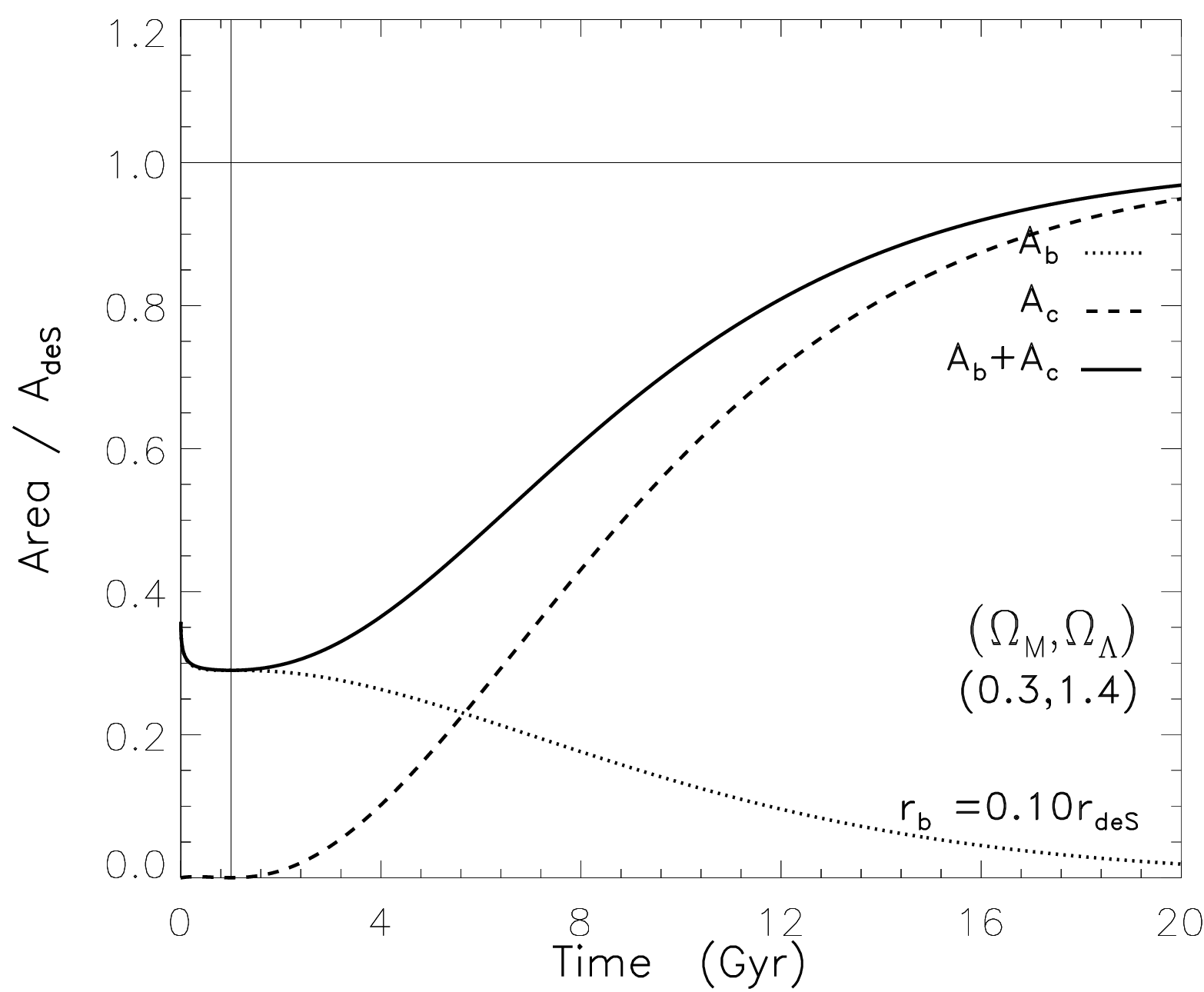}
\caption[Black holes cross the cosmological event horizon (no corrections)]{\small The evolution of total horizon area is shown as a function of time for three FRW models filled with a pressureless gas of black holes.  
In all models the black hole radius is $r_{\rm b}=0.1r_{\rm deS}$.   The vertical axis has been scaled to the de Sitter horizon area, $A_{\rm deS}$, in each model.    The dotted line shows the total area of black hole horizons within the cosmological event horizon.  The dashed line shows the area of the cosmological event horizon.  The thick solid line shows the sum of the black hole and cosmological horizon areas.  The thin, solid vertical lines mark turning points in the total horizon area curve.
This series of graphs shows the results when we do not correct for the effects described in Sect.~\ref{sect:corrections}.  Significant departures from the GSL can be seen at early times in all models.  
}
\label{fig:area-uncorrected}\ectr
\end{figure*}

The de Sitter horizon at $r_{\rm deS}=\sqrt{3/\Lambda}$ is the horizon that would exist if the matter density were zero in each model.  As such it is the asymptotic limit in time of the cosmological event horizon.  We express the results of the numerical calculations in terms of the radius and area of the de Sitter horizon.  The results of these numerical calculations are shown in Fig.~\ref{fig:area-uncorrected}.  Black hole event horizon area, cosmological event horizon area and the total horizon area are plotted against time for a variety of models. 

\begin{figure*}\bctr
\includegraphics[width=48.5mm]{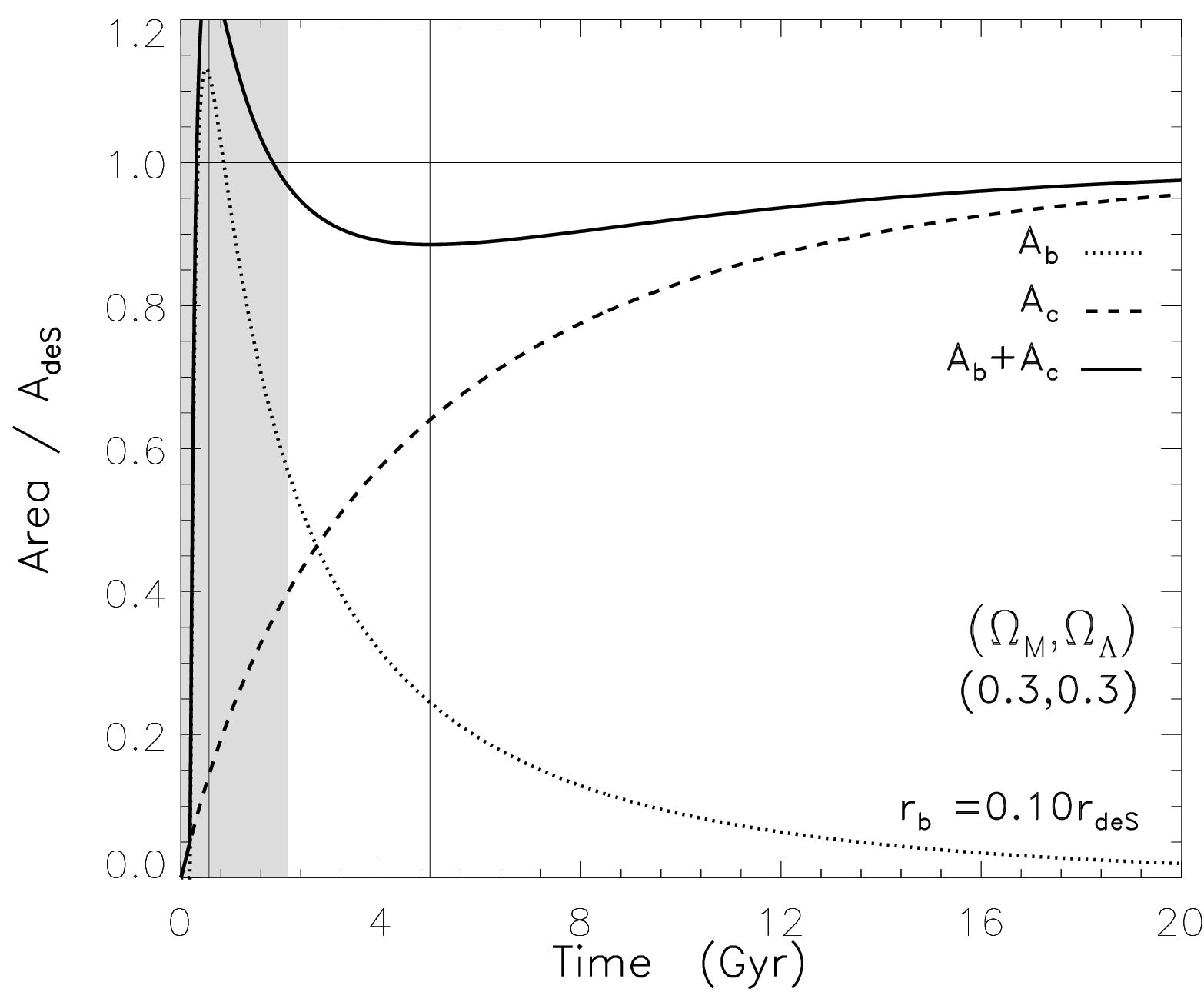}
\includegraphics[width=48.5mm]{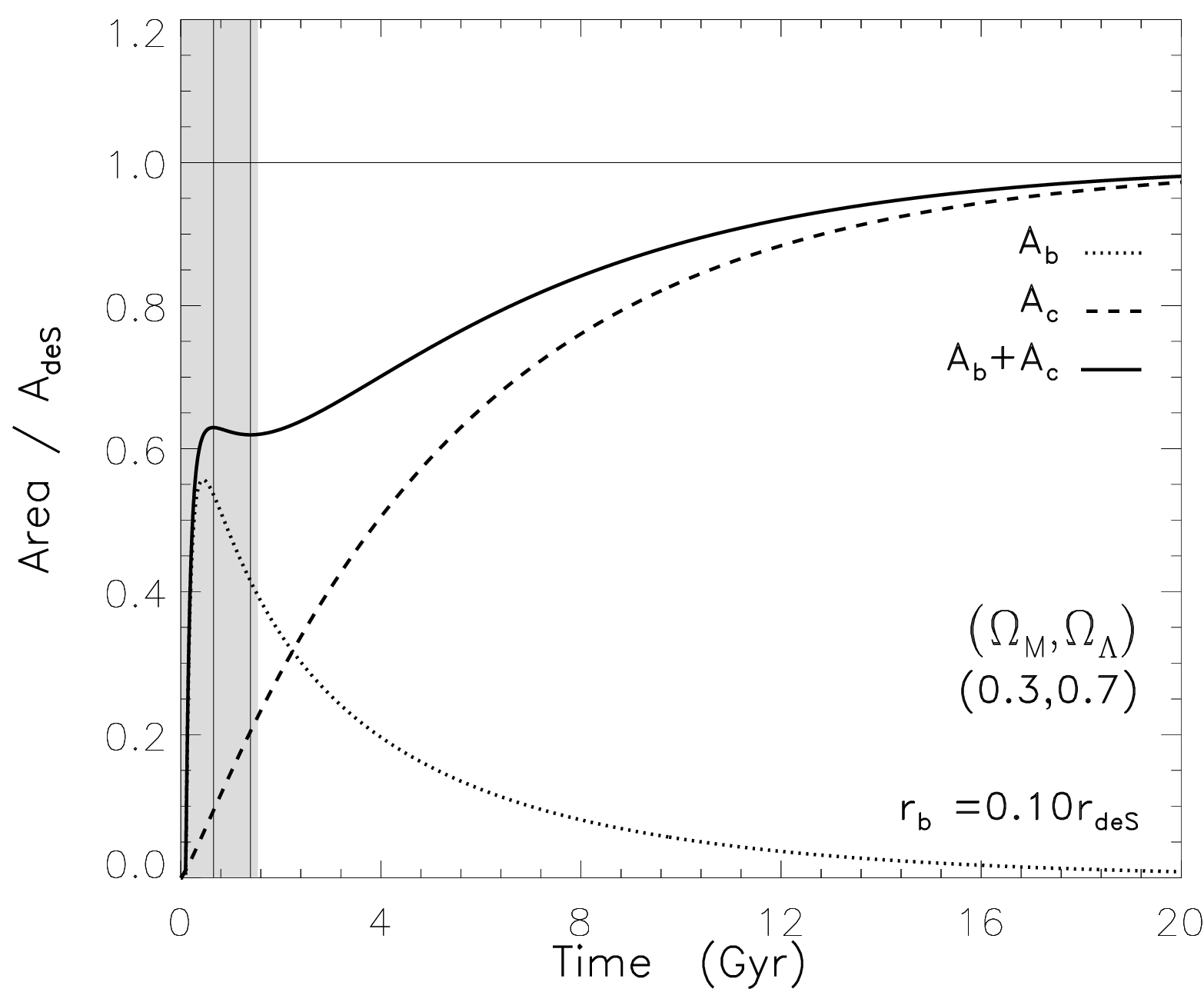}
\includegraphics[width=48.5mm]{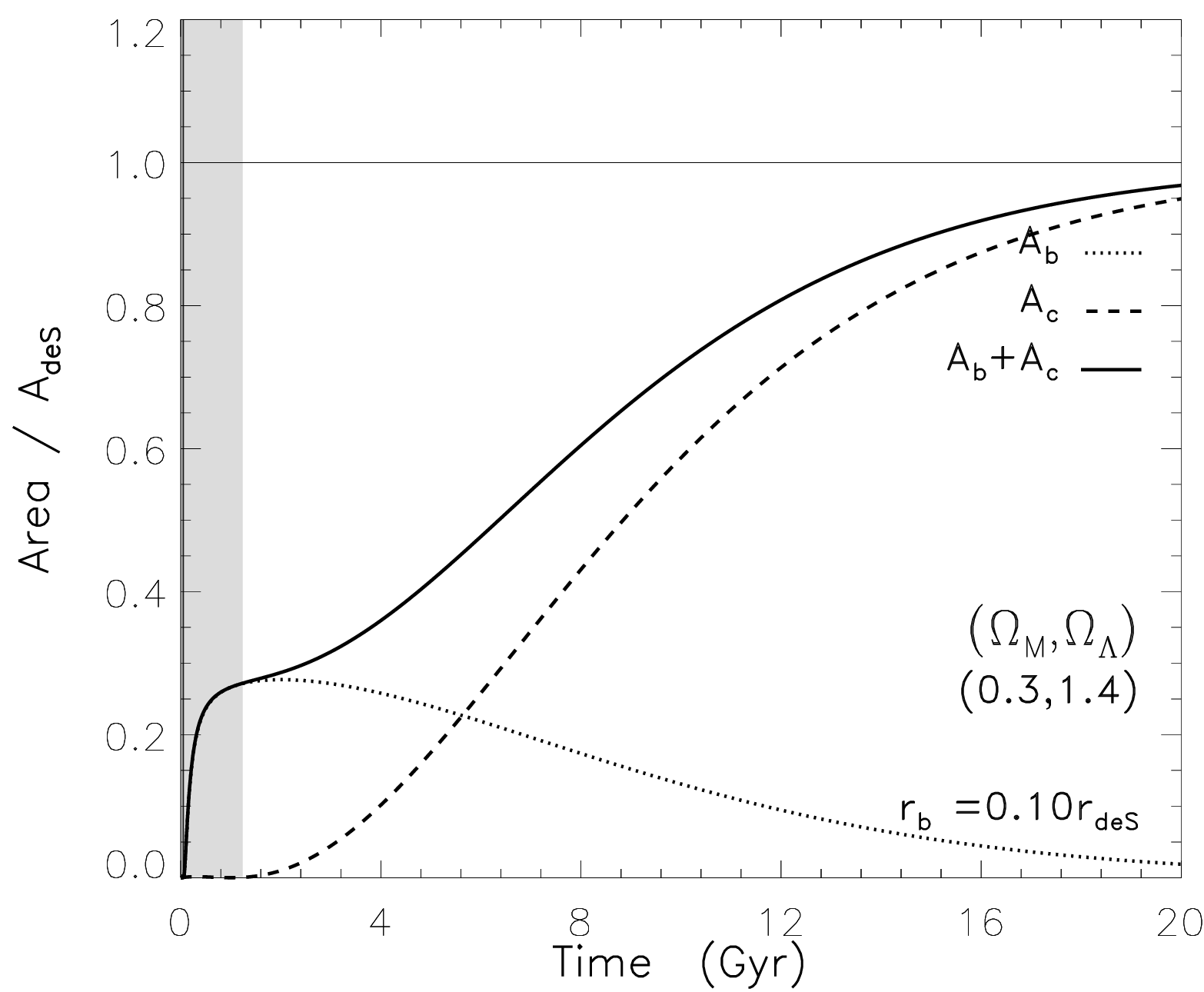}
\includegraphics[width=48.5mm]{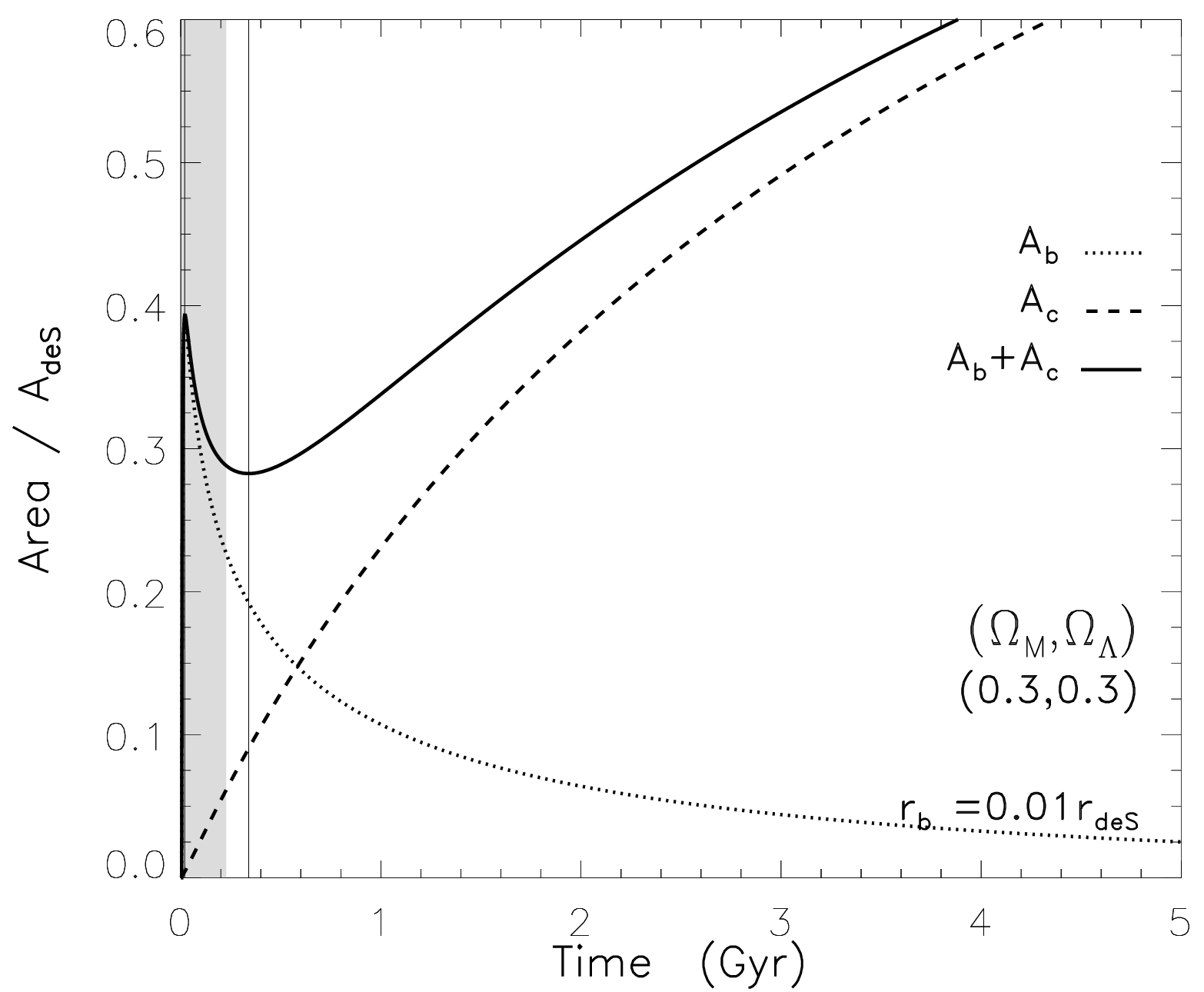}
\includegraphics[width=48.5mm]{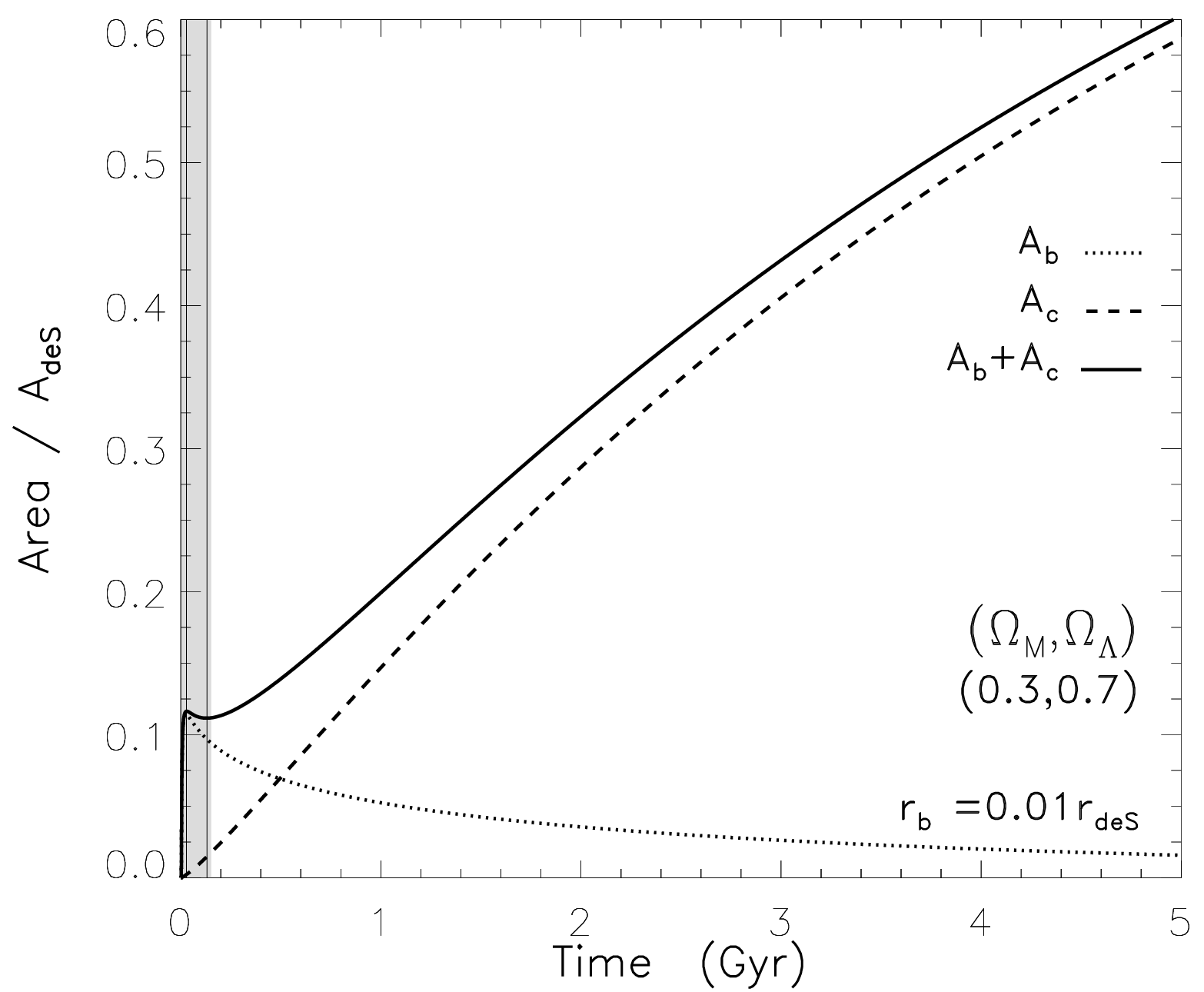}
\includegraphics[width=48.5mm]{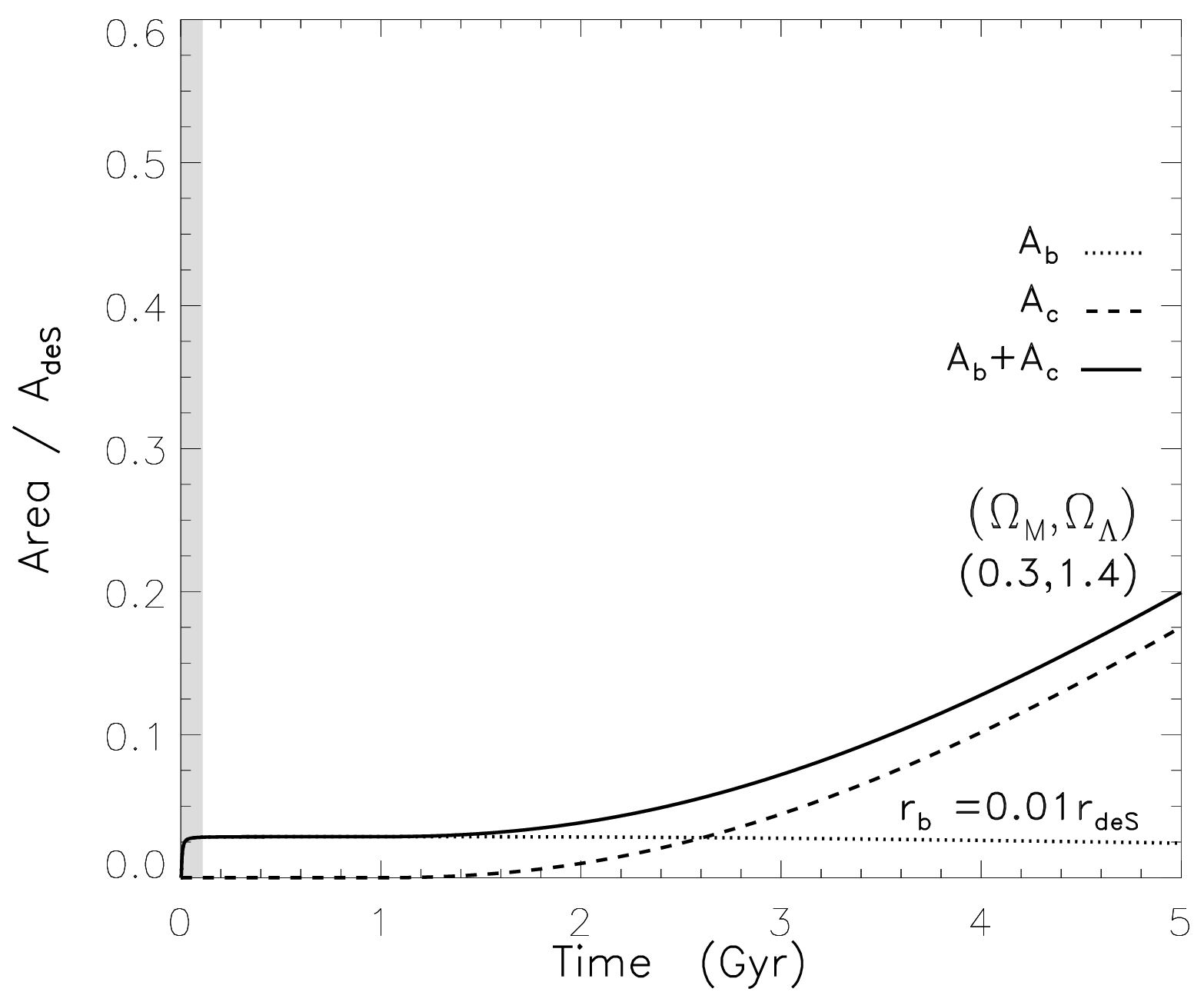}
\caption[Black holes cross the cosmological event horizon (with corrections)]{\small Corrections have been made for the three assumptions listed in Sect.~\ref{sect:corrections}.  The gray shading indicates the region that should be neglected because black holes overlap.  The black hole contribution to area starts from zero and peaks because black holes initially have a radius larger than the cosmological horizon radius and so are excluded from the area calculation by Eq.~\ref{eq:rc-secant}.  The upper row has $r_{\rm b}=0.1r_{\rm deS}$ while the lower row has $r_{\rm b}=0.01r_{\rm deS}$.
Here the areas of black holes have been calculated in the geometry of the type of universe they are embedded in (using Eq.~\ref{eq:Abhsingle}). The results are qualitatively unchanged when $A_{\rm b}=4\pi r_{\rm b}^2$ is used.
With these corrections the GSL violation in flat and closed models has been removed, but a GSL violation remains in the open model at early times.}
\label{fig:area}\ectr
\end{figure*}

\subsection{Corrections needed to naive calculation}\label{sect:corrections}
Treating the problem as stated so far we find significant departures from the GSL at early times in all models and at late times for large black holes. The majority of these departures are an obvious artefact of the approximations we have used.
Firstly, by treating the black holes as dilute dust (and as solid spheres) we have neglected interactions between them.  At very early times the black holes in the simulation are so densely packed that they overlap, which is clearly unphysical (see~\ref{sect:overlap}).    Secondly, we have assumed that the disappearance of a black hole across the cosmological horizon is instantaneous, but for black holes of size comparable to the cosmological horizon this is unrealistic.  A proper GR treatment of the merging of horizons, which will involve significant departures from homogeneity and isotropy, is beyond the scope of this thesis.  However, as a first approximation to compensating for this effect, we use a simple geometric argument (see~\ref{sect:secant}).  Taking both the above considerations into account removes almost all the departures from the GSL.  The corrected area-time plots are shown in Fig.~\ref{fig:area} with the unphysical black-hole-overlap regions shaded gray.
\begin{figure}\bctr
\includegraphics[width=78mm]{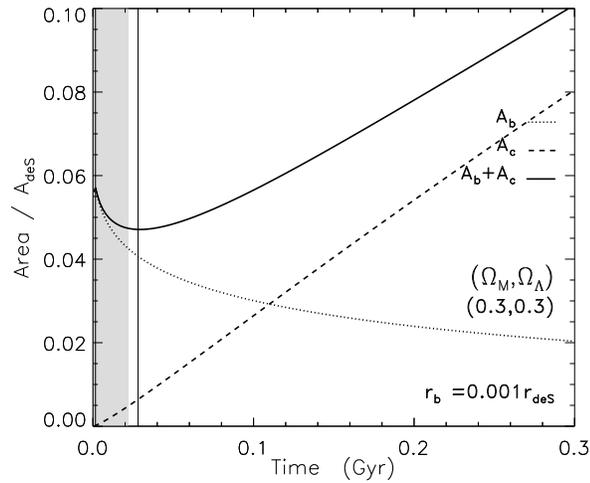}
\caption[GSL violation due to break down of assumptions (mass outside black hole)]{\small An example in which the assumption of the Schwarzschild-de Sitter solution for black hole area breaks down because of the presence of matter density outside the black holes.  The GSL appears to be violated by the entropy decrease at early times even for small black holes.}
\label{fig:area-r0001rdeS-curved}\ectr
\end{figure}

A third approximation which we have used but cannot correct for is the assumption that the Schwarzschild-de Sitter solution for the black hole radius holds.  This neglects the presence of matter density outside the black hole.   This approximation is therefore suspect at early times in FRW universes while the universe is dominated by matter rather than dark energy ($\Lambda$).  The effect can be minimized by concentrating on small black holes.  An example of a GSL violation which we attribute to the breakdown of the Schwarzschild-de Sitter assumption is shown in Fig.~\ref{fig:area-r0001rdeS-curved} for the spatially open ($k=-1$) model where departures from the GSL are indicated at early times. 

\begin{figure}\bctr
\includegraphics[width=78mm]{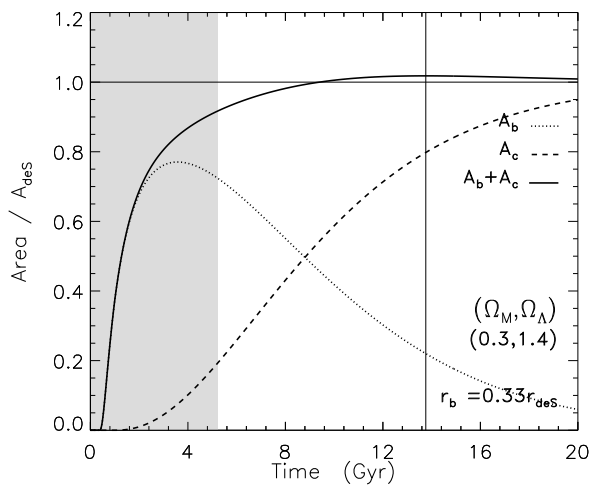}
\caption[GSL violation due to break down of assumptions (large black holes)]{\small A model universe filled with large black holes for which the assumption of the Schwarzschild-de Sitter solution breaks down.  The GSL appears to be violated by the entropy decrease at late times.}
\label{fig:area-r033rdeS-curved}\ectr
\end{figure}

The Schwarzschild-de Sitter approximation also breaks down when the radius of the black hole is comparable to the radius of the cosmological event horizon.  This is because the effect of the embedding spacetime on the mass-radius relationship of a black hole becomes larger for larger black holes (see the term in brackets in Eq.~\ref{eq:mbsdeS}).    
An example is shown in the spatially closed ($k=+1$) model illustrated in Fig.~\ref{fig:area-r033rdeS-curved}, where departures from GSL are indicated at late times.  

A more accurate resolution of these departures from the GSL awaits the derivation of horizon solutions for black holes embedded in arbitrary evolving FRW spacetimes.  In the next section we provide brief details of how we have dealt with these corrections and the effect they have on the total entropy results.

\section{Correction details}

\subsection{Exclude overlapping and superhorizon-sized black holes}\label{sect:overlap}
We rule out the times when black holes are so close that they overlap as being unphysical.  The separation between black holes is given by $n_{\rm b}^{-1/3}$.  So we rule out any regions for which,
\beq 2r_{\rm b} \le n_{\rm b}^{-1/3}. \label{eq:nb}\eeq
The unphysical region defined by Eq.~\ref{eq:nb} is shaded gray in Figs.~\ref{fig:area}--\ref{fig:area-r033rdeS-curved}.  This also implicitly excludes the region for which the black holes are larger than the cosmological event horizon.  The black holes become smaller than the cosmological horizon before they cease to overlap.  To ensure we never inadvertantly include an unphysical region our program also includes a clause excluding superhorizon-sized black holes.

\subsection{Geometric considerations}\label{sect:secant}
By considering a black hole to have crossed the cosmological horizon when its centre passes over it we calculate too much black hole horizon area (averaged over all black holes) to be inside the cosmological horizon.
To fix this we need to calculate the point at which exactly half the black hole horizon is outside the cosmological horizon.  This occurs when the black hole's diameter makes a secant to the cosmological horizon.  

\setlength{\unitlength}{5mm}
\begin{figure}[h!]\bctr
\begin{picture}(16,6.5)
	\thicklines
	\put(3,3){\circle{6}} 
	\put(5.11,5.11){\circle{2}} 
	\put(5.11,5.11){\line(-1,1){1.5}}
	\put(5.11,5.11){\line(1,-1){1.5}}
	\put(3,3){\line(1,1){2.1}}
	\put(3.4,3.9){\makebox(0,0){$r_c$}}
	\put(7.1,5.5){\makebox(0,0){black}}
	\put(6.9,4.9){\makebox(0,0){hole}}
	\put(7.5,0.6){\makebox(0,0){cosmological}}
	\put(7.5,0.0){\makebox(0,0){event horizon}}
	\put(5.0,0.0){\vector(-1,1){0.5}}
	\put(10.0,0.0){\vector(1,1){0.5}}
	\put(12,3){\circle{6}} 
	\put(14.4,3){\circle{3.5}} 
	\put(14.4,1.25){\line(0,1){3.5}}
	\put(12,3){\line(1,0){3}}
	\put(12,3){\line(4,3){2.4}}
	\put(13.2,2.3){\makebox(0,0){$r_c-\delta$}}
	\put(12.4,3.8){\makebox(0,0){$r_c$}}
	\put(14.65,2.6){\makebox(0,0){$\delta$}}
	\put(14.0,3.6){\makebox(0,0){$r_{\rm b}$}}
	\put(12,2.8){\vector(1,0){2.4}}
	\put(14.4,2.8){\vector(-1,0){2.4}}
	\put(16.3,4.8){\makebox(0,0){black}}
	\put(16.6,4.2){\makebox(0,0){hole}}
\end{picture}\ectr
\end{figure}

Therefore we should consider black holes to have left the horizon when they are a distance $\delta$ from the horizon where $\delta$ is the length of the perpendicular bisector of the secant between the secant and the perimeter of the event horizon.  That is, when we calculate the volume within the event horizon that contains black holes we should use $r_c-\delta$. 
\beq    r_{\rm c} - \delta = \sqrt{r_{\rm c}^2 -r_{\rm b}^2}. \label{eq:rc-secant}\eeq
This corrected calculation is shown in Fig.~\ref{fig:area}. 

\subsection{Radius calculation}\label{sect:radius}
An indication of the magnitude of the effect of different embeddings can be gained by comparing the Schwarzschild-de Sitter solution to the Schwarzschild solution.  For a particular black hole radius the difference in mass for the two embeddings is $\Delta m_{\rm b}/m_{\rm b}=(m^{\rm deS}_{\rm b}-m^{\rm Sch}_{\rm b})/m^{\rm Sch}_{\rm b} = -\Lambda r_{\rm b}^2/3c^2$.  That means that for $H_0=70\, kms^{-1}Mpc^{-1}$ and  $\oll = 0.7$ the difference between the two solutions is less than $\Delta m/m = 0.01$ as long as black holes are smaller than 1.7 billion light years across (the de Sitter horizon for a Universe with $\oll=0.7$ sits at $r_{\rm deS}=\sqrt{3/\Lambda}=16.7\, Glyr$ so $1.7\, Gyr$ represents $r_{\rm b} = 0.1 r_{\rm deS}$, c.f.~Fig.~\ref{fig:area}).   Therefore to minimize the effect of the embedding spacetime on the radius of a black hole we simply need to use ``small'' black holes (a ``small'' black hole of $0.17\, Glyr$ radius is still on the order of $10^{21}$ solar masses).

The only GSL violation that does not disappear when black holes are restricted to small sizes is the early time entropy decrease that occurs in open universes because of the breakdown of the Schwarzschild-de Sitter solution in this regime.  It may be possible to infer the direction of the correction from the difference between the Schwarzschild-de Sitter and Schwarzschild solutions.  When the space surrounding the black hole is accelerating, as in the Schwarzschild-de Sitter case, the black hole event horizon radius is larger than in the pure Schwarzschild case.  If the opposite were to happen and black hole radius decreased when a black hole was surrounded not with the accelerating effect of cosmological constant but with the decelerating effect of matter density, then it would reduce the entropy contribution from black holes.
Such a correction is in the direction required to remove the total entropy decrease at early times in these open models.

We emphasize that any other apparent departures from GSL are manifested only in the extreme cases where the size of the black holes approach the size of the observable universe.  In a realistic cosmological model, the largest black holes formed by merger will still be orders of magnitude smaller than the cosmological horizon.  In those cosmological models that permit primordial black hole formation from density perturbations, the size of the holes is still generally much less than the cosmological horizon size at the epoch of formation.

\subsection{Area calculation}
\begin{figure*}\bctr
\includegraphics[width=48.5mm]{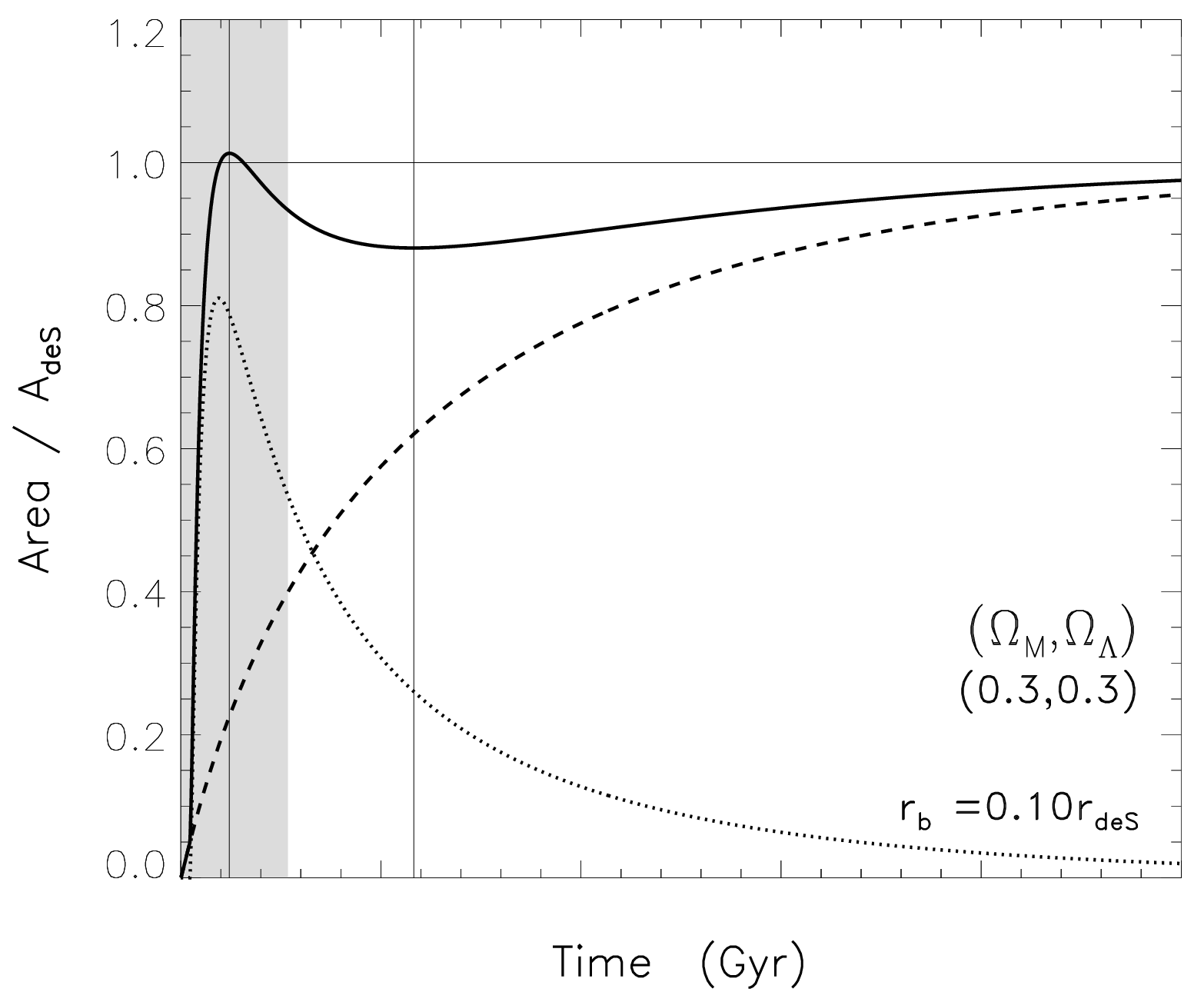}
\includegraphics[width=48.5mm]{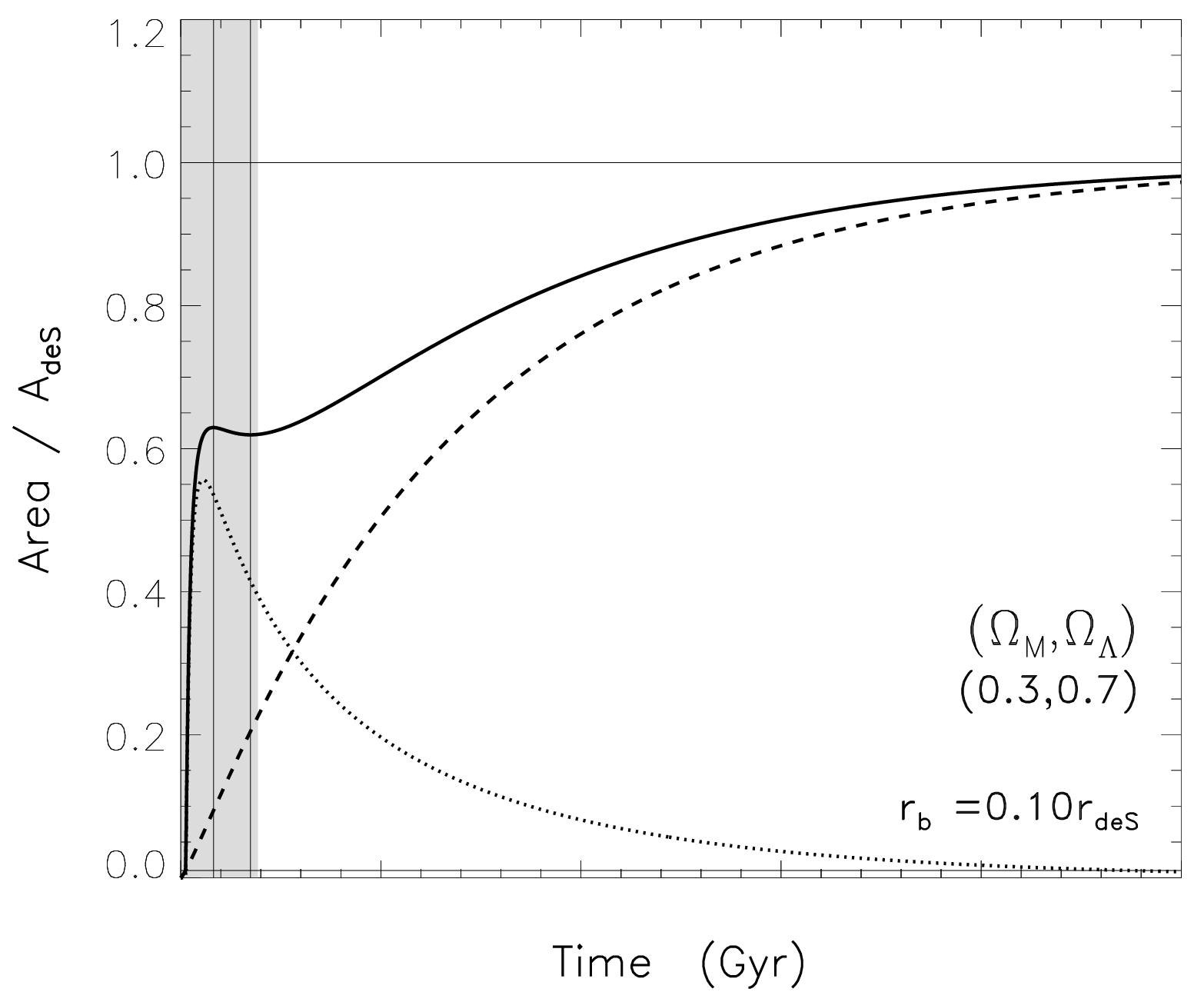}
\includegraphics[width=48.5mm]{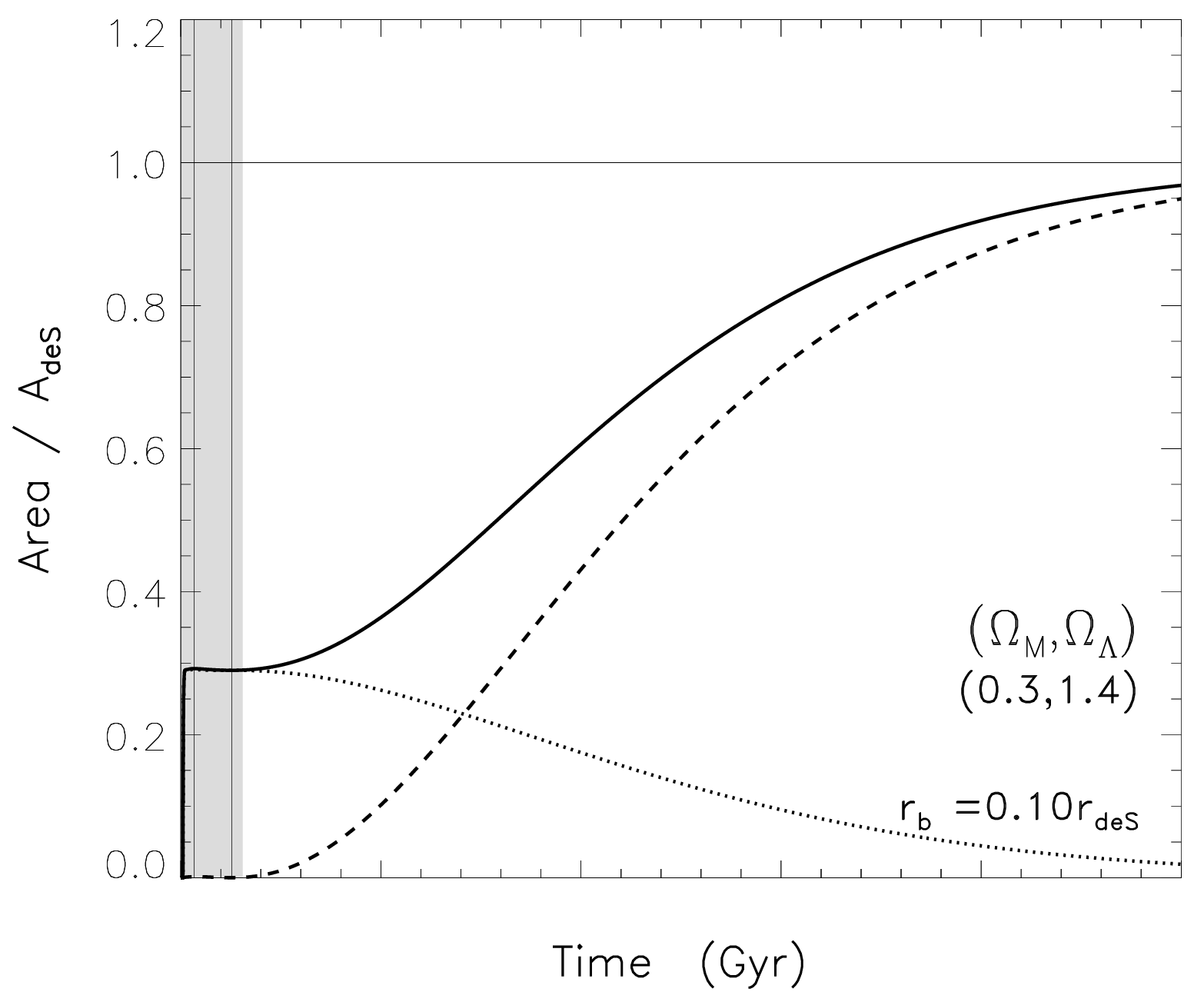}
\includegraphics[width=48.5mm]{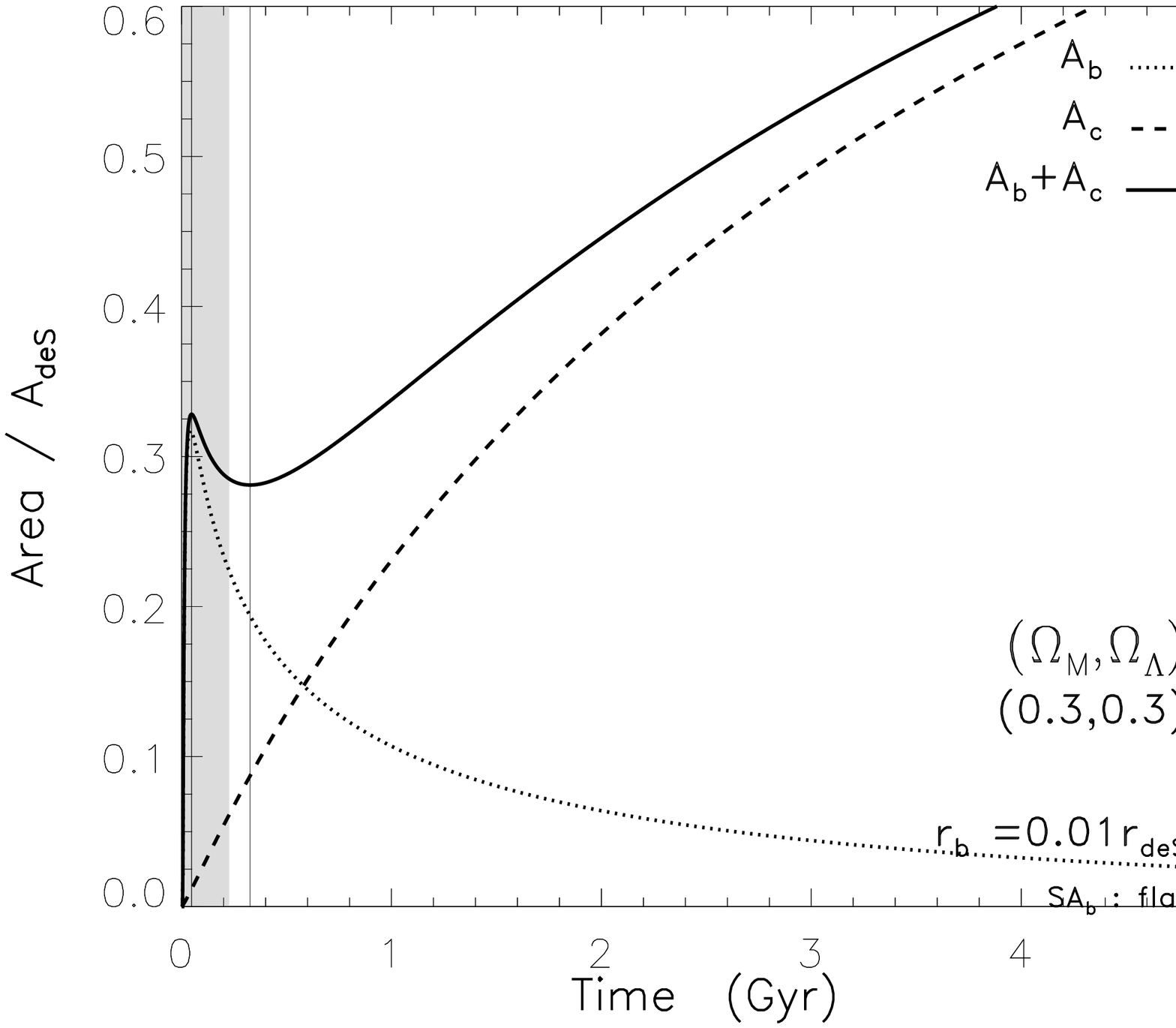}
\includegraphics[width=48.5mm]{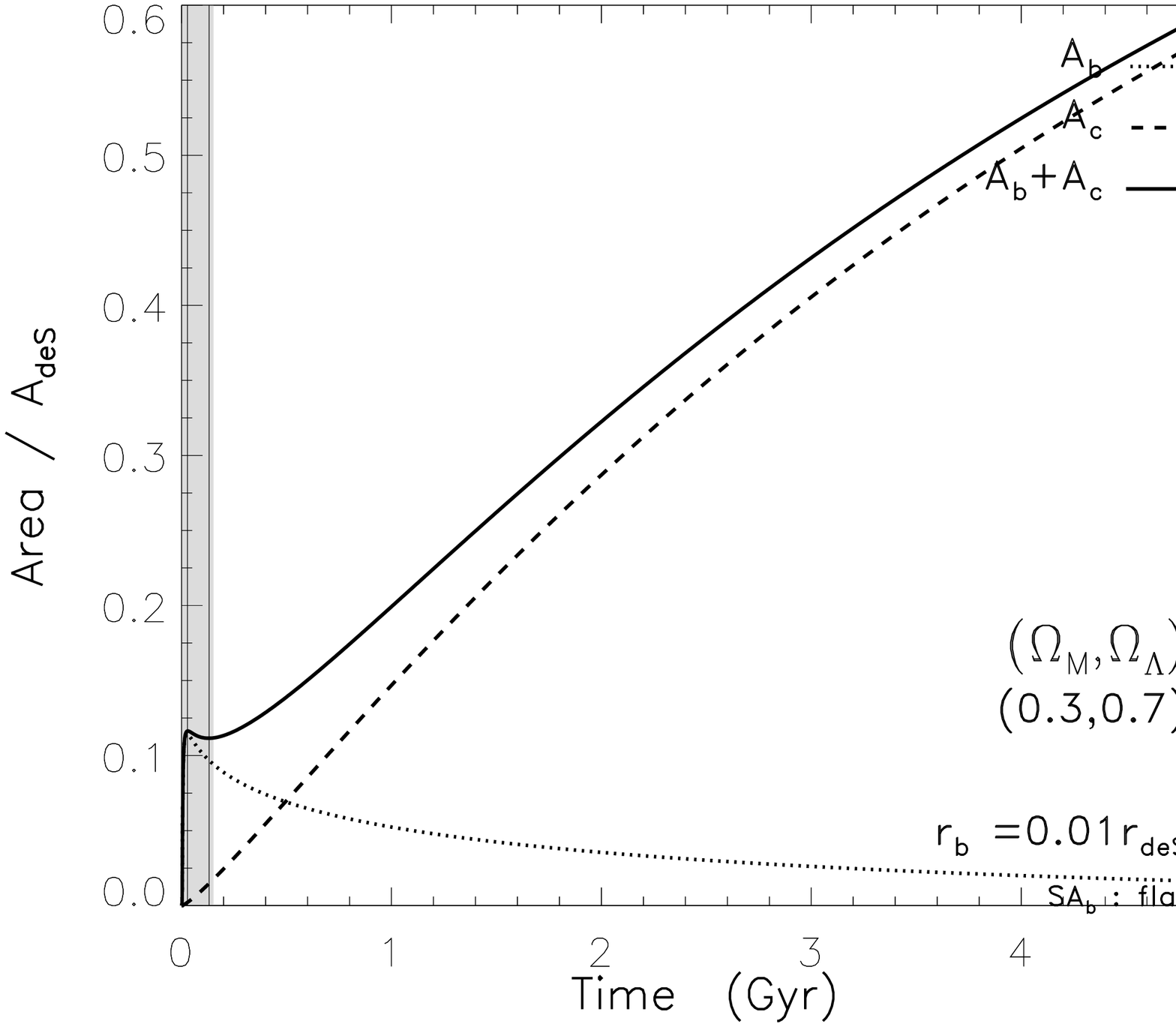}
\includegraphics[width=48.5mm]{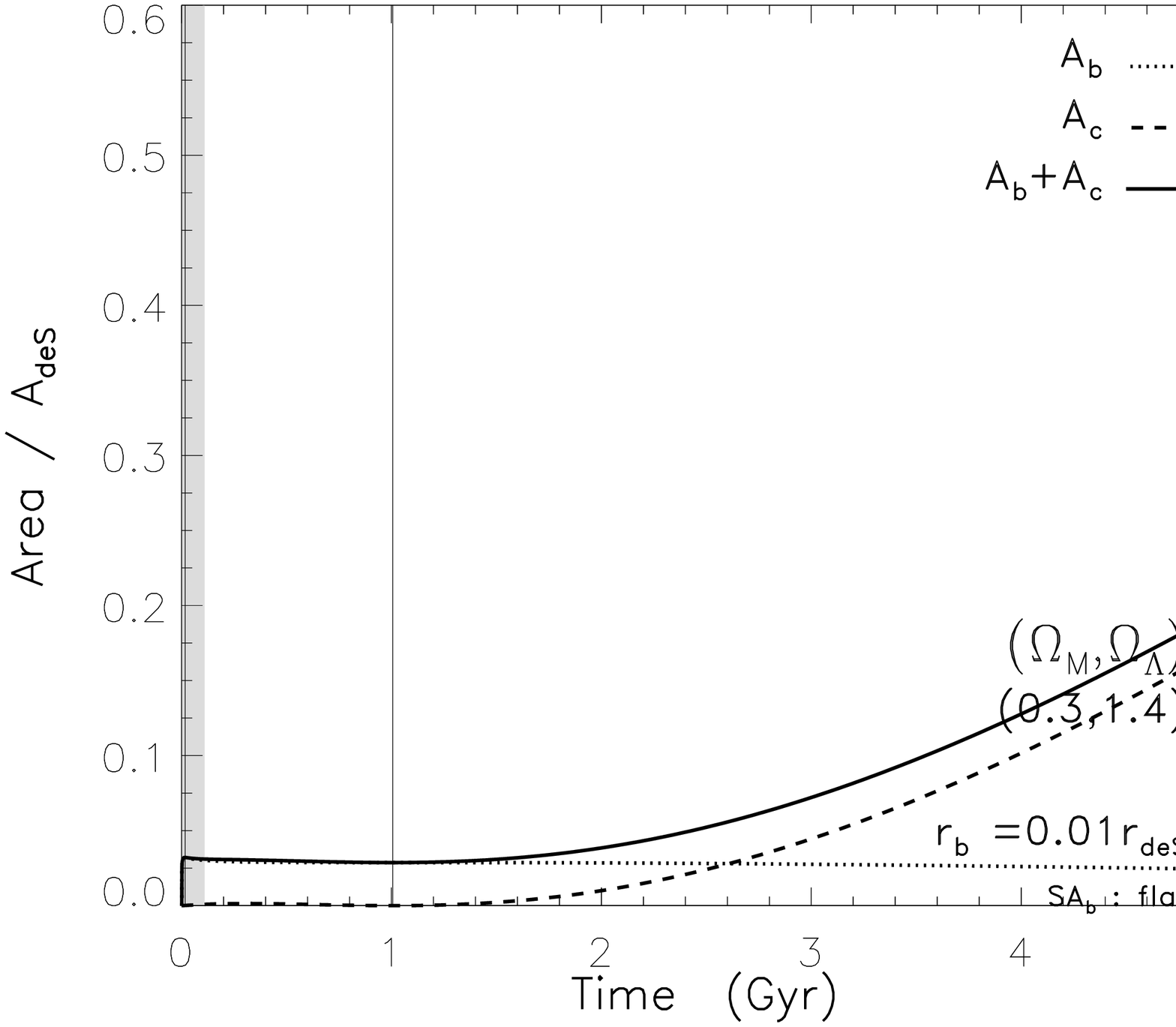}
\caption[Entropy using basic area calculation]{\small We show the corrected results as calculated using $A_{\rm b}=4\pi r_{\rm b}^2$ instead of Eq.~\ref{eq:Abhsingle} for area.  The difference is significant but does not remove the GSL violations.}
\label{fig:area-flat}\ectr
\end{figure*}

Since we have used Eq.~\ref{eq:Abhsingle} without proof we provide in Fig.~\ref{fig:area-flat} graphs of the calculation using, $A_{\rm b}=4\pi r_{\rm b}^2$, to show the magnitude of the correction for various models.  The results are qualitatively unchanged.

\vspace{-4mm}
\section{Relationship to other work}
\vspace{-2mm}
We have concentrated on several explicit examples, including some numerical solutions, of extensions of the generalized second law of thermodynamics to cosmological horizons.  This work complements some recent theorems that prove more general but less explicit results that have a bearing on the GSL. For example, for black hole-de Sitter spacetimes, 
\cite{shiromizu93} show that the black hole event horizon area is non-decreasing in asymptotically de Sitter space times,  while 
\cite{hayward94} show that the black hole event horizon area is bounded by $4\pi/\Lambda$.

The cosmological horizon area in models with a positive cosmological constant has been considered by 
\cite{boucher84} who show that the horizon area is bounded by $12\pi/\Lambda$ on a regular time-symmetric hypersurface, while 
\cite{shiromizu93} show that the horizon area is bounded by $12\pi/\Lambda$ on a maximal hypersurface.  Neither of these proves a bound in a non-stationary asymptotically de Sitter universe such as an FRW universe.

\cite{maeda98} extend the work of 
\cite{davies88,davies88b} by showing that the cosmological event horizon area does not decrease in any asymptotically de Sitter spacetimes.  They also show that the de Sitter horizon is the upper limit of horizon size for any cosmological model with nonzero $\Lambda$. Our results both illustrate these theorems and demonstrate that, for certain specific models, the GSL is satisfied not just asymptotically, but at all times.

\vspace{-3mm}
\section{Summary}
\vspace{-3mm}

We define total entropy to be the entropy of a cosmological event horizon plus the entropy within it.  \cite{davies88} 
showed that the entropy of the cosmological event horizon in FRW universes, subject to the dominant energy condition, never decreases.  We examined radiation filled FRW universes and showed that total entropy never decreases for a wide range of models by testing the parameter space using numerical calculations.  We then assessed the entropy lost as black holes disappeared over the cosmological event horizon.  The lack of a black hole solution for arbitrary spacetime embeddings restricts the application of this technique.  Limiting the size of black holes to those small enough that the difference in embedding in empty space compared to de Sitter space is less than 0.1\% allowed us to show that no GSL violation occurs in any of the closed or flat models tested, but an apparent violation occurs at early times in open FRW universes, probably due to the breakdown of the Schwarzschild-de Sitter assumption in the presence of matter density outside a black hole.  Further progress in resolving this matter will require more realistic approximations of black hole solutions in cosmological backgrounds.  An associated issue that needs to be addressed is what constitutes the appropriate surface that characterizes horizon entropy when black holes are situated in a time-dependent background.

\clearemptydoublepage

\chapter{Black hole thermodynamics may constrain theories of varying constants}\label{chap:varies}

\section{Introduction}
Recent evidence suggests the fine structure constant $\alpha =e^2/\hbar c\;$ may have been smaller in high redshift quasar absorption clouds \citep{webb01,murphy01,murphy02a,webb03}.  This raises the question of which fundamental quantities are truly constant and which might be time- or environment-dependent. Several candidate theories have been proposed in which different fundamental constants vary.   We propose that black hole thermodynamics may provide a means to discriminate between alternative theories, because changes in $\alpha$ may affect the horizon area of a charged black hole. Since the event horizon area is widely accepted as a measure of the entropy of the black hole \citep{bekenstein73, bekenstein74,hawking76} some variations in the fundamental `constants' could lead to a violation of the generalized second law of thermodynamics.  

In this Chapter we outline our proposal (\nocite{davies02}Davies, Davis and Lineweaver, 2002) for how black hole entropy could provide a theoretical test of candidate varying constant theories. 
In response to criticisms we include a discussion of the operational meaning of variations in dimensional parameters and provide explicitly dimensionless constraints.  We discuss the limitations of this idea, in particular that the black hole entropy needs to be calculated from within a theory of varying constants for which few solutions are currently available.  

\section{Observational evidence for a varying fine structure constant}\label{sect:varies-obs}

Physicists have long been questioning the constancy of the fundamental constants of nature.  \cite{milne37} and \cite{dirac37} independently suggested that gravitational and electromagnetic clocks may tick at different rates and proposed theories in which the gravitational constant, $G$, varies with time\footnote{Milne's theory had an increasing $G$ while Dirac's theory had a decreasing $G$.}.  
  Both theories have been ruled out observationally \citep[at least in their original form, see review by][]{uzan02}, but the basic idea remains: the constancy of the fundamental quantities of nature is a feature that must be tested experimentally. 

\cite{webb03} give observational evidence for a smaller fine structure constant in high redshift quasar absorption clouds \citep[see also][]{murphy01,murphy02a,webb01}.  They report a variation in the fine structure constant of $\Delta \alpha/\alpha = -0.57 \pm 0.10 \times 10^{-5}$ averaged over the redshift range $0.2 < z < 3.7$ (corresponding to about 2.5 to 11.8 billion years ago). 
The observations are statistically significant (preferred over no variation at the $5.7\sigma$ level) but remain tentative as the observers cannot rule out some as-yet-undiscovered systematic error. 
Nevertheless the many-multiplet method used was chosen because it is fairly robust to systematics and several possible systematic errors have been ruled out. 


 For a detailed examination of possible systematics and how they have been dealt with we refer the reader to \cite{murphy01b,murphy03}.  So far all of the observations have been made from Keck with the HIRES instrument, so the first step in reducing the chance of instrumental systematic error is to repeat the observations on another telescope with a different spectrograph.  Exploring the same technique with different transitions in other wavelength bands would provide a further test.  

Other methods of constraining $\Delta \alpha / \alpha$ have been consistent with no variation in the fine structure constant \citep[e.g.][]{fujii00,olive02}.  The quasar absorption line measurements improve on previous techniques by constraining $\Delta \alpha / \alpha$ directly without needing to assume the constancy of any other parameters.  For a review of experimental and observational tests of varying constants we refer the reader to \cite{uzan02} and the recent work reported in \cite{martins03}.

\section{Theoretical motivation}

Modern theoretical motivation for the search for varying constants comes in two main forms.  Firstly there are several scalar theories in which one or more of the fundamental parameters is coupled to a varying scalar field.   
In an attempt to make a gravitational theory that complied with Mach's principle \cite{brans61} 
suggested such a theory in which $G$ was replaced by a scalar field that can vary in both space and time.  The first fully covariant, gauge invariant theory of varying constants was proposed by \cite{bekenstein82} who derived a theory of varying electric charge motivated by the need for a framework against which to compare experimental constraints on a varying fine structure constant.   Over the last decade quintessence theories have questioned the constancy of the cosmological constant \citep[see review by][]{wetterich02}.  The latest resurgence of interest has been motivated by the possibility of solving some of the problems with the current big bang model by invoking a varying speed of light \citep{moffat93,albrecht99,barrow98,magueijo00,sandvik02,moffat03b,moffat03} and has been fueled by cosmological observations which indicate that the fine structure constant may have been different in the past (see Sect.~\ref{sect:varies-obs}).

The second theoretical motivation arises from attempts to quantize gravity.  String and M-brane theories provide a natural regime for constant variation in which the constants depend on the properties of the extra dimensions.   Kaluza-Klein theory \citep[see for example review by][]{overduin97}, which is considered an early prototype for many-dimensional unification theories, attempts to unify gravity with the other fundamental forces through the addition of an extra dimension.  (It has been rejected, at least in its initial form, partially due to its classical rather than quantum nature.)  Kaluza-Klein theory predicts that fundamental constants would vary with the size of the extra dimensions.  Since the three spatial dimensions of our experience do seem to be expanding it seems natural that the sizes of the extra dimensions, and thus the values of the fundamental constants, could also change.   Attempts to quantize gravity and link it to the other fundamental forces result in similar predictions.   For an excellent review of theoretical motivations of the search for variations in fundamental constants we refer the reader to \cite{uzan02} Sect.~VI.  For examples see \cite{dent03}, \cite{gregori02} and \cite{youm02a,youm01b,youm01a}.

The current state of affairs sees many different theories predicting (or allowing) different variations in the fundamental constants and a need for experimental results to decide between them.   \cite{magueijo02} derive observational tests that can be used to distinguish between two manifestations of scalar field theories -- one a varying-$e$ theory, the other a varying-$c$ theory.  
Although these observations may be attained by the next generation of experimental programs, they remain for the moment out of reach.  For many of the quantum theories experimental confirmation is even further away.

\vspace{-2mm}
\section[How black hole entropy may constrain theories of varying constants]{How black hole thermodynamics may constrain theories of varying constants}
\vspace{-2mm}
Observational evidence of varying fundamental constants would have far reaching implications for physics and the quest for the unification of the fundamental forces.
Should the variation in the fine structure constant be confirmed it would force a profound shift in our understanding of the Universe and perhaps help direct our search for a more fundamental theory.  
However, the lack of other observational tests is a persistent problem that means theories are currently selected for on theoretical or aesthetic grounds. Our aim is to find a theoretical test that could distinguish between varying constant theories to decide which are more viable.      One theoretical criterion that has held throughout the development of modern physics is the second law of thermodynamics.  

One expects that the second law of thermodynamics is likely to hold in any new theories because a violation is equivalent to getting something for nothing.  Sir Arthur Eddington\footnote{In {\em The Nature of the Physical World}, Maxmillan: New York (1948), p.~74, ``The law that entropy always increases -- the second law of thermodynamics -- holds I think, the supreme position among the laws of Nature. If someone points out to you that your pet theory of the Universe is in disagreement with Maxwell's equations - then so much worse for Maxwell equations. If it is found to be contradicted by observation - well these experimentalists do bungle things sometimes. But if your theory is found to be against the second law of thermodynamics, I can give you no hope; there is nothing for it but to collapse in deepest humiliation.''} writes ``if your theory is found to be against the second law of thermodynamics, I can give you no hope; there is nothing for it but to collapse in deepest humiliation.''  Albert Einstein\footnote{In {\em Thermodynamics in Einstein's Universe} by M.~J.~Klein, Science, 157, (1967) p~509, ``[A law] is more impressive the greater the simplicity of its premises, the more different are the kinds of things it relates, and the more extended its range of applicability. Therefore, the deep impression which classical thermodynamics made on me. It is the only physical theory of universal content, which I am convinced, that within the framework of applicability of its basic concepts will never be overthrown.''} is quoted to have claimed ``[The second law of thermodynamics] is the only physical theory of universal content, which I am convinced, that within the framework of applicability of its basic concepts will never be overthrown.''

Such emotive language indicates a level of philosophy rather than observation which seems inappropriate as a basis for physical theory.  However, the second law of thermodynamics has already survived drastic theoretical paradigm shifts such as the transition to general relativity and quantum mechanics and there is no reason to expect it to fail in the next theory. 
In this chapter we assess some of the implications of the second law of thermodynamics on varying constant theories under the {\em assumption} that it holds.    Black holes offer a very clean, geometric measure of entropy.  
We therefore suggest that when black hole entropy calculated in a varying-constant theory produces a violation of the second law of thermodynamics then that theory should be considered suspect.  

Unfortunately most of the theories under investigation are not sufficiently well developed to have black hole entropy solutions.  Some have solutions for the area of a black hole \citep[e.g.][]{magueijo01}, but more work needs to be done to check that in these cases area remains a good measure of entropy.  In the absence of a full varying-constant solution we take the value of black hole area in general relativity, assume that this is a slow-variation limit of a more fundamental varying-constant theory, and assess how horizon area varies as the constants change.  

The fine structure constant is given by\footnote{In this formula $e$ is in atomic units (measured in $kg\,m^3s^{-2}$).  To convert from Coulombs to atomic units we redefine $e^2/4\pi \epsilon_0\rightarrow e^2$ (and $\epsilon_0$, the permittivity of free space, is a conversion factor between the Coulomb and $kg\,m^3s^{-2}$).} $\alpha = e^2/\hbar c$.  
An increase in the fine structure constant can therefore be attributed to an increase in electric charge, or a decrease in either the speed of light or Planck's constant (with all other dimensional constants unchanged -- we discuss the meaningfulness of variations in dimensional constants in Sect.~\ref{sect:dim}.)
In the case of a non-rotating black hole with an electric charge $Q$, and mass $M$, the area of the black hole event horizon, $A_{\rm H}$, is given in conventional general relativity theory from the Reissner-Nordstr\"om solution of Einstein's field equations (\citeauthor{misner73}, 1973, Box 33.2C), 
\beq A_{\rm b} = 4\pi r^2, \eeq
where,
\beq r = \frac{G}{c^2}\left[M + \sqrt{M^2 - Q^2/G}\right]. \eeq
We can write $Q$ as an integer, $n$, multiplied by the fundamental electric charge, $e$.  The expression for the entropy of a black hole becomes, 
\bea S_{\rm b}&=&\frac{k_{\rm b}c^3}{G\hbar}\;\frac{A_{\rm b}}{4}, \\
      &=&\frac{k_{\rm b}\pi G}{\hbar c}\left[M + \sqrt{M^2 - n^2e^2/G}\right]^2,\label{eq:entropy-bhMQ}\eea
where $k_{\rm b}$ is Boltzmann's constant. 
It is immediately obvious that an increase in the magnitude of the electric charge $Q$, with $c$, $\hbar$, $G$, $k_{\rm B}$ and $M$ remaining constant, implies a decrease in the event horizon area. By contrast, a decrease in the speed of light, $c$, or Planck's constant, $\hbar$, would lead to an increase in event horizon area. Thus two contending alternatives for an increase in $\alpha$ produce opposite outcomes as far as black hole entropy is concerned.

One could argue that a decrease of horizon area implies a violation of the generalized second law of thermodynamics and so the fundamental electric charge can not increase. However, before we can be secure in that interpretation, a number of conditions need to be satisfied. The black hole will radiate heat into its environment via the Hawking process, and as $Q$ changes the temperature will vary. For a violation of the second law of thermodynamics to occur the black hole must not raise the entropy of the environment more than its own entropy decreases. This condition is readily satisfied by immersing the black hole in a heat bath of equal temperature, and allowing the heat radiation to change isentropically as the charge varies (for a highly charged black hole the specific heat is positive \citep{davies77}, 
and it can reside in stable equilibrium in an infinite heat bath).

Moreover, Eq.~\ref{eq:entropy-bhMQ} is based on standard gravitation theory. In a non-standard theory involving varying $e$, $\hbar$ or $c$, the formula for the event horizon area will likely differ \citep[e.g.][]{magueijo01}.
Also, the Hawking process may be modified in a way that alters the relationship between temperature, entropy and horizon area.  Eq.~\ref{eq:entropy-bhMQ} must then be considered as an approximation in the limit of small variation of `constants.'  The validity of GR in this limit may not be secure.  Newton's laws are the low velocity, low mass (energy) limit of general relativity.  We have suggested that the next step in the ladder means general relativity is the slow variation limit of a varying-constant theory.  However, the analogy may not hold.  The concepts of mass and velocity are intrinsic to Newton's laws so it is natural that they agree with the limit of general relativity.  The concept of constant variations, on the other hand, does not appear in general relativity.  So a varying-constant theory is not necessarily just a step up in magnitude from general relativity, it may be qualitatively different.

However, it seems unlikely that minor modifications of Eq.~\ref{eq:entropy-bhMQ} will reverse the sign of the relationship between charge and horizon area. Moreover, in the standard theory there is a maximal electric charge given by $Q^2 = M^2$ above which the horizon disappears and the black hole is replaced by a naked singularity. A modified theory might alter the value of this maximal charge, but there may still be a limit above which any increase in charge will create a naked singularity, in violation of the cosmic censorship hypothesis \citep{penrose69}. 
Thus both cosmic censorship and the second law of thermodynamics are threatened by theories in which $e$ increases with time.

One might also consider theories in which major modifications occur to the structure of charged black holes. For example, in the theory of \cite{bekenstein82}
 and of \cite{barrow98}  
 electric charge can vary with position as well as time, and depends on the energy in the Coulomb field. In the case of a highly charged black hole, the spacetime geometry and causal structure close to the hole might depart greatly from the standard theory. Whether these departures would lead to the horizon area increasing, rather than decreasing, as a function of electric charge, is unclear in the absence of an exact solution. But in such theories one could consider the case of a black hole with a small electric charge, and hence low Coulomb field, for which Eq.~\ref{eq:entropy-bhMQ} remains a good approximation. Thus, a violation of the second law seems probable if variations in $e$ are responsible for the variation in $\alpha$. Furthermore, a violation of cosmic censorship would also seem probable by combining a small electric charge with sufficiently rapid rotation of the black hole to produce the equivalent of an extreme Kerr-Newman black hole, for which any increase in charge would create a naked singularity.

Our arguments, although only suggestive, indicate that varying $e$ theories run a serious risk of being in violation of both the second law of thermodynamics and cosmic censorship. Thus, black hole thermodynamics may provide a criterion against which contending theories for varying `constants' should be tested.

\section{The meaning of varying dimensional constants}\label{sect:dim}

Our paper \citep{davies02} sparked the latest round in an ongoing debate over whether it is meaningful to discuss variations in {\em dimensional} constants.  Generalizing broadly, protagonists such as \cite{dent02,duff02,dzuba02,flambaum02,wolfe02} claim that it is meaningless to talk about variations in dimensional constants. While champions of varying constant theories such as \cite{albrecht99,magueijo00,barrow02,moffat02,sandvik02} disagree.    However, there is a wide area of common ground.  Both parties agree that it is meaningless to discuss varying dimensional constants without defining what they vary with respect to.  Both parties agree that varying constant theories are meaningful discussions of varying dimensional constants.  Both parties agree that with or without varying constant theories it is impossible to unambiguously discover {\em which} dimensional constant is varying.  In fact, there is very little that the two parties disagree on, and some of the points of disagreement lie in miscommunication.  We believe the remaining point of debate concerns whether it is meaningful to discuss varying dimensional constants outside a theory of varying constants.  Since these points have raised so much vehement debate we take the time to summarize the arguments and clarify what we believe is the solution.

In its simplest form, the claim that variations in dimensional constants is meaningless relates to the fact that only dimensionless quantities are measurable.  All measurements are ratios of the measured quantity to the appropriate measuring stick.  For example, when we measure a person's height we use a ruler.  What we measure is the dimensionless ratio of height divided by ruler length.  If we find that in a subsequent measurement this ratio is larger we know that something has changed but we cannot say whether the person grew taller or whether the ruler shrank.  

We note that since our current standards of length and time\footnote{Our current standard for the metre is the length of the path traveled by light in vacuum during a time interval of 1/299 792 458 of a second \citep{PDBook02}.  The current standard for the second is defined to be the duration of 9 192 631 770 periods of the radiation corresponding to the transition between the two hyperfine levels of the ground state of the caesium-133 atom \citep{PDBook02}, and thus depends on the value of the fine structure constant.} both depend upon the speed of light, $c$, it is impossible to use them to measure any variation in $c$.
However, if we use the speed of light to calibrate a ruler today and another ruler tomorrow we could find that the two rulers are different.  Once calibrated, the length of the rulers depends on molecular or atomic forces and is therefore only indirectly sensitive to a variation in the speed of light (molecular forces may depend on $c$ but in a different functional form than the original calibration).

Varying the speed of light (in a particular system of units) while all other fundamental constants (in that system of units) remain constant has real physical, measurable effects.  For example, as in George Gamow's classic book ``Mr Tompkins in Paperback'' \citep{gamow65}, if the speed of light were greatly reduced relativistic effects such as length contraction would become evident in everyday life.  Varying speed of light theories are of interest because a faster speed of light in the early Universe could remove the horizon problem and thus remove one of the primary motivations of inflation.  
However, we {\em can not} unambiguously say that the physical effects we are seeing are due to a variation in the speed of light because the ``speed of light'' can always be redefined using a different set of units so that it does not vary.  The physical effects are clear --- but their cause is up for debate.

To elucidate this more clearly I will elaborate on an example provided in \cite{duff02}.  Consider SI units.  Length is measured in metres, 
and time is measured in seconds. 
We can define a number of alternate sets of units, one of which is the familiar Planck units.  The conversions between SI and Planck units for length and time are respectively, 
\bea L_P^2 &=& \frac{G\hbar}{c^3}, \\
     T_P^2 &=& \frac{G\hbar}{c^5}.\eea 
A length, $L$, measured in metres is equal to $L/L_P$ Planck lengths.  So to convert a speed from metres per second to Planck lengths per Planck time we divide by $L_P/T_P = c$.  Thus the speed of light measured in Planck units is identically one, $c_P=c/c=1$.  We have converted to a set of units in which the speed of light can never vary because our definition of units depends upon it.
However, consider another set of units that \cite{duff02} calls Schr\"odinger units.  The conversions between SI and Schr\"odinger units for length and time are respectively,
\bea L_\psi^2 &=& \frac{G\hbar^4}{e^6},\\
     T_\psi^2 &=& \frac{G\hbar^6}{e^{10}}. \eea
Again, to convert a speed from metres per second to Schr\"odinger length per Schr\"odinger time we divide by $L_\psi/T_\psi = e^2/\hbar$.  Therefore the speed of light measured in Schr\"odinger units is $c_\psi=c/(e^2/\hbar) = 1/\alpha$.  In this set of units the speed of light changes in inverse proportion to the fine structure constant.  From this we can conclude that if $c$ changes but $e$ and $\hbar$ remain constant then the speed of light in Schr\"odinger units, $c_\psi$ changes in proportion to $c$ but the speed of light in Planck units, $c_P$ stays the same.  Whether or not the ``speed of light'' changes depends on our measuring system (three possible definitions of the ``speed of light'' are $c, c_P$ and $c_\psi$).  Whether or not $c$ changes is unambiguous because the measuring system has been defined.

All the above systems of units are interchangeable and so together define a unique concept.  How fast a photon travels between two points is a real physical observable that is unchanged by any of humanity's choices of units.  Once we have chosen a system of units (e.g. metre and second) it is meaningful to discuss changes amongst the dimensional fundamental parameters (e.g. the speed of light) in these units.  We simply have to recognize that if we chose another set of units we might find that different fundamental constants vary.

Moreover any variation in dimensional constants can be re-expressed as a variation in dimensionless constants.  To demonstrate this final point more clearly lets return to the entropy of a charged black hole as defined in Eq.~\ref{eq:entropy-bhMQ}.

\cite{dent02,duff02,flambaum02} and \cite{wolfe02} all suggested that this entropy be written in dimensionless form.  To do this we must divide through by Boltzmann's constant,
\beq s_{\rm b} = \frac{S_{\rm b}}{k_{\rm b}} = \pi \left[m+\sqrt{m^2 - n^2\alpha}\right]^2,\label{eq:S-dimless}\eeq
where $m=M/M_{\rm P}$ and the Planck mass $M_P^2 = \hbar c/G$.  Arguments from quantum theory suggest that it is more natural to expect that $m$ would remain constant than $M$ \citep{flambaum02,carlip02}.  Under this assumption Eq.~\ref{eq:S-dimless} suggests that any increase in $\alpha$ would violate the second law of thermodynamics, independent of which of $e$, $c$ or $\hbar$ varies.  This seems to be in contradiction to our previous result in which an increase in $e$ decreased $S_{\rm b}$ but a decrease in $c$ or $\hbar$ increased $S_{\rm b}$.  However, in our initial formulation we had assumed that $M$ remained constant whereas here we are assuming $m$ remains constant.  
If we allow $m$ to vary such that $M$ remains constant the result for black hole entropy is unchanged from the previous version. 

This demonstrates how strongly dependent any conclusions are on the assumptions of what is held constant.  \cite{bekenstein79} points out why several attempts \citep{baum76,solheim76} to measure the variation of dimensional constants gave a null result because they had implicitly {\em assumed} the constancy of the quantity they were trying to measure.
Certain combinations of varying and constant constants may result in a violation of the second law of thermodynamics, but it is not sufficient to say which constants vary, you must also specify those that do not change.  This is where theories of varying constants come into play. 
(As an example, \citet[Sect.~II]{albrecht99} 
provides a clear description of what exactly it means to have a theory of varying speed of light.)   Varying constant theories make definite physical predictions that we can in principle measure \citep[e.g.][]{magueijo02}.  Nevertheless, these varying constant theories are unable to unambiguously say which dimensional constants are varying, even if all their physical predictions are borne out.  

\begin{figure}
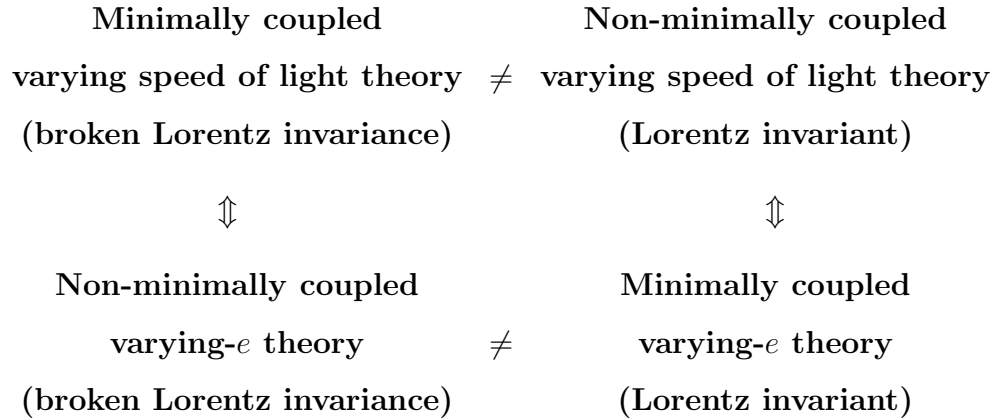

\bea
\parbox{60mm}{\bctr {\bf Minimally coupled\\ varying speed of light theory\\ (broken Lorentz invariance)}\ectr} &\neq & \parbox{60mm}{\bctr {\bf Non-minimally coupled\\ varying speed of light theory\\(Lorentz invariant)}\ectr}\nn \\
\Updownarrow \hspace{30mm}&&\hspace{30mm} \Updownarrow\nonumber\\ 
\parbox{60mm}{\bctr{\bf Non-minimally coupled}\\ {\bf varying-$e$ theory}\\ {\bf (broken Lorentz invariance)}\ectr} &\neq & \parbox{60mm}{\bctr{\bf Minimally coupled}\\ {\bf varying-$e$ theory}\\ {\bf (Lorentz invariant)}\ectr}\nn
\eea
\caption[Transforming between varying $e$ and varying $c$ theories]{\small This diagram represents the fact that a varying speed of light theory can be transformed into a varying-$e$ theory and vice versa.  However, the theory created by transforming a simple varying speed of light theory into a varying-$e$ theory is not equivalent to a simple varying-$e$ theory.   This diagram follows a similar one by \cite{dzuba02}.}
\label{fig:dzuba}
\end{figure}

The reason for this is that any theory of varying constants can be transformed by a change of units into a theory in which a different constant varies (see Fig.~\ref{fig:dzuba}).  For example, \cite{barrow98} showed how a varying $c$ theory can be transfomed into a varying $e$ theory by a change in units.  
It is not important that these theories do not answer the question of which constant is really varying because they make clear physical predictions that we can test.  If we find that one of these theories predicts correctly all observational tests we can perform then it will supercede previous theories.  It is then up to the user to choose to solve problems in whichever mode of the theory they find convenient.
 
\subsection{A question of philosophy}
The discussion over the meaningfulness of varying dimensional constants is partially a question of philosophy of science.  Does physics endeavour to best describe ``reality'' or does it endeavour to reveal ``reality''?  Proponents of the different camps answer differently when asked of the meaningfulness of varying dimensional constants. 
It is impossible to answer definitively that {\em it is} $e$ that varies or {\em it is} $c$ that varies.  
On the other hand, to say that the observables are {\em best described} by a varying $e$ or the observables are {\em best described} by a varying $c$ {\em is} meaningful, though it relies on a value judgment that is usually drawn from arguments of simplicity.

Some may argue that value judgments such as which description of reality is simplest should be kept out of physics, but value judgments are made every time we choose a frame of reference in which to perform a calculation (e.g. the time coordinate in an expanding universe, see Chapter~\ref{chap:coord}).  We are not saying that that is the only frame of reference available, but we choose the frame so that the calculations are simplest.

\section{Extension to particular varying constant theories}

Theories in which fundamental constants are allowed to vary provide solutions to black hole entropy that supersede our assumption that Eq.~\ref{eq:S-dimless} derived from GR provides a slow variation limit to varying constant theories.  Since publication of \nocite{davies02}Davies, Davis and Lineweaver (2002) several papers have applied the black hole entropy criterion using various varying constant theories.  Here we briefly summarize some of their results.

\cite{barrow03,barrow03b} gives the example of Brans-Dicke theory, in which black hole horizon entropy,
\beq S_{\rm b} \propto GM^2.\eeq 
He notes that the entropy decrease criterion would rule out any theories in which G decreased.  Essentially all Brans-Dicke theories have this behaviour.  
However, by definition $G$ is constant on the Schwarzschild horizon in this theory.  When $G$ is allowed to vary the static, spherically symmetric solution in Brans-Dicke theory is no longer a black hole.  For this reason he argues that black holes are unable to constrain Brans-Dicke theories. 

\cite{fairbairn03} consider the entropy of a class of charged dilaton black holes related to string theory.  They consider both adiabatic and non-adiabatic variations of the fine structure constant and find that black hole entropy does not change in the former and increases in the latter.  
\cite{vagenas03} shows that entropy decreases as $e$ decreases for 2-d stringy black holes.  He notes that there are no model independent constraints on black hole entropy and concludes that there are no {\em theories} that are out of favour with black hole thermodynamics.  However, he has demonstrated a specific {\em model} that is out of favour with black hole thermodynamics.

The above analyses concentrate on black hole solutions in varying constant theories.  Others have concentrated on analyzing any concomitant increase in the entropy of the Reissner-Nordstr\"om black hole environment when constants vary.
\cite{carlip03} consider the full thermal environment of a black hole inside a heat bath -- a box of fixed radius, temperature and charge (the canonical ensemble) or electrostatic potential (the grand canonical ensemble).  Presuming that the features of the box remain fixed as $\alpha$ changes they show that black hole entropy increases as $\alpha$ increases, independent of whether the change is due to decreasing $c$ or $\hbar$, or due to increasing $e$.  They conclude that black hole thermodynamics `militates' against models in which $e$ decreases, but places no constraints on increasing $e$, and is insufficient to constrain theories in which $\alpha$ {\em increases}.  

Other constraints relate to existing efforts to quantize black holes.
\cite{carlip02} analyzes the problems varying constants introduce if you try to quantize the entropy of black holes. 
\cite{flambaum02} also deals with quantizing black holes and suggests that quantum black holes give reason to believe that entropy is conserved.

\section{Conclusion}

The subject of varying fundamental constants is an exciting one that could prove to be the next step in physical theories.  Observational tests are needed to determine which forms of varying constant theories best describe reality.  We have proposed black hole thermodynamics as a tool to theoretically rule out some varying constant theories.  This stimulated much debate and subsequent work by a variety of authors applies the black hole thermodynamics test to several different varying constant theories.  

\clearemptydoublepage

\chapter{Conclusions}\label{chap:conclusions}

We have clarified some common misconceptions surrounding the expansion of the Universe, and shown with numerous references how misleading statements manifest themselves in the literature.  
Superluminal recession is a feature of all expanding cosmological models that are homogeneous and isotropic and therefore obey Hubble's law, $\vrec=HD$.  This does not contradict special relativity because the superluminal motion does not occur in any observer's inertial frame.  All observers measure light locally to be travelling at $c$ and nothing ever overtakes a photon.  Inflation is often called ``superluminal recession'' but even during inflation objects with $D<c/H$ recede subluminally while objects with $D>c/H$ recede superluminally.  Precisely the same relationship holds for non-inflationary expansion.  We showed that the Hubble sphere is not a horizon --- we routinely observe galaxies that have, and always have had, superluminal recession velocities.  All galaxies at redshifts greater than $z\sim 1.46$ today are receding superluminally in the $\Lambda$CDM concordance model.  We have also provided a more informative way of depicting the particle horizon on a spacetime diagram than the traditional worldline method.  

An abundance of observational evidence supports the general relativistic big bang model.  The duration of supernovae light curves shows that models predicting no expansion are in conflict with observation.
Using magnitude-redshift data from supernovae 
we were able to rule out the SR interpretation of cosmological redshifts at the $\sim23\sigma$ level.   
Together these observations provide strong evidence that 
the general relativistic interpretation of the cosmological redshifts is preferred over tired light and special relativistic interpretations.  The general relativistic description of the expansion of the Universe agrees with observations, and does not need any modifications for $\vrec > c$. 

We have pointed out and interpreted some additional counter-intuitive results
of the general relativistic description of our Universe. We have
shown that the unaccelerated expansion of the Universe has no
effect on whether a galaxy set up at rest with respect to our position approaches or recedes from us. 
In a decelerating universe the untethered galaxy approaches us, while in an accelerating universe it recedes from us. 
The expansion, however, {\em is} responsible for the galaxy joining
the Hubble flow, and we have shown that this happens irrespective of whether the untethered galaxy approaches or recedes from us.

The expansion of the Universe is a natural feature of general
relativity that also allows us to unambiguously convert observed
redshifts into proper distances and recession velocities and to
unambiguously define approach and recede. We have used this
foundation to predict the existence of receding blueshifted and
approaching redshifted objects in the universe. To our knowledge
this is the first explicit derivation of this counter-intuitive
behavior.

We used the example of the empty universe to relate SR expansion (the Milne universe) to FRW expansion.  Two choices of time coordinate are intuitive.  The first chooses the time coordinate of a fundamental observer's inertial frame (Milne universe).  This has the advantage of allowing Lorentz transformations, time dilation and length contraction to hold and be calculable in the familiar SR formalism.  However, this choice means the universe is not homogeneous, and the formalism cannot easily be translated into a non-empty universe.  The second chooses the time coordinate as the proper time of comoving observers (FRW universe).  This choice makes the Universe homogeneous and allows us to use Friedmann's equations to describe the evolution of the Universe in the non-empty case.  The SR Doppler shift equation relates redshift to velocity only in the Milne universe.  This is not the velocity that appears in Hubble's law.

We define total entropy to be the entropy of a cosmological event horizon plus the entropy within it.  Davies (1988) showed that the entropy of the cosmological event horizon in FRW universes, subject to the dominant energy condition, never decreases.  We examined radiation filled FRW universes and showed that total entropy never decreases for a wide range of models by testing the parameter space using numerical calculations.  We then assessed the entropy lost as black holes disappeared over the cosmological event horizon.  The lack of a black hole solution for arbitrary spacetime embeddings restricts the application of this technique.  Limiting the size of black holes to those small enough that the difference in embedding in empty space compared to de Sitter space is less than 0.1\% allowed us to show that no GSL violation occurs in any of the closed or flat models tested, but an apparent violation occurs at early times in open FRW universes, probably due to the breakdown of the Schwarzschild-de Sitter assumption in the presence of matter density outside a black hole.  Further progress in resolving this matter will require more realistic approximations of black hole solutions in cosmological backgrounds.  An associated issue that needs to be addressed is what constitutes the appropriate surface that characterizes horizon entropy when black holes are situated in a time-dependent background.

We have also used black hole thermodynamics to suggest constraints on theories of varying constants.  We propose that if a theory of varying constants predicts a decrease in black hole entropy without a concomitant increase in horizon entropy then that theory should be considered suspect.  Constraints on varying constant theories are increasingly important as new observational evidence suggests a possible variation in the fine structure constant over cosmological time scales.

\clearemptydoublepage

\addcontentsline{toc}{chapter}{Appendices}
\appendix

\chapter[Standard results]{Standard results}\label{sect:math}
\section{General relativistic definitions of expansion and horizons}

\subsection{The metric}
The metric for an homogeneous, isotropic universe is the Robertson-Walker (RW) metric,
\beq  
ds^2 = -c^2dt^2 + R(t)^2[d\chi^2+S_k^2(\chi)d\psi^2], 
\label{eq:frwmetric-app}
\eeq
where $c$ is the speed of light, $dt$ is the time separation, $d\chi$ is the comoving coordinate separation and $d\psi^2=d\theta^2+\sin^2\theta d\phi^2$, where $\theta$ and $\phi$ are the polar and azimuthal angles in spherical coordinates. The scalefactor, $R$, has dimensions of distance.   The function $S_k(\chi)=\sin\chi$, $\chi$ or $\sinh\chi$ for closed ($k=+1$), flat ($k=0$) or open ($k=-1$) universes respectively~\cite[p.~69]{peacock99}.  The time, $t$, is the proper time of a comoving observer, also known as cosmic time (see Section~\ref{sect:conversion}).
The proper distance $D$, at time $t$, in an expanding universe, between an observer at the origin and 
a distant galaxy is defined to be along a surface of constant time ($dt=0$).  
We are interested in the radial distance so $d\psi=0$. The RW metric then reduces to $ds = R d\chi$ which, upon 
integration yields,
\beq {\rm Proper} \; {\rm distance,}\quad\quad\quad \;D(t) = R(t)\chi.  \label{eq:Hubble}\eeq
Differentiating this yields the theoretic form of Hubble's law~\citep{harrison93}, 
\bea{\rm Recession}\; {\rm velocity,}\,\;\;v_{\rm rec}(t,z) &=& \dot{R}(t)\chi(z), \label{eq:vchi}\\
&=& H(t) D(t),  \label{eq:hubbleslaw}
\eea
where $v_{\rm rec}=\dot{D}$ (for $\dot{\chi}=0$) and $\chi(z)$ is the fixed comoving coordinate associated with a galaxy observed today at redshift $z$.  Note that the redshift of an object at this fixed comoving coordinate changes with time\footnote{In addition, objects that have a peculiar velocity also move through comoving coordinates, therefore more generally Eq.~\ref{eq:vchi} above should be written with $\chi$ explicitly time dependent, $v_{\rm rec}(t,z)=\dot{R}(t)\chi(z,t)$.} (Eq.~\ref{eq:dz}).
A distant galaxy will have a particular recession velocity when it emits the photon at $t_{\rm em}$ and a different recession velocity when we observe the photon at $t_0$. Eq.~\ref{eq:hubbleslaw} evaluated at $t_0$ and $t_{\rm em}$ gives the recession velocities plotted in  Fig.~\ref{fig:vz} (top and bottom panels respectively).

The recession velocity of a comoving galaxy is a time dependent quantity because the expansion rate of the 
universe $\dot{R}(t)$ changes with time.  The current recession 
velocity of a galaxy is given by $v_{\rm rec}= \dot{R}_0\chi(z)$.  
On the spacetime diagram of Fig.~\ref{fig:dist} this is the velocity taken at points along the line of constant 
time marked ``now''.
The recession velocity of an emitter at the time it emitted the light we observe is the velocity at points taken along the past light cone\footnote{The recession velocity at the time of emission is $v_{\rm rec}(t_{\rm em}) = R(t_{\rm em})\chi(z)$ where $R(t_{\rm em})=R(t)$ as defined in Eq.~\ref{eq:z}.}.  However, we can also compute the recession 
velocity a comoving object has at {\em any time} during the history of the universe, having initially calculated 
its comoving coordinate from its present day redshift.

Allowing $\chi$ to vary when differentiating Eq.~\ref{eq:Hubble} with respect to time gives two distinct velocity terms~\citep{landsberg77,silverman86,peacock99,davis03chain},
\bea \dot{D} &=& \dot{R}\chi + R\dot{\chi}, \\
     v_{\rm tot} &=& v_{\rm rec} + v_{\rm pec}.\label{eq:vtot}\eea
This explains the changing slope of our past light cone in the upper panel of Fig.~\ref{fig:dist}.  The peculiar velocity of light is always $c$ (Eq.~\ref{eq:dchi}) so the total velocity of light whose peculiar velocity is towards us is $v_{\rm tot}=v_{\rm rec}-c$ 
which is always positive (away from us) when $v_{\rm rec}>c$.  Nevertheless we can eventually receive photons that initially were receding from us because the Hubble sphere expands and overtakes the receding photons so the photons find themselves in a region with $v_{\rm rec}<c$ (Section~\ref{sect:notobserve}).

Photons travel along null geodesics, $ds=0$.  To obtain the comoving distance, $\chi$, between an observer at the origin 
and a galaxy observed to have a redshift $z(t)$, set $ds = 0$ (to measure along the path of a photon) and $d\psi = 0$ 
(to measure radial distances) in the RW metric yielding, 
\beq c\;dt = R(t) d\chi.  \label{eq:dchi} \eeq
This expression confirms our previous statement that the {\em peculiar} velocity of a photon, $R\dot{\chi}$, is $c$.  
Since the velocity of light through comoving coordinates is not constant ($\dot{\chi}=c/R$), to 
calculate comoving distance we cannot simply multiply the speed of light through comoving space by time, we have 
to integrate over this changing comoving speed of light for the duration of propagation.  Thus, the comoving coordinate of a 
comoving object that emitted the light we now see at time $t$ is attained by integrating Eq.~\ref{eq:dchi},
\beq 
{\rm Past} \;{\rm Light} \;{\rm  Cone,}\: \:
\chi_{\rm lc}(t_{\rm em}) = c\int_{t_{\rm em}}^{t_0}\frac{dt'}{R(t')}.\label{eq:lightconet}
\eeq
We can parametrize time using redshift and thus recast Eq.~\ref{eq:lightconet} in terms of observables.
The cosmological redshift of an object is given by the ratio of the scalefactor at the time of observation, $R(t_0)=R_0$, to the scalefactor at the time of emission, $R(t)$,
\beq 
{\rm Redshift,}\;\;\;    1+z = \frac{R_{0}}{R(t)}. 
\label{eq:z}\eeq
Differentiating  Eq.~\ref{eq:z} with respect to $t$ gives $dt/R(t)=-dz/R_{\rm 0}H(z)$ where redshift is used instead of time to parametrize Hubble's constant.  $H(z)$ is Hubble's constant at the time an object with redshift, $z$, emitted the light we now see.  Thus for the limits of the integral in Eq.~\ref{eq:lightconet} the time of emission becomes $z=0$ while the time of observation becomes the observed redshift, $z$.
The comoving coordinate of an object in terms of observables is therefore,
\beq
\chi(z) = \frac{c}{R_0}\;\int_{o}^{z} \frac{dz^{\prime}}{H(z^{\prime})}.  \label{eq:chiz}
\eeq
Thus there is a direct one to one relationship between observed redshift and comoving coordinate.   Notice that in 
contrast to special relativity the redshift does not give you the velocity, it gives you the distance\footnote{\footnotesize Distance is proportional to recession velocity at any particular time, but a particular 
redshift measured at different times will correspond to different recession velocities.}.  That is, the redshift tells us not the velocity of the emitter, but where the emitter sits (at rest locally) in the coordinates of the universe.  The recession velocity is obtained by inserting Eq.~\ref{eq:chiz} into Eq.~\ref{eq:vchi} yielding Eq~\ref{eq:vGR}. 

A conformal time interval, $d\tau$, is defined as a proper time interval $dt$ divided by the scalefactor,
\beq 
{\rm Conformal}\;\;{\rm time,}\;\;\;   d\tau = dt/R(t). \label{eq:conformalt} \eeq
This allows us to rewrite the RW metric as,
\beq  
ds^2 =  R(t)^2[-c^2d\tau^2 +d\chi^2+S_k^2(\chi)d\psi^2], \label{eq:conformalmetric} 
\eeq
so the part of the metric inside the square brackets looks like the Minkowski metric of special relativity.

\subsection{The Friedmann equations}

The time dependence of the scalefactor, $R(t)$, is given by the Friedmann equations,
\bea \dot{\rho} &=& -3H(\rho + p),\label{eq:Fried1-app}\\
3H^2&=&8\pi\rho + \Lambda - 3k/R^2,\label{eq:Fried2-app}\eea 
where $\rho$ and $p$ are the density and pressure of the cosmological fluid respectively.  
The radiation density and cosmological constant can be normalized to,
\beq \om =  \frac{8\pi \rho_0}{3H_0^2}, \;\;\;\;\mbox{and}\;\;\;\; \oll = \frac{\Lambda}{3H_0^2}, \eeq 
respectively so that $\om+\oll = 1$ represents flat space at the present day.  The dimensionless scalefactor $a(t)$ is defined as $a(t) = R(t)/R_0$ where $R_0$ is the present day radius of curvature of the Universe,
\beq R_0 = \frac{c}{H_0}\; \left|\frac{1}{1-\om-\oll}\right|^{1/2}.\eeq  
Equation~\ref{eq:Fried2-app} can then be rewritten as,
\beq \dot{R}(t) = R_0\dot{a}=R_0 H_0\left[1+\om\left(\frac{1}{a}-1\right) + \oll(a^2-1)\right]^{1/2},\label{eq:dotR}\eeq
which we use with the identity $dt/R(t) = dR/(\dot{R}R)$ to evaluate Eqs.~\ref{eq:lightconet},~\ref{eq:chipht} and~\ref{eq:eventhorizont}.
Rearranging the Friedmann Equation also gives the time dependence of the Hubble's constant,
\beq H(z)=H_0 \: (1+z)\left[1+\om z+\oll\left(\frac{1}{(1+z)^2}-1\right)\right]^{1/2}.\label{eq:H} \eeq    
Expressing Hubble's constant this way is useful because it is in terms of observables, but it restricts our calculations to objects with redshift $z<\infty$.  That is, objects we can currently see.  There is no reason to assume the Universe ceases beyond our particle horizon and expressing Friedmann's equation in the form of Eq.~\ref{eq:dotR} allows us to extend the analysis to $t\rightarrow\infty$ which is beyond what we can observe.

\subsection{Cosmological horizons}

Altering the limits on the integral in Eq.~\ref{eq:lightconet} gives the horizons we have plotted on the spacetime diagrams.
The time dependent particle horizon we plot in Fig.~\ref{fig:dist} uses $D_{\rm ph}=R(t)\chi_{\rm ph}(t)$ with,
\beq 
{\rm Particle}\:\:{\rm Horizon,}\: \:\chi_{\rm ph}(t)= c\int_0^{t}\frac{dt^\prime}{R(t^\prime)}. 
\label{eq:chipht}
\eeq
The traditional depiction of the particle horizon as a worldline uses $D_{\rm ph}=R(t)\chi_{\rm ph}(t_0)$.  
The comoving distance to the cosmological event horizon is given by,
\beq 
{\rm Event}\:\:{\rm  Horizon,} \;\;\:\chi_{\rm c}(t)= c\int_{t}^{t_{\rm end}}\frac{dt^{\prime}}{R(t^{\prime})},
\label{eq:eventhorizont}
\eeq
where $t_{\rm end}=\infty$ in eternally expanding models or the time of the big crunch in recollapsing models.

A universe expands forever provided: 
\beq \oll \ge \left\{\begin{array}{ll} 
0    & 0\le\om\le 1\\
4\om \left\{ \cos\left[ \frac{1}{3}\cos^{-1}\left(\frac{1-\om}{\om}\right) + \frac{4\pi}{3}\right] \right\}^3 & \om>1,
\end{array}\right. 
\eeq
(Carroll, Press and Turner 1992).  All FRW universes with a positive cosmological constant $\oll>0$ that expand forever have an event horizon. 
These universes tend towards de Sitter space $\omol=(0,1)$ as $t\rightarrow\infty$ and their event horizons therefore tend toward the Hubble sphere.  De Sitter space has a deceleration parameter of $q=-\ddot{R}R/\dot{R}^2=-1$.  Universes that tend towards $q=-1$ from above have an event horizon that increases in size as it approaches the Hubble sphere from below.  In these universes we can observe some superluminally receding objects because photons in superluminal regions later find themselves in subluminally receding regions and can therefore approach us.  On the other hand universes that tend towards $q=-1$ from below have an event horizon that decreases in size as it approaches the Hubble sphere.  In these universes we cannot observe some subluminally receding galaxies because photons in subluminal regions later find themselves in superluminally receding regions and never again approach us.  Figure~\ref{fig:q-t} shows the evolution of the deceleration parameter for several cosmological models.

 Universes that `bounce' (collapse then expand) also have event horizons as most of these tend towards exponential expansion\footnote{Bounce universes (universes that stop their collapse at a finite scalefactor and re-expand) occur when,
\beq \oll \ge 4\om \left\{{\rm coss} \left[\frac{1}{3}{\rm coss}^{-1}\left(\frac{1-\om}{\om}\right)\right] \right\}^3, \eeq
where ``coss'' is defined as being cosh when $\om<1/2$, cos when $\om>1/2$ and either when $\om=1/2$ (Carroll, Press and Turner 1992).}.  
There is no event horizon in eternally expanding universes without a cosmological constant. 
Collapsing universes that end in a big crunch can have an event horizon.   In big crunch universes the event horizon occurs not because light travels only a finite distance in an infinite time, but because light only has a finite time to travel.    


In closed, eternally-expanding universes the comoving distance that light can travel in infinite time (the comoving distance to the event horizon) can exceed the distance to the antipode.  The antipodal point is where light rays we emit again converge on the opposite side of the universe (the south pole if you are the north pole).  Events beyond the event horizon when the event horizon is beyond the antipode are visible from light they emit in the other direction around the closed universe.  In these cases the event horizon appears out of the antipode at a finite time\footnote{When you reduce $\Lambda$ enough you find closed universes in which the event horizon does not extend beyond the antipode.  It is more difficult to achieve (and for some $\Lambda$ is impossible to achieve)  by increasing $\om$, but increasing $\om$ brings the emergence of the horizon from the antipode to earlier times (for a particular $\Lambda$).  Increase $\om$ enough and the time of event horizon emergence from the antipode starts to increase again. }.  



The volume within a cosmological event horizon is given by:
\bea V_{\rm c} &=& 4\pi\,R^3  \int_0^{\chi_{\rm c}}S^2_k(\chi)d\chi\\
&=&\left\{ \begin{array}{ll}
	2\pi\,R^3\, (\chi_{\rm c}-\sin\chi_{\rm c}\cos\chi_{\rm c}) & \mbox{closed,} \\ & \\
	\frac{4}{3} \pi R^3 \chi_{\rm c}^3 & \mbox{flat,}\\ & \\
	2\pi\,R^3\, (-\chi_{\rm c}+\sinh\chi_{\rm c}\cosh\chi_{\rm c}) & \mbox{open.}\end{array}\right.\label{eq:volume}
\eea

\subsection{Peculiar velocity decay}\label{sect:pecdecay}

Here we show that peculiar momentum decays as $1/a$ in the expanding FRW universe.  The following derivation arises as a combination of the derivations in \citet[][Sect.~15.3]{peacock99} and \citet[][Sect.~6.2(c)]{padmanabhan96}.

From \citet[][Sect.~15.3]{peacock99}:
``Consider first a galaxy that moves with some peculiar velocity in an otherwise uniform universe.  Even though there is no peculiar gravitational acceleration acting, its velocity will decrease with time as the galaxy attempts to catch up with successively more distant (and therefore more rapidly receding neighbours. If the proper peculiar velocity is $v$, then after time $dt$ the galaxy will have moved a proper distance $x=v\,dt$ from its original location.  Its near neighbours will now be galaxies with recessional velocities $Hx = Hv\,dt$, relative to which the peculiar velocity will have fallen to, 
\beq v^\prime=v-Hx. \label{eq:pec-vel-relative}\eeq  
The equation of motion is therefore just,
\beq \dot{v} = -Hv = -\frac{\dot{a}}{a} v,\eeq
with the solution $v\propto a^{-1}$: peculiar velocities of non-relativistic objects suffer redshifting by exactly the same factor as photon momenta.''

Because the galaxies' velocities were not summed relativistically in Eq.~\ref{eq:pec-vel-relative} the above does not apply for relativistic objects.  We now provide the more general derivation following the treatment of \cite[][Sect.~6.2(c)]{padmanabhan96}.
Relativistic velocity addition requires us to replace Eq.~\ref{eq:pec-vel-relative} with,
\beq v^\prime=\frac{v-Hx}{1-vHx/c^2} \;\approx\; v-(1-v^2/c^2)Hx, \eeq
where the approximation assumes that the relative velocity of the two frames, $Hx$, is infinitesimal (this is a requirement of our setup since we integrate over these infinitesimal increments to calculate the decay of peculiar velocity.)  
The equation of motion is therefore,
\beq \dot{v} = -(1-v^2/c^2)Hv, \eeq
which can be integrated to give,
\beq \frac{v}{\sqrt{1-v^2/c^2}} \propto \frac{1}{a}. \eeq
The left hand side of this equation is proportional to the relativistic equation for momentum, $p=\gamma mv$ and so $p\propto 1/a$ follows.


\subsection{Infinitesimal Doppler shifts}\label{sect:infinitesimal}

We have shown in Sect.~\ref{sect:analytic-empty} that the SR Doppler shift equation does not relate redshift to the recession velocity that appears in Hubble's law $\vrec=HD$, but does relate redshift to velocity in the Milne description of the empty universe.  There is another way that the SR Doppler shift is linked to the FRW description.  We know that SR holds in an infinitesimal region around each comoving observer.  So how does the SR Doppler shift, which holds locally, become the cosmological redshift globally?  The following derivation relies heavily on ~\cite{padmanabhan96}, Sect.~6.2(a). 

Take an object an infinitesimal comoving distance $\delta \chi$ away from an observer.  The proper distance of that object is $\delta r = a(t) \delta \chi$.  The velocity is given by $\delta v = \dot{\delta r} = \dot{a} \delta \chi = (\dot{a}/{a})\delta r$, Hubble's law.  The time taken for a photon to transit between these infinitesimally separated observers is $\delta t = \delta r/c$, so the velocity can be re-expressed as 
\beq \delta v = \frac{\dot{a}}{a}c\delta t = c\frac{\delta a}{a}.\eeq  
So far we have just rearranged the expression for velocity in Hubble's law.  Now we assume that the special relativistic Doppler shift applies for this infinitesimal redshift.  In fact, since velocity tends towards zero as distance tends towards zero, the low redshift approximation $v=cz$ of the SR Doppler shift is adequate in the infinitesimal limit.  If the original wavelength of light is $\lambda$ and the redshifted wavelength is $\lambda + \delta \lambda$ then the definition of redshift gives us,
\beq \frac{\delta \lambda}{\lambda} = z =\frac{\delta v}{c} = \frac{\delta a}{a}.\eeq
This relationship holds for all of the infinitesimal shifts between every point between emission and observation of a photon, therefore we can integrate this equation over the duration of propagation of a photon to calculate the cosmological redshift.  Performing the integration gives,
\bea \ln \lambda &=& \ln (a \times \mbox{constant})\\
         \lambda &\propto& a,\eea
as expected.  This shows how infinitesimal SR Doppler shifts are useful, and that these should be extended to large distances by integration.

\clearemptydoublepage
\chapter{Examples of misconceptions in the literature}\label{sect:quotes}
In text books and works of popular science it is often standard practice to simplify arguments for the reader.  Some of the quotes below fall into this category.  We include them here to point out the difficulty encountered by someone starting in this field and trying to decipher what is really meant by `the expansion of the Universe'.  \vspace{3mm}

\noindent
{\footnotesize 
\xcite{1}{Feynman, R.~P. 1995, {\em Feynman Lectures on Gravitation (1962/63)}, (Reading, Mass.: Addison-Wesley) p.~181, ``It makes no sense to worry about the possibility of galaxies receding from us faster than light, whatever that means, since they would never be observable by hypothesis.''}
\xcite{2}{Rindler, W. 1956,  \mnras, {\bf 6}, 662-667, {\em Visual Horizons in World-Models}, Rindler acknowledged that faster than $c$ expansion is implicit in the mathematics, but expresses discomfort with the concept: ``\dots certain physical difficulties seem to be inherent in models possessing a particle-horizon: if the model postulates point-creation we have material particles initially separating at speeds exceeding those of photons.''}
\xcite{3}{McVittie, G.~C. 1974, \qjras, {\bf 15}, 246-263, {\em Distances and large redshifts}, Sect.~4, ``These fallacious arguments would apparently show that many quasars had `velocities of recession' greater than that of light, which contradicts one of the basic postulates of relativity theory.''}
\xcite{4}{Weinberg, S. 1977, {\em The First Three Minutes}, (New York: Bantum Books), p.~27, ``The conclusion generally drawn from this half century of observation is that the galaxies are receding from us, with speeds proportional to the distance (at least for speeds not too close to that of light).'', see also p.~12 and p.~25.  Weinberg makes a similar statement in his 1972 text {\em Gravitation and Cosmology} (New York: Wiley), p.~417, ``a {\em relatively close} galaxy will move away from or toward the Milky Way, with a radial velocity [$v_{\rm rec}=\dot{R}(t_0)\chi$].'' (emphasis ours).  Shortly thereafter he adds a caution about SR and distant sources: ``it is neither useful nor strictly correct to interpret the frequency shifts of light from very distant sources in terms of a special-relativistic D\"oppler shift alone. [The reader should be warned though, that astronomers conventionally report even large frequency shifts in terms of a recessional velocity, a ``red shift'' of $v$ km/sec meaning that $z=v/(3\times 10^5)$.]''}
\xcite{5}{Field, G. 1981, {\em This Special Galaxy}, in Section II of {\em Fire of life, the book of the Sun}, (Washington, DC: Smithsonian Books) ``The entire universe is only a fraction of a kilometer across [after the first millionth of a second], but it expands at huge speeds --- matter quite close to us being propelled at almost the speed of light.''}
\xcite{6}{Schutz, B.~F. 1985, {\em A first course in General Relativity}, (\cupadr: \cup) p.~320, ``[v=HD] cannot be exact since, for $D>1.2\times10^{26}m=4000$ Mpc, the velocity exceeds the velocity of light!  These objections are right on both counts.  Our discussion was a {\em local} one (applicable for recession velocity $<<1$) and took the point of view of a particular observer, ourselves.  Fortunately, the cosmological expansion is slow...''}
\xcite{7}{Peebles, P.~J.~E., Schramm, D.~N., Turner, E.~L. and Kron, R.~G. 1991, Nature {\bf 352}, 769, {\em The case for the relativistic hot Big Bang cosmology}, ``There are relativistic corrections [to Hubble's Law, $v=H_0D$,] when $v$ is comparable to the velocity of light $c$.''  However, Peebles, in his 1993 text {\em Principles of Physical Cosmology}, (Princeton: Princeton University Press), p.~98, explains: ``Since equation [$D=R\chi$] for the proper distance [$D$] between two objects is valid whatever the coordinate separation, we can apply it to a pair of galaxies with separation greater than the Hubble length... Here the rate of change of the proper separation, [$\dot{D}=HD$], is greater than the velocity of light.  This is not a violation of special relativity;''  Moreover, in the next paragraph Peebles makes it clear that, dependent upon the cosmological parameters, we can actually observe objects receding faster than the speed of light.}
\xcite{8}{Peacock, J.~A. 1999, {\em Cosmological Physics}, (\cupadr: \cup) p.~6, ``\dots objects at a vector distance {\bf r} appear to recede from us at a velocity ${\bf v}=H_0{\bf r}$, where $H_0$ is known as Hubble's constant (and is not constant at all as will become apparent later.)  This law is only strictly valid at small distances, of course, but it does tell us that objects with $r\simeq c/H_0$ recede at a speed approaching that of light.  This is why it seems reasonable to use this as an upper cutoff in the radial part of the above integral.''  However, Peacock makes it very clear that cosmological redshifts are not due to the special relativistic Doppler shift, p.~72, ``it is common but misleading to convert a large redshift to a recession velocity using the special-relativistic formula $1+z = [(1+v/c)/(1-v/c)]^{1/2}$.  Any such temptation should be avoided.'' }
%
\xcite{9}{Davies, P.~C.~W. 1978, {\em The Runaway Universe} (London: J. M. Dent \& Sons Ltd) p.~26, ``\dots galaxies several billion light years away seem to be increasing their separation from us at nearly the speed of light.  As we probe still farther into space the redshift grows without limit, and the galaxies seem to fade out and become black.  When the speed of recession reaches the speed of light we cannot see them at all, for no light can reach us from the region beyond which the expansion is faster than light itself.  This limit is called our horizon in space, and separates the regions of the universe of which we can know from the regions beyond about which no information is available, however powerful the instruments we use.''}
\xcite{10}{Berry, M. 1989, {\em Principles of Cosmology and Gravitation}, (Bristol, U.K.: IOP Publishing) p.~22 ``\dots if we assume that Euclidean geometry may be employed, \dots galaxies at a distance  $D_{max} = c/H \sim 2 \times 10^{10}$ light years $\sim 6\times 10^9$ pc are receding as fast as light.  Light from more distant galaxies can never reach us, so that $D_{max}$ marks the limit of the observable universe; it corresponds to a sort of {\em horizon}.''}
\xcite{11}{Raine, D.~J. 1981, {\em The Isotropic Universe}, (Bristol: Adam Hilber Ltd) p.~87, ``One might suspect special relativistic effects to be important since some quasars are observed to exhibit redshifts, $z$, in excess of unity.  This is incompatible with a Newtonian interpretation of the Doppler effect, since one would obtain velocities $v=cz$ in excess of that of light.  The special relativistic Doppler formula $1+z=[(c+v)/(c-v)]^{1/2}$ always leads to sub-luminal velocities for objects with arbitrarily large redshifts, and is at least consistent.  In fact we shall find that the strict special relativistic interpretation is also inadequate.  Nevertheless, at the theoretical edge of the visible Universe we expect at least in principle to see bodies apparently receding with the speed of light.''}
\xcite{12}{Liddle, A.~R. 1988, {\em An introduction to Modern Cosmology}, (Sussex: John Wiley \& Sons Ltd) p.~23, Sect 3.3, ``\dots ants which are far apart on the balloon could easily be moving apart at faster than two centimetres per second if the balloon is blown up fast enough.  But if they are, they will never get to tell each other about it, because the balloon is pulling them apart faster than they can move together, even at full speed.'' This is a very useful analogy, and this statement is strictly true.  However, to ensure the balloon is blown up ``fast enough'' we must ensure the expansion rate accelerates such that the balloon's radius increases at least exponentially with time.}
\xcite{13}{Krauss, L.~M. and Starkman, G.~D. 1999, \apj, 531(1), 22--30, {\em Life, the universe and nothing:  Life and death in an ever-expanding universe}, ``Equating this recession velocity to the speed of light $c$, one finds the physical distance to the so-called de Sitter horizon... This horizon, is a sphere enclosing a region, outside of which no new information can reach the observer at the center''.  This would be true if only applied to empty universes with a cosmological constant - de Sitter universes.  However this is not its usage: ``the universe became $\Lambda$-dominated at about 1/2 its present age.  The `in principle' observable region of the Universe has been shrinking ever since.  ... Objects more distant than the de Sitter horizon [Hubble Sphere] now will forever remain unobservable.''}
\xcite{14}{Harrison, E.~R. 1991, \apj, 383, 60--65, {\em Hubble spheres and particle horizons}, ``All accelerating universes, including universes having only a limited period of acceleration, have the property that galaxies at distances $L<L_H$ are later at $L>L_H$, and their subluminal recession in the course of time becomes superluminal.  Light emitted outside the Hubble sphere and traveling through space toward the observer recedes and can never enter the Hubble sphere and approach the observer.  Clearly, there are events that can never be observed, and such universes have event horizons.'' The misleading part of this quote is subtle -- there will be an event horizon in such universes (accelerating universes), but it need not coincide with the Hubble sphere.  Unless the universe is accelerating so quickly that the Hubble sphere does not expand (exponential expansion) we will still observe things from beyond the Hubble sphere, even though there is an event horizon (see Fig.~\ref{fig:dist}).}
\xcite{15}{Harwit, M. 1998, {\em Astrophysical Concepts}, 3rd Ed., (New York: Springer-Verlag) p.~467, ``Statement (i) [In a model without an event horizon, a fundamental observer can sooner or later observe any event.] depends on the inability of particles to recede at a speed greater than light when no event horizon exists.''}
\xcite{16}{Hubble, E. and Humason, M.~L. 1931, \apj, 74, 443--480, {\em The Velocity-Distance relation among Extra-Galactic Nebulae}, pp.~71--72, ``If an actual velocity of recession is involved, an additional increment, equal to that given above, must be included in order to account for the difference in the rates at which the quanta leave the source and reach the observer$^1$.'', Footnote 1: ``The factor is $\sqrt{\frac{1+v/c}{1-v/c}}$ which closely approximates $1+d\lambda/\lambda$ for red-shifts as large as have been observed.  A third effect due to curvature, negligible for distances observable at present, is discussed by R. C. Tolman...''}
\xcite{17}{Lightman, A.~P., Press, W.~H., Price, R.~H. and Teukolsky, S.~A. 1975, {\em Problem book in relativity and gravitation}, (\pupadr: \pup) Prob.~19.7}
\xcite{18}{Halliday, D., Resnick, R. and Walker, J. 1974, {\em Fundamentals of Physics}, (USA: John Wiley \& Sons) 4th Ed., Question 34E, ``Some of the familiar hydrogen lines appear in the spectrum of quasar 3C9, but they are shifted so far toward the red that their wavelengths are observed to be three times as large as those observed for hydrogen atoms at rest in the laboratory.  (a) Show that the classical Doppler equation gives a relative velocity of recession greater than $c$ for this situation.  (b) Assuming that the relative motion of 3C9 and the Earth is due entirely to recession, find the recession speed that is predicted by the relativistic Doppler equation.'' See also Questions 28E, 29E and 33E.}
\xcite{19}{Seeds, M.~A. 1985, {\em Horizons - Exploring the Universe}, (Belmont, California: Wadsworth Publishing) pp.~386--387, ``If we use the classical Doppler formula, a red shift greater than 1 implies a velocity greater than the speed of light.  However, as explained in Box 17-1, very large red shifts require that we use the relativistic Doppler formula.... {\em Example:} Suppose a quasar has a red shift of 2.  What is its velocity?  {\em Solution:} [uses special relativity]''}
\xcite{20}{Kaufmann, W.~J. and Freedman, R.~A. 1988, {\em Universe}, (New York: W.~H.~Freeman \& Co.) Box 27-1, p.~675, ``\dots quasar PKS2000-330 has a redshift of $z=3.78$.  Using this value and applying the full, relativistic equation to find the radial velocity for the quasar, we obtain $v/c=(4.78^2-1)/(4.78^2+1)=...=0.92$.  In other words, this quasar appears to be receding from us at 92\% of the speed of light.''}
\xcite{21}{Taylor, E.~F. and Wheeler, J.~A. 1991, {\em Spacetime Physics: introduction to special relativity}, (New York: W.~H.~Freeman \& Co.) p.~264, Ex.~8-23}
\xcite{22}{Hu, Y., Turner, M.~S. and Weinberg, E.~J. 1993, \prd, 49(8), 3830--3836, {\em Dynamical solutions to the horizon and flatness problems}, ``\dots many viable implementations of inflation now exist.  All involve two key elements:  a period of superluminal expansion..." They define superluminal expansion later in the paper as ``Superluminal expansion might be most naturally defined as that where any two comoving points eventually lose causal contact.''  Their usage of superluminal conforms to this definition, and as long as the reader is familiar with this definition there is no problem.  Nevertheless, we should use this definition with caution because even if the recession velocity between two points is $\dot{D}>c$ this does not mean those points will eventually lose causal contact.}
\xcite{23}{Khoury, J., Ovrut, B.~A., Steinhardt, P.~J. and Turok, N. 2001, \prd, 64(12), 123522, {\em Ekpyrotic universe: Colliding branes and the origin of the hot big bang}, ``The central assumption of any inflationary model is that the universe underwent a period of superluminal expansion early in its history before settling into a radiation-dominated evolution.''  They evidently use a definition of `superluminal' that is common in inflationary discussions, but again, a definition that should be used with caution.}
\xcite{24}{Lovell, B. 1981, {\em Emerging Cosmology}, (New York: Columbia U. Press) p.~158, ``\dots observations with contemporary astronomical instruments transfer us to regions of the universe where the concept of distance loses meaning and significance, and the extent to which these penetrations reveal the past history of the universe becomes a matter of more sublime importance''}
\vspace{-1mm}
\xcite{25}{McVittie, G.~C. 1974, \qjras, 15(1), 246--263, {\em Distance and large redshifts},``\dots conclusions derived from the assertion that this or that object is `moving with the speed of light' relative to an observer must be treated with caution.  They have meaning only if the distance and time used are first carefully defined and it is also demonstrated that the velocity so achieved has physical significance.''}
\vspace{-1mm}
\xcite{26}{Bowers, R.~L. and Deeming, T. 1984, {\em Astrophysics II, Interstellar Matter and Galaxies}, (Jones and Bartlett Publishers, Inc.), Sect.~26.5, p.~478, ``Unfortunately, although the concept of distance is relatively trivial in ordinary experience, particularly in flat space, where the velocity of light is, to all practical purposes, infinite, distance is much harder to handle in cosmology....  The answer is that (a) any of these definitions of distance is adequate, but (b) they do not all give the same result for the distance of a galaxy, and therefore (c) the concept of a unique distance that has any absolute physical meaning must be abandoned''}
}

\clearemptydoublepage

\addcontentsline{toc}{chapter}{References}
\bibliography{all}
\bibliographystyle{thesis-tam}
\clearemptydoublepage

\end{document}